\documentclass[12pt]{report}

\usepackage{a4,graphics,doublespace,epsfig,float,latexsym}

\usepackage{axodraw}

\usepackage{subfigure}

\usepackage{amssymb}

\usepackage{amsbsy}

\makeindex 

\begin{document} 
\parskip 4mm 
\parindent 0mm


%
%
\pagestyle{empty} 
\begin{center} 
    {\LARGE UNIVERSITY OF SOUTHAMPTON} \\
\vspace{4cm} 
    {\Huge{\bf The  }} \vspace{12pt} \\ 
    {\Huge{\bf Local Potential Approximation }} \vspace{12pt} \\
    {\Huge{\bf of the }} \vspace{12pt} \\
    {\Huge{\bf Renormalization Group}} \vspace{12pt} \\ 
\vspace{1.5cm} 
    by \\
\vspace{1.5cm} 
    {\LARGE Christopher Simon Francis Harvey-Fros} \\
\vspace{2.0cm} 
    A thesis submitted for the degree of \\
\bigskip 
    Doctor of Philosophy \\
\vspace{1cm} 
\bigskip 
    Department of Physics and Astronomy \\
\bigskip 
March 1999
\end{center} 
\vspace{5cm}
This thesis was submitted for examination in March 1999. It does not
necessarily represent the final form of the thesis as  deposited in the
University after examination.
%

%
\newpage 
\pagestyle{empty} 
\begin{center} 
\vspace*{8cm} 
\hspace{-1cm}
\emph{Dedicated to my family}\\
\hspace{1cm}
\end{center} 
%
%
\setstretch{1.5}
\newpage 
\pagestyle{empty} 
\begin{center} 
      {\Large UNIVERSITY OF SOUTHAMPTON}  \\
\bigskip 
      \underline{\large ABSTRACT} \\
\bigskip 
      {\Large FACULTY OF SCIENCE} \\
\bigskip 
      {\Large PHYSICS} \\
\bigskip 
      \underline{\large Doctor of Philosophy} \\
\bigskip 
      {\Large The Local Potential Approximation } \\
      {\Large of the Renormalization Group} \\
\bigskip 
      {\large Christopher Simon Francis Harvey-Fros} \\
\end{center} 
We introduce Wilson's, or Polchinski's, exact renormalization group, and review
the Local Potential Approximation as applied to scalar field theory. Focusing
on the 
Polchinski flow equation, standard methods are investigated, and by choosing
restrictions to some sub-manifold of coupling constant space we arrive at a
very promising variational approximation method. Within the Local Potential  
Approximation, we construct a function, $C$, of the coupling constants;
it has the 
property that (for unitary theories) it decreases monotonically along flows
and is stationary only at fixed points - where it `counts degrees of freedom',
\emph{i.e.} is extensive, counting one for each Gaussian scalar.

In the latter part of the thesis, the Local Potential Approximation is used to
derive a non-trivial 
Polchinski flow equation to include Fermi fields. Our flow equation does not
support chirally invariant solutions and does not reproduce
the features associated with the corresponding invariant
theories. We solve both for a finite number of components, $N$, and within the
large $N$ limit. The Legendre flow equation provides a comparison with exact
results in the large $N$ limit. In this limit, it is solved to yield both
chirally invariant and non-invariant solutions.
%
%
\newpage 
\pagenumbering{roman} 
\pagestyle{plain} 
\tableofcontents 
\newpage 
\listoffigures 
\newpage 
\listoftables 
%
%
\newpage 
\chapter*{Preface} 
\addcontentsline{toc}{chapter} 
{\protect\numberline{Preface\hspace{-96pt}}} 
Original work appears in the latter parts of chapters two and three and
throughout chapter four. Some of the work presented has been published in

J.Generowicz, C.Harvey-Fros and T.R. Morris, Phys. Lett. \textbf{B407}, (1997)
27.

Other work awaits publication.
%
%
\newpage 
\chapter*{Acknowledgements} 
\addcontentsline{toc}{chapter}  
                {\protect\numberline{Acknowledgements\hspace{-96pt}}} 
Foremost I would like to thank my immediate family, Yvonne, John and Natalie,
for incalculable support, both emotionally and financially. 

Additionally I feel
grateful for the tuition and guidance of my supervisor, Tim Morris, without
whom this thesis would never have reached completion. My thanks are also due
to my colleagues with whom I have worked, to varying degrees, during my three
years at Southampton. In particular, Scott Griffiths, Jenny Sanderson, Kevin
Anderson, Miguel Oliveira, Jacek Generowicz, Massimo Di Pierro, Victor Lesk
and Chris Dawson have all contributed to a friendly and lively atmosphere at
the group.

Outside work I acknowledge friends who have made life better during my time at
Southampton. These include Andrea, Mirella, Sandra, Sarah, Richard, Spencer,
Calum and Jonty among others. These acknowledgements would be incomplete
without a special mention of the lads, Paul and Kevin.

I feel grateful for the life of my son Julian and particularly thank his mother
Elisabeth.


\typeout{final: Numbering introductory pages sequentially}
\pagenumbering{arabic}


\renewcommand{\floatpagefraction}{0.8}
\renewcommand{\topfraction}{0.8}
\renewcommand{\bottomfraction}{0.8}
\renewcommand{\textfraction}{0.2}

\chapter{Introduction}

\section{Motivation and Structure}

The purist aspires to a reliable method of performing non-perturbative
calculations in quantum field theory whereas the practically minded physicist
desires an efficient cost effective procedure of accurately generating
physical quantities. We will argue that the methods presented here offer us
hope of attaining these desires and aspirations. The strength of the scheme
discussed exhibits itself in two important features. First the method can be
improved in a systematic way producing a sequence of converging
approximations and second, there is no small parameter to control the
approximation (unlike perturbation theory), allowing us to reach and test new
and interesting areas of physics. In particular we demonstrate that the method
produces competitive results in the context of scalar field theory, both
numerically and theoretically via critical exponents and Zamalodchikov's
$C$-theorem respectively. Many authors use these methods for alternative
reasons, \emph{e.g.} to calculate effective potentials. The successes
presented here motivate our attempts to extend the formalism to include
Fermions. 

We begin the thesis by introducing the essential concepts of the
renormalization group methods from a field theory perspective. The inevitable
parallels with theories of critical phenomena are drawn with the usual lengthy
and detailed discussions left to reference [1-5]. Then the methods applied to
scalar field theory are reviewed and compared, paying particular attention to
a $Z_{2}$-invariant one component field theory in three
dimensions. We develop a new variational approach which is applied to
the scalar field theory under consideration and compared to the more standard
techniques. We give a brief introduction to conformal field theory and
review Zamalodchikov's $C$-theorem. Then, completing the defence of the Local
Potential Approximation, we show how an appropriate generalisation
of Zamalodchikov's $C$-function can be found within this scheme. The thesis
finishes with an outline of progress made and problems associated with
attempts to extend the work beyond purely scalar field theory. In particular
we compare the Local Potential Approximation in the context of Fermionic field
theories with exact results obtained in the large $N$ limit. This leads us to
an understanding of the corresponding results at finite $N$. We will work in
Euclidean space throughout, unless stated otherwise. 

\section{Renormalization}

Prior to the main discussion of the thesis we wish to stress the modern point
of view of renormalization as distinct to that presented in older
textbooks. Here, we define renormalization as the procedure of
re-expressing the parameters which define a problem in terms of some other,
perhaps simpler set, while keeping unchanged those physical aspects of the
problem which are of interest [4].

\section{The Ising model}

A classical theory of spin-spin coupling in a ferromagnet can be
described by the Hamiltonian,
\begin{equation}
H(s,h)=-\sum_{n,m} V_{n,m}s_{n}.s_{m} + h\sum_{n}s_{n}
\end{equation}
where $h$ is the external magnetic field and $V_{n,m}$ is the coupling of
spins $s_{n}$ and $s_{m}$ at sites $x_{n}$ and $x_{m}$
respectively. Neighbouring sites in the lattice of spins are separated
by a distance $a$. If we assume the lattice is $D$-dimensional hyper-cubic and
we only have nearest neighbour interactions, we can write
\begin{equation}
-\sum_{n,m} V_{n,m} s_{n}.s_{m}=K\sum_{n} \left( \sum_{\mu}
(s_{n+\mu}-s_{n})^{2}-2Ds_{n}^{2} \right)
\end{equation}
where $\mu=1,\cdots,D$. Within the Ising model $s$ is a scalar which takes the
values $s=\pm 1$. An effective\footnote{we replace the discrete spins by
continuous variables which by universality will produce the critical behaviour
of the Ising model} description of the partition function is given by 
\begin{equation}
Z(h,\beta )=\int \prod_{n} ds_{n}\rho (s_{n})e^{-\beta H(s,h)}
\end{equation}
where $\beta=\frac{1}{K_{B}T}$ and $\rho (s_{n})$ is a weight factor
describing the averaged local, microscopic properties of the spin. 
A choice like 
\begin{equation}
\rho (s_{n})\propto e^{-cs_{n}^{2}-\lambda s_{n}^{4}}
\end{equation}
gives a convenient description. The partition function (1.3) may now be
written as
\begin{equation}
Z(K,\nu,\lambda,h)=\int\prod_{n} ds_{n}e^{-H(s;K,\nu,\lambda,h)}
\end{equation}
where the effective Hamiltonian is given by
\begin{equation}
H(s;K,\nu,\lambda,h)=\sum_{n}\left( K(\beta)\sum_{\mu}(s_{n+\mu}-s_{n})^{2} +
\nu(\beta)s_{n}^{2} + \lambda (\beta )s_{n}^{4}+hs_{n}\right)
\end{equation}
and $\nu(\beta)=c(\beta)-2D$. Consider the potential
$\nu(\beta)s^{2}+\lambda(\beta)s^{4}$ (figure 1.1). A ferromagnetic transition
(symmetry breaking) occurs at $\nu(\beta_{c})=0$. The value of $T$, for which
$\nu(\beta_{c})=0$, is called the critical temperature, $T_{c}$.
\begin{figure}
\vspace{-133mm}
\hspace{-3mm}
\scalebox{0.67}{\includegraphics*[0pt,0pt][1000pt,900pt]{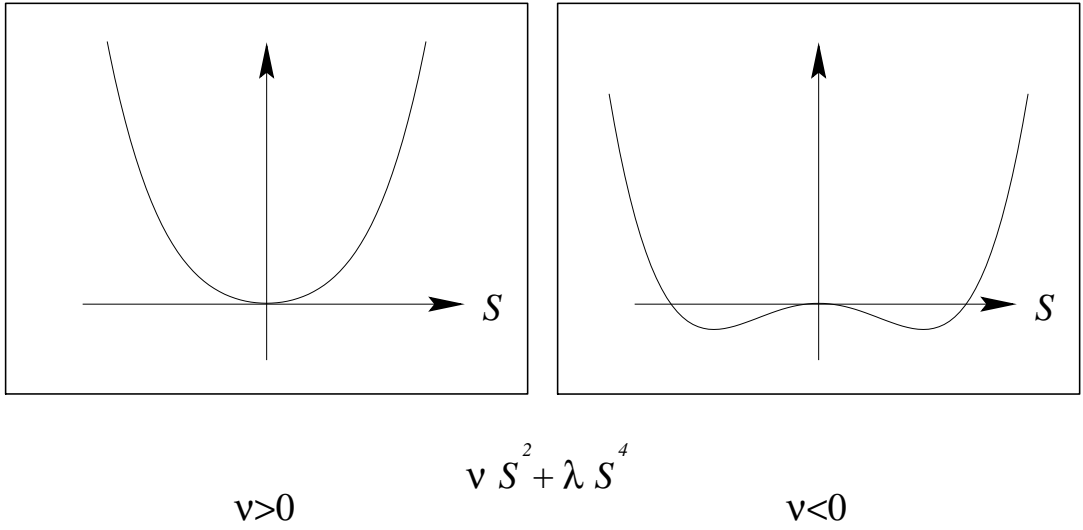}}\caption{Symmetry breaking}
\end{figure}

Minimising the effective Hamiltonian in (1.6) with $\nu(\beta)>0$ leads to a
ground state where all the $s_{n}=0$, leading to zero magnetisation,
\begin{equation}
<s>\equiv \frac{1}{V}\sum_{n}s_{n}=0.
\end{equation} 
However the ground state for $\nu(\beta)<0$ corresponds to all $s_{n}$ aligned
with $s_{n}=\sqrt{\frac{-\nu}{2\lambda}}$, leading to magnetisation,
\begin{equation}
<s>=\sqrt{\frac{-\nu}{2\lambda}}.
\end{equation}
The Ginzburg Landau theory of ferromagnetic transitions assumes that
$K(\beta)$, $\nu(\beta)$ and $\lambda(\beta)$ are smooth functions of
temperature. Thus we might write
\begin{equation}
\nu(\beta)\propto(\beta-\beta_{c})
\end{equation}
for $\beta\approx\beta_{c}$, leading to
\begin{equation}
<s>\propto\sqrt{\beta-\beta_{c}}
\end{equation}
for $\beta>\beta_{c}$. This shows the typical non-analytic behaviour at a phase
transition, as indicated at $\beta = \beta_{c}$ in figure 1.2. Near the
critical point, translational invariance implies that the correlation function
will behave as 
\begin{equation}
<(s_{n}-<s>)(s_{m}-<s>)>\sim |x_{n}-x_{m}|^{-(D-2+\eta)}
\end{equation}
for the range $a\ll |x_{n}-x_{m}|\ll \xi(h,\beta)$. The lack of any dependence
on a fundamental length scale reflects that at criticality the system is scale
invariant (to be discussed in greater depth later). The correlation length,
$\xi$, can be thought of as the size to which we can reduce the system without
qualitatively changing its properties. The anomalous dimension, $\eta$, is
effectively defined by (1.11). We will discuss this quantity in greater detail
later.
\begin{figure}
\vspace{-44mm}
\hspace{28mm}
\scalebox{0.35}{\includegraphics*[0pt,0pt][1000pt,900pt]{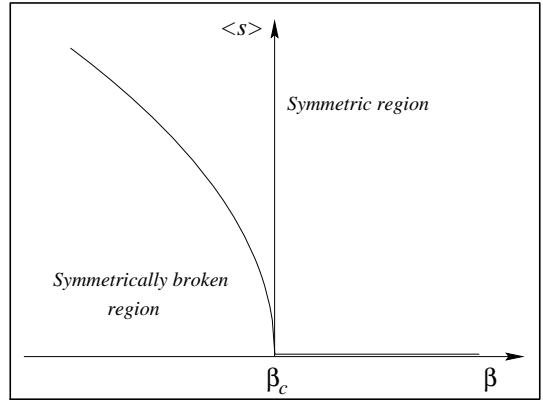}}
\caption{Magnetisation as a function of temperature for the Ising model with
zero external magnetic field}
\end{figure}

\section{Lattice Field theory}

Consider the Lagrangian for one component $\phi^{4}$ scalar field theory in
$D$-dimensional Euclidean space,
\begin{equation}
\mathcal{L}=\frac{1}{2}(\partial_{\mu}\phi)^{2}+\frac{1}{2}m^{2}\phi^{2} +
\frac{\lambda}{4!}\phi^{4}. 
\end{equation}
We choose to regularise the field theory by
discretising spacetime using a finite volume hypercubic lattice. This leads to
a direct comparison with the Ising model of a ferromagnet. We let
\begin{equation}
x \rightarrow x_{n} \equiv a n \hat{e}_{\mu},
\end{equation}
\begin{equation}
\phi (x) \rightarrow \phi_{n} \equiv \phi (x_{n}),
\end{equation}
\begin{equation}
\partial_{\mu}\phi \rightarrow \frac{1}{a}\left( \phi(x_{n}+\hat{e}_{\mu}) -
\phi(x_{n})\right)\equiv \frac{1}{a}\left(\phi_{n+\mu}-\phi_{n}\right)
\end{equation}
and the measure
\begin{equation}
\mathcal{D} \phi \rightarrow \prod^{V}_{n} d\phi_{n}
\end{equation}
where $\hat{e}_{\mu}$ denotes $D$ orthonormal vectors and $a$ is the lattice
spacing. 

By scaling the fields, sources and coupling constants, such that they are
dimensionless, 
\begin{equation}
\lambda=g^{2}a^{D-4},\ \ \phi'=ga^{\frac{1}{2}(D-2)}\phi,\ \
J'=ga^{\frac{1}{2}(D+2)}J,
\end{equation}
the discretised action can be written
\begin{equation}
S(\phi', J')=\sum_{n}\left(\frac{1}{2}\sum_{\mu}(\phi'_{n+\mu}-\phi'_{n})^{2} +
\frac{1}{2}m^{2}a^{2}\phi_{n}^{'2} + \frac{1}{4!}\phi_{n}^{'4} +
J'_{n}\phi'_{n}\right).
\end{equation}
If we compare this with the Hamiltonian for the spin system (1.6) we see that
at the classical (or Ginzburg Landau) level the 
continuum limit of the dimensionless lattice field theory ($a\rightarrow 0$)
corresponds to 
approaching the critical point of the ferromagnetic transition ($\nu(\beta)
\rightarrow 0$) because of the identification
\begin{equation}
m^{2}a^{2}\sim\nu(\beta).
\end{equation}
Furthermore, we associate the even symmetry of the lattice field theory
($\phi\rightarrow -\phi$)\footnote{later we will be interested in a
$Z_{2}$-invariant one component scalar field theory} with a `flipped spin' symmetry of the spin system
($s\rightarrow -s$). Thus the continuum field theory can be viewed as a
critical classical spin system. 
 
\section{Universality}

The association between the lattice $\phi^{4}$ theory and the Ising model may
appear contrived, however 
it is found that different physical systems form `universality classes', in
the sense that quantum field theories can be used to describe critical
phenomena for a range of models. Although there is no proof, there is
considerable evidence that $O(N)$ symmetric scalar field
theories correspond to various universality classes according to the value of
$N$ [1];

\underline{\textit{N=0:}} \textit{Critical behaviour of polymers.}
This is only formally defined in the limit $N \rightarrow 0$ as noted by
Gennes [6].

\underline{\textit{N=1:}} \textit{Liquid-vapour transition and alloy 
order-disorder transition.}
Observe that we have no internal symmetries and the liquid-vapour
transition can be modelled by particles living on a lattice, allowing the
occupation of each site to be either zero or unity (equivalent to the Ising
model) [7].

\underline{\textit{N=2:}} \textit{He$^{2}$ superfluid phase transition and 
planar ferromagnets.} 
The planar ferromagnet has symmetry in a plane corresponding to
an $O(2)$ invariant scalar field theory.

\underline{\textit{N=3:}} \textit{Ferromagnetic phase transitions.} Here we
observe a three dimensional symmetry corresponding to an $O(3)$ invariant
field theory.

\underline{\textit{N=4:}} \textit{Chiral phase transition.} It has been
postulated to correspond to the chiral phase transition for two flavours of
quarks.

Arguments presented in the following analysis provide a phenomenological
explanation of universality.

\section{Kadanoff blocking}

We now drop the distinction between the Ising model and the lattice field
theory and use a renormalization procedure known as Kadanoff blocking [8] to
reduce the number of degrees of freedom. Our toy model Hamiltonian will be
\begin{equation}
H=\sum_{n}\left(\sum_{\mu}\frac{1}{2}(\phi^{\phantom{2}}_{n+\mu} -
\phi^{\phantom{2}}_{n})_{\phantom{n}}^{2} + \mu\phi^{2}_{n} +
\lambda\phi^{4}_{n}\right).
\end{equation}
We begin by dividing the original lattice into
blocks of size $s^{D}$ where $s$ is
an integer, and define an average field in each block $B_{s}(n')$
labelled $n^{\prime}$ (figure 1.3) by
\begin{equation}
\phi^{\prime}_{n'} = \frac{1}{s^{D}}  \sum_{n \epsilon B_{s}(n^{\prime})}
\phi_{n}.
\end{equation}
We end by scaling the blocks back to the original size
\begin{equation}
x_{s} = \frac{x}{s}
\end{equation}
and
\begin{equation}
\phi_{s} = s^{\frac{1}{2} (D-2+\eta)} \phi^{\prime}
\end{equation}
to ensure that the short distance properties of the correlation functions
(physics) are left unchanged. This can be seen by considering the effect of
the block transformation on the correlation functions:
\begin{eqnarray}
<\phi^{\prime}_{n'}\phi^{\prime}_{0'}> &=& \frac{1}{s^{2D}}<\sum_{n\epsilon
B_{s}(n')} \phi_{n} \sum_{m\epsilon B_{s}(0')} \phi_{m}> \\
&\approx & <\phi_{sn'}\phi_{0}>
\end{eqnarray}
since all the $s^{2D}$ correlation functions in (1.24) are at essentially the
same distance if $n'$ is very large. Then, from (1.11) we see that 
\begin{equation}
<\phi_{n}\phi_{0}>\sim \frac{1}{n^{D-2+\eta}}
\end{equation}
and thus for the correlation functions to be left unchanged by blocking we
must simultaneously scale the fields as indicated by (1.23).

The new Hamiltonian, $H_{s}(\phi_{s})$, will not be identical to the original,
(1.20), but will contain infinitely many additional arbitrarily complicated
terms including for example
\begin{equation}
(\phi^{\phantom{2}}_{n+\mu} -
\phi^{\phantom{2}}_{n})_{\phantom{n}}^{2}\phi_{n}^{2}, 
\ \ \ \ \ \phi_{n+\mu}^{2}\phi_{n}^{4}\ \ \ \mathrm{and}
\ \ \ (\phi_{n+\mu}-2\phi_{n}+\phi_{n+\mu})^{2}.
\end{equation}
Thus, given that we will consider successive blockings, it is natural to start
with a completely general action including all terms which obey the symmetries
of the theory under consideration. In general these terms will be infinite in number. Then, each blocking  
can be viewed as a mapping of the
`coupling constant space' onto itself, called the renormalization group
transformation, denoted $T(s)$. We can see the group structure explicity
through $T(s_{1})T(s_{2})=T(s_{1}s_{2})$, although strictly speaking we only
have a semi-group, since there will not necessarily be an inverse
transformation. 
\begin{figure}
\vspace{-47mm}
\hspace{13mm}\scalebox{0.43}{\includegraphics*[0pt,0pt][900pt,900pt]{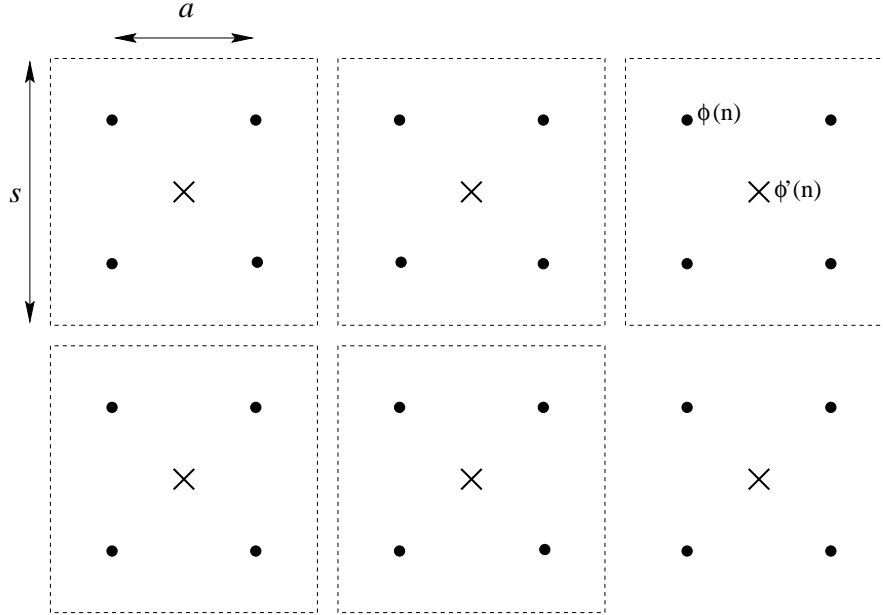}} 
\caption{Kadanoff Blocking}
\end{figure}

In terms of the ferromagnet we imagine looking at a sample through a
microscope such that our eyes can see spin variations down to a certain
size. Then Kadanoff blocking represents the operation of decreasing the
magnification of the microscope by the factor $s$. 

\section{Effective Lagrangians}

The lattice we discussed earlier is simply a method of regularisation which
corresponds (more or less) to introducing an overall momentum cutoff
($\Lambda_{o}\sim\frac{1}{a}$). Thus, Kadanoff blocking can 
be interpreted as integrating out high momentum modes\footnote{introducing a
secondary momentum cutoff, $\Lambda$} and scaling the cutoff
back to its original size, $\Lambda=\Lambda_{o}$, such that the physics remains
unchanged. However, it is simpler and equivalent to ensure that all variables
are `measured' in units of $\Lambda$, \emph{i.e.} we change variables to ones
that are 
dimensionless, by dividing by $\Lambda$ raised to the power of their scaling
dimensions. Then the action of lowering $\Lambda$ reproduces the scaling steps
in (1.22) and (1.23).

Here we are considering directly a continuum effective Lagrangian describing
physics over a limited range of energies, \emph{i.e.} momentum less than
$\Lambda_{o}$. The most famous example of an effective theory is the
Standard Model which produces good results below the scale of Grand
Unification, where new physics occurs. Throughout this thesis we will be
dealing primarily with the Legendre and Wilsonian (or Polchinski) effective
actions. Begin by considering a general effective action, $S^{eff}$, which we
assume to be valid below the overall momentum cutoff, $\Lambda_{o}$. Below
$\Lambda_{o}$ the effective Lagrangian will be described by an infinite set
of couplings. Then, to look at physics at some energy scale
$E \ll\Lambda_{o}$  we lower the cutoff to a scale $\Lambda\sim E$ and
allow the couplings to flow in order to keep the physics constant (partition
function invariant with respect to $\Lambda$). Then the action will evolve
according to a flow equation of the general form
\begin{equation}
\Lambda \frac{\partial S^{eff}}{\partial\Lambda}=f[S^{eff}].
\end{equation}
Conventionally we parametrise flow equations in terms of the `renormalization
time', defined by
\begin{equation}
t=\ln\left(\frac{\Lambda_{o}}{\Lambda}\right)
\end{equation}
so that $t$ runs from 0 to $\infty$ as $\Lambda$ runs from $\Lambda_{o}$ to
0. Generally we will be interested in solving in
the continuum limit, corresponding to $\Lambda_{o}\rightarrow\infty$, where we
require the solutions of interest to be insensitive to $\Lambda_{o}$.

The effective theory is said to be renormalizable if we can calculate physical
processes once we have determined a finite number of couplings, known as
relevant. Relevant couplings , denoted $g^{R}$, flow away from
some `fixed point' value, $g^{R}_{*}$. The other, irrelevant couplings,
$g^{I}$, flow toward some `fixed point' value, $g^{I}_{*}$. The
renormalization (semi-) group moves the action along the trajectories in the
coupling space. 

\section{Fixed points and Perturbations}  

If the bare action (at $\Lambda = \Lambda_{o}$) is chosen in the neighbourhood
of the fixed point in such a way that the relevant couplings are their fixed
point values ($g^{R}_{o} = g^{R}_{*}$) and the irrelevant couplings are close
to their fixed point values then the action will flow into that fixed point,
$S^{eff}_{*}$, defined by 
\begin{equation}
\Lambda \frac{\partial S^{eff}_{*}}{\partial \Lambda} = f[S^{eff}_{*}] = 0.
\end{equation}
A fixed point is associated with a critical surface in the coupling space
which contains all the bare actions which flow into that fixed point. Figure  
1.4 illustrates the neighbourhood of a fixed point with the critical surface
and renormalization group trajectory associated with it. In practice this
space is infinite dimensional. Similarly, if we begin with the
ferromagnet at its critical point, $T=T_{c}$, and decrease the magnification
sufficiently by Kadanoff blocking, eventually we shall observe no further
change (\emph{i.e.} we reach a fixed point). 

We enquire about the stability of the fixed points by perturbing about 
the fixed point solution. For example, later we will expand a dimensionless 
potential (for dimensionless scalar fields) as
\begin{equation}
V(\phi,t)=V_{*}(\phi) + \epsilon\sum_{n} f_{n}(\phi)e^{\lambda_{n}t}
\end{equation}
where $V_{*}$ is the fixed point solution under consideration and
the perturbation is considered for small $\epsilon$. We immediately
notice that $\lambda_{n} > 0$ are associated with relevant directions in the
coupling space, in which the perturbations grow with $t$ and similarly
$\lambda_{n} < 0$ are associated with irrelevant directions. Fixed points are
characterised by the $\lambda_{n}$ associated with them. If $\lambda_{n} = 0$
we can to first order add any amount of the corresponding `marginal
perturbation' to the fixed point without introducing any $t$ dependence. Such
a scenario can result in lines of fixed points, and does in a few specific
cases to be discussed. 

Substituting (1.29) into (1.31) we find that the dimensionless perturbation
around a fixed point is given by
\begin{equation}
\delta V=\epsilon\sum_{n}f_{n}(\phi)
\left(\frac{\Lambda_{o}}{\Lambda}\right)^{\lambda_{n}}.
\end{equation}
However, more generally the flow equation (1.28) implies that we can write
\begin{equation}
\Lambda\frac{\partial}{\partial\Lambda}\delta V_{\Lambda}=-L(\delta
V_{\Lambda})
\end{equation}
where $L$ is a linear operator acting on $\delta V_{\Lambda}$. Thus from
(1.32) we see that the $\lambda_{n}$ form a discrete set of eigenvalues for
$L$, where the $f_{n}(\phi)$ are the corresponding eigenfunctions.
\begin{figure}
\vspace{-8mm}
\hspace{20mm}\scalebox{0.4}{\includegraphics*[0pt,0pt][700pt,700pt]{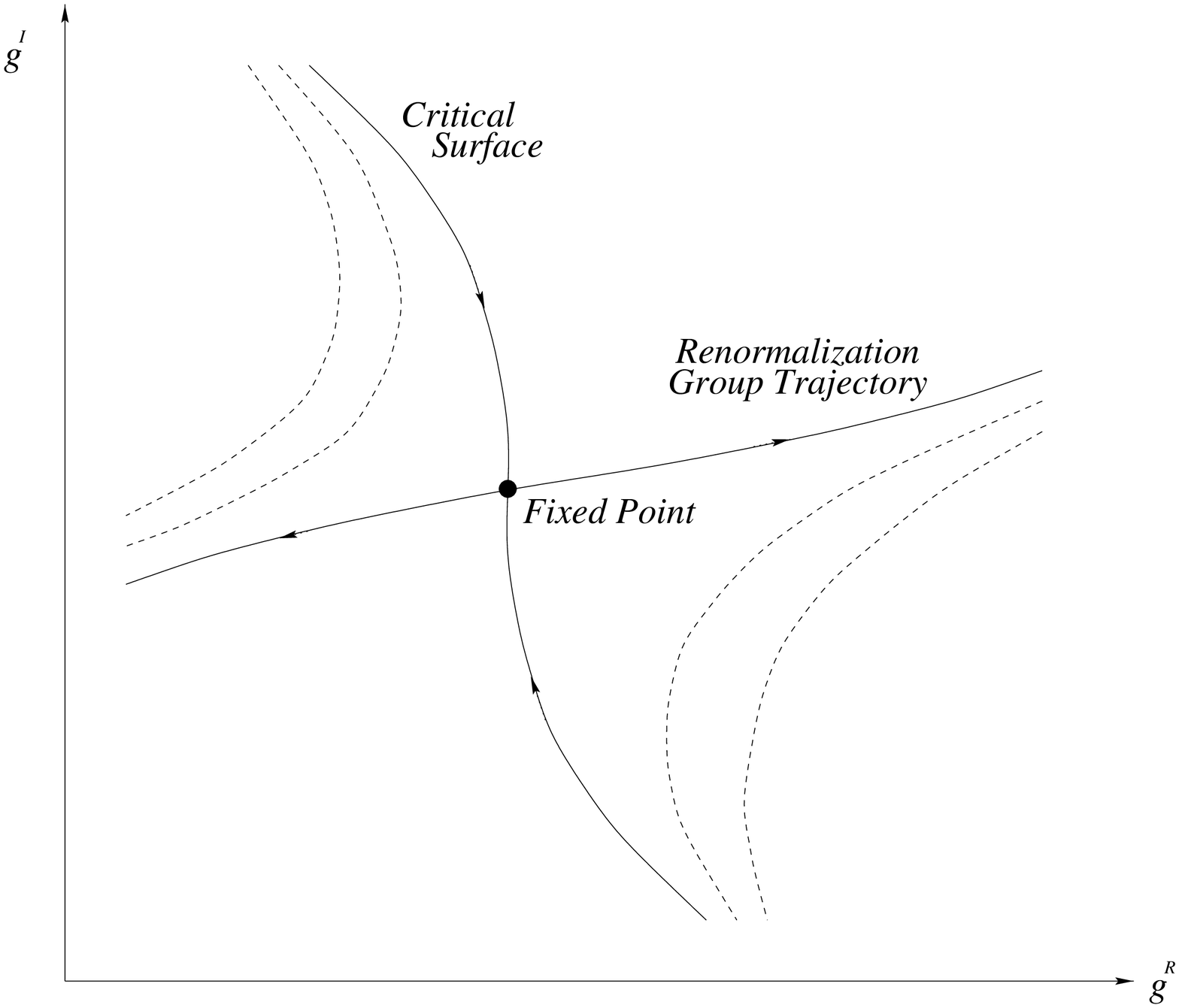}} 
\caption{A two dimensional projection of the flow near a fixed point}
\end{figure}

Since we exchange all dimensionful variables for dimensionless ones using
$\Lambda$, independence of $\Lambda$ would imply lack of dependence on any
scale at all. Thus a fixed point, defined by (1.30), corresponds to a scale
invariant theory and thus is associated 
with either divergent or vanishing correlation lengths. Hence a fixed point is
massless (or infinitely massive) respectively. A massive theory
leads to no propagation and is thus considered less
interesting\footnote{such a fixed point is known as non-critical or high
temperature}. Later, we will discover a trivial fixed point known as the
Gaussian  which represents a massless quantum field theory. The value of the
coefficient 
of $\phi^{2}$ will generally be non-zero at a massless fixed point (with the
exception of the Gaussian), but will be
precisely the required `counter-term' to cancel the remaining quantum
corrections, leaving the theory overall massless. Furthermore, the massless
fixed point must correspond to the continuum limit
($\Lambda_{0}\rightarrow\infty$), for otherwise $\Lambda_{o}$ would set the
scale. 

Now we are in a position to make the concept of universality more
concrete. Universality classes are characterised by their eigenvalues,
$\lambda_{n}$,  
(and more generally by the dimensions of the operator spectrum and their
symmetries) and thus the eigenvalues calculated from field theory may, at
least in principle, be 
compared with measurements taken in the laboratory. We interpret universality
as arising due to the fact that near the critical surface the correlation
length becomes large. As a result the behaviour of the system under
consideration is to a large extent unaffected by the detailed structure of the
model under consideration [9]. Thus models `converge' in the critical domain
leading to universality. 

\section{Critical exponents}

Within the Local Potential Approximation we will use the renormalization group
to compute experimentally measurable critical exponents, which 
are related to the $\lambda_{n}$. Specifically, we define for the
largest positive 
$\lambda$ 
\begin{equation}
\nu=\frac{1}{\lambda},
\end{equation}
and for the negative $\lambda$ closest to zero
\begin{equation}
\omega=|\lambda |.
\end{equation}

To calculate these critical exponents we need to consider the behaviour near
the fixed point. Suppose we are near a fixed point but not on the critical
surface. The flow of the action will be dominated by the largest relevant
eigenvalue, and thus by (1.32) we can write
\begin{equation}
\delta V_{\Lambda}\sim\left(\frac{\Lambda_{o}}{\Lambda}\right)^{\lambda_{L}}
f_{L}(\phi)
\end{equation}
where $\lambda_{L}$ is the largest eigenvalue and $f_{L}(\phi)$ is the
eigenperturbation associated with it. We see that $f_{L}(\phi)$ has scaling
dimension $-\lambda_{L}$ and is associated with a coupling of dimension
$\lambda_{L}$\footnote{recall that $\Lambda$ has been used to trade
dimensionful couplings for dimensionless ones}. Now consider the situation at
$\Lambda=\Lambda_{o}$. There will 
be an eigenperturbation corresponding to the deviation from criticality. This
will be the dominant eigenperturbation determining the statistical state of
the system. Therefore, we associate this operator with $f_{L}(\phi)$:
\begin{equation}
\delta V_{\Lambda_{o}} = (T-T_{c})f_{L}(\phi)+O((T-T_{c})^{2}).
\end{equation}
As we know the coefficient of $f_{L}(\phi)$ has scaling dimension
$\lambda_{L}$, we see that the scaling of the temperature difference is also
$\lambda_{L}$. Therefore as the correlation length $\xi$ has scaling dimension
$-1$, we must have that
\begin{eqnarray}
\xi & \sim & (T-T_{c})^{-\frac{1}{\lambda_{L}}} \\
    & = & (T-T_{c})^{-\nu}.
\end{eqnarray}
Corrections to this scaling behaviour are given by the subleading exponents,
and are dominated by $\omega$. Specifically
\begin{equation} 
\xi \sim (T-T_{c})^{-\nu}+a_{\xi}(T-T_{c})^{(\omega -1)\nu}.
\end{equation}

Strictly speaking this is only pertinent to a critical point with one relevant
direction. With more than one relevant eigenvalue one normally considers a
universal scaling function of the relevant parameters and then again the least
negative eigenvalue quantifies the first corrections to scaling.

\section{$\beta$-functions}

Let us remark that the flow equation in (1.28) can be rewritten in terms of a
complete set of parameters $g^{i}$ (\emph{a.k.a.} the coupling constants). In
general (away from a continuum limit) this will require an infinite number of
couplings. Then (1.28) may be rewritten in terms of the $\beta$-functions
defined as
\begin{equation}
\beta^{i}(g)=\frac{d g^{i}}{d t}.
\end{equation}
By the definition (1.30), the $\beta$-functions vanish at a fixed point where
$g^{i}_{\phantom{*}}=g^{i}_{*}$ for all $i$. Later
we will perform this reformulation within an approximation. 

\section{Anomalous dimension}

As described it is convenient to rewrite the flow equations in terms of
dimensionless quantities, denoted here with a subscript $dim$. We do this by
scaling the couplings and fields by appropriate powers of $\Lambda$. Hence,
for classical scalar field theory we would let
\begin{equation}
\phi = \phi_{dim} \Lambda^{\frac{1}{2}(D-2)}
\end{equation}
such that the kinetic term in the action is overall dimensionless. However, to
take account of wave function renormalization in the corresponding quantum
theory we use the anomalous dimension, $\eta$, by letting
\begin{equation}
\phi = \phi_{dim} \Lambda^{\frac{1}{2}(D-2+\eta)}.
\end{equation}
Compare this with (1.23) to deduce that this is indeed the anomalous dimension
introduced by (1.11). Likewise we scale the other operators 
($\phi^{2i}$ for integer $i=0,1,\cdots$)\footnote{here we are
primarily interested in a $Z_{2}$-invariant theory and concentrate on
terms which maintain this symmetry} using a set of parameters
which we denote $\gamma_{i}$ which may be deduced from the renormalization
group (for example
$\phi^{2}\sim\phi^{2}_{dim}\Lambda^{D-2+\gamma_{1}}$). Similarly the
couplings, $g^{i}$, scale in an $\gamma_{i}$ dependent way leaving the
action dimensionless. Then considering the scaling behaviour of the couplings
with $\Lambda$ we deduce which correspond to relevant operators and irrelevant
operators. For example, the mass is found to scale as
\begin{equation}
m^{2}\sim m^{2}_{dim} \Lambda^{2-\gamma_{1}}
\end{equation}
which is clearly relevant provided $\gamma_{1} < 2$. In fact, in this way it
may be possible for 
classically relevant operators to become irrelevant and visa-versa when quantum
corrections are included. From (1.44) it
is clear that a fixed point with $\gamma_{1} < 2$ corresponds to choices of
couplings and interactions in the Lagrangian such that it produces a theory
which is massless, regardless of the value of the dimensionless
mass\footnote{recall that at a fixed point, the dimensionless mass
$m^{2}_{dim}$ is fixed under variations in $\Lambda$} ($m^{2}_{dim}$), whereas 
$\gamma_{1} > 2$ corresponds to an infinitely massive theory, and 
$\gamma_{1} = 2$ is marginal.

\section{The Momentum expansion}

In practice, exact flow equations prove very difficult to handle. Here we will
be interested in using the so called Local Potential Approximation (LPA
hereafter) as a method of deriving flow equations which can be solved with
relative ease. We consider the effective Lagrangian as an expansion in powers
of momentum and within the LPA truncate at the zeroth order, hence discarding 
all momentum dependence. For example, for scalar field theory we could write
\begin{equation}
S^{eff}=\int d^{D}x\left( V(\phi,t)+\frac{1}{2}(\partial_{\mu}\phi)^{2}\right)
\end{equation}
in $D$ dimensions. This approximation has been extensively investigated 
[10-13] and has produced some competitive results, as will be
discussed. Clearly as all quantum corrections to the kinetic term are
neglected, the LPA is equivalent to assuming vanishing anomalous dimension.  

By including successive momentum dependent terms it is expected that a
sequence of converging improvements over the LPA will be generated [14]
allowing for non-zero anomalous dimension. For example at first order we may
paramaterise $S^{eff}$ by
\begin{equation}
S^{eff} = \int d^{D}x \left(V(\phi,t)+\frac{1}{2}K(\phi,t)(\partial_{\mu}
\phi^{a})^{2}+\frac{1}{2}Z(\phi,t)(\phi^{a}\partial_{\mu}\phi^{a})^{2}\right)
\end{equation}
where $a=1,\cdots ,N$ for an $N$-component theory. This leads
to three partial differential flow equations for $V, K$ and $Z$. For one
component field theory we can simplify the effective action by eliminating $K$
and $Z$ in favour of a single quantity, $K + \phi^{2} Z$ [14]. We should note,
that in general the expansion (beyond LPA) is found 
to destroy `reparameterisation symmetry' [15], the requirement that physics
should not depend 
on the normalisation of $\phi$ (\emph{i.e.} $\phi \rightarrow \Omega \phi$ and
$J \rightarrow \frac{J}{\Omega}$). In some cases this problem can be
circumvented by a careful choice of cutoff.

\section{Discussion}

The title `Renormalization Group' is somewhat unfortunate because it may be
confused with the perturbative (Callen Symanzik) renormalization group that
appears in most textbook treatments of quantum field theory. It should be
stressed that more generally the renormalization group is merely a framework,
a set of ideas which may also
be applied to problems quite unrelated to field theory. Thus, Kadanoff blocking
simply represents the traditional introduction to renormalization from a field
theory perspective. However, whatever the motivation, these methods have the
common feature that they lead to mathematical equations describing
renormalization group flows in some complicated parameter space. It is the
study of these flows, and what information they yield about the system, which
is the essence of renormalization group theory. With
quantum field theory, the renormalization group methods become particularly
advantageous when we consider the continuum limit. We have found that a
fixed point of the flow equation corresponds to the continuum limit. Thus, by
solving for a fixed point we are solving directly in the continuum. 

In this chapter and throughout the rest of this thesis we will be primarily
interested 
in calculating critical exponents, which provide a convenient testing ground
for the Local Potential Approximation. However, the ultimate motivation for a
fixed point search runs much deeper. To make a quantum field theory well
defined, traditionally we impose an ultra-violet cutoff, $\Lambda_{o}$. Then,
the bare parameters in the Lagrangian are chosen functions of $\Lambda_{o}$,
such that as the cutoff is removed, we obtain
a theory with only finite mass scales. On the natural scale $\Lambda_{o}$
this means that all the mass scales must vanish as
$\Lambda_{o}\rightarrow\infty$. This in turn implies that the choice of bare
parameters must be such that it results in a fixed point (as discussed in
section 1.8). The Standard Model of particle physics is based entirely around
the Gaussian fixed point, which has problems. The scalar fields of the Higgs
sector may define a trivial theory about the Gaussian fixed point, in the
sense that as $\Lambda_{o}$ is removed (\emph{i.e.}
$\Lambda_{o}\rightarrow\infty$), all choices of bare couplings result in a 
free theory. Also, there has been no explanation for why the masses and
couplings take the values observed experimentally. One possibility is that the
Standard Model might be defined about a non-trivial fixed point. Such a fixed
point would have couplings at some non-zero values and would be an interacting
scale free theory (conformal field theory).

\chapter{Scalar field theory} 

In the following analysis we make more concrete some of the appealing
arguments and concepts presented in chapter one.
As discussed, the ultimate goal is to derive and solve flow equations for
non-trivial theories. However, it is useful to begin with the simplest scalar
field theory to develop techniques and understanding. As indicated, we will
use critical exponents as a testing ground for the approximation scheme under
consideration. In the first part of this chapter, we
investigate standard methods (\emph{i.e.} truncations and shooting) as applied
to a $Z_{2}$-invariant one component field theory. In particular we will be
interested in the non-trivial Wilson-Fisher fixed point found in three
dimensions, however some analysis between two and four dimensions is also 
presented. Subsequently, we develop a new variational procedure and compare
with the standard methods presented. The large $N$ limit of the Local
Potential Approximation is solved analytically and found to yield the
eigenvalue spectrum exactly. In the closing section we compare these results
with those due to other leading methods. 

\newpage

\section{The Wilsonian effective action}

We begin by making some
more precise statements about the Wilsonian effective action [16] discussed in
section 1.7. The generating functional
\begin{equation}
Z=\int_{0}^{\Lambda_{o}}\!\! \mathcal{D}\phi\ e^{-S_{bare}}
\end{equation}
is replaced with
\begin{equation}
Z=\int_{0}^{\Lambda}\!\! \mathcal{D}\phi\ e^{-S_{ren}}.
\end{equation}
(In practice, we ignore a multiplicative $\Lambda$-dependent factor which does
not contribute in any correlation function). We are then free to lower
$\Lambda$ as represented schematically in figure 2.1.
\begin{figure}
\vspace{-79mm}
\hspace{26mm}\scalebox{0.45}{\includegraphics*[0pt,0pt][1000pt,1000pt]{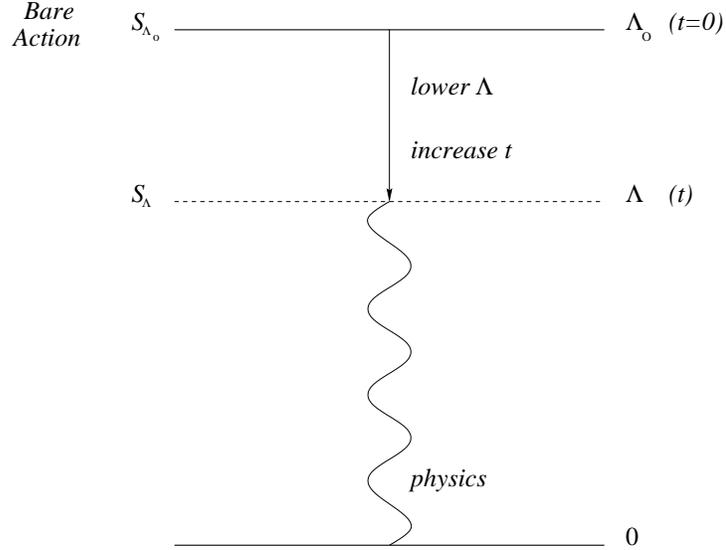}}
\caption{A schematic representation of the momentum cutoffs used}
\end{figure}

Consider an $N$-component scalar field theory with an ultra-violet (UV
hereafter) cutoff imposed using the kinetic term. We write the
$\Lambda$-dependent action as
\begin{equation}
S_{ren}=\frac{1}{2}\phi_{a}.\Delta^{-1}_{UV}.\phi_{a}+S_{\Lambda}
\end{equation}
where the dots have the usual interpretation as representing an integral over
position space\footnote{\emph{e.g.} $f.g=f_{x}g_{xy}=\int d^{D}xf(x)g(x,y)$
and $\mathrm{tr}(g)=g_{xx}=\int d^{D}xg(x,x)$} and
where $a$ runs from 1 to $N$, in momentum space
$\Delta_{UV}=\frac{C_{UV}}{q^{2}}$ and 
$C_{UV}(\frac{q^{2}}{\Lambda^{2}})$ is an UV cutoff function. We insist
$C_{UV}$ 
is a function of $\frac{q^{2}}{\Lambda^{2}}$ ensuring that it is dimensionless
and Lorentz invariant, and that it falls off sufficiently rapidly such that the
theory is efficiently regulated. Furthermore, we require that $C_{UV}$ is a
profile that 
acts as a UV cutoff, \emph{i.e.} $C_{UV}(0)=1$ and $C_{UV}\rightarrow 0$
for $q\rightarrow\infty$. It might take the form shown in figure 2.2,
however it sometimes proves convenient to choose $C_{UV}$ as a smooth
[14] (\emph{e.g.}, power law) function. 
\begin{figure}
\vspace{-56mm}
\hspace{25mm}\scalebox{0.35}{\includegraphics*[0pt,0pt][1000pt,1000pt]{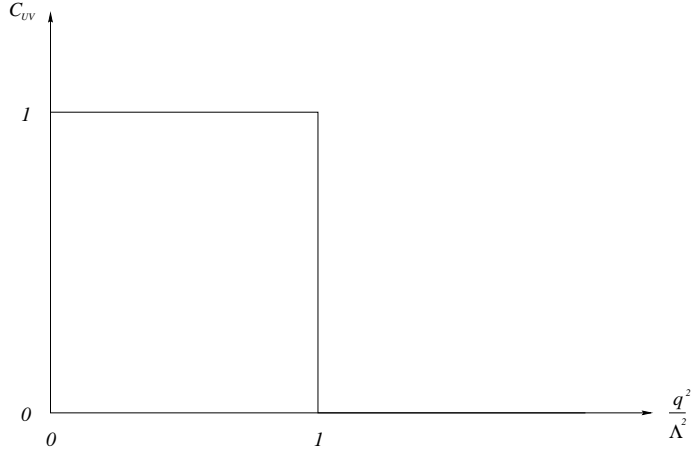}}
\caption{An ultra-violet cutoff function}
\end{figure}
Note that as $C_{UV}\rightarrow 0$ for $q>\Lambda$ in figure 2.2, $S_{ren}$ in
(2.3) diverges providing the required cutoff for the partition function in
(2.2). Similarly we introduce an infra-red (IR hereafter) cutoff by modifying
the propagator
$\frac{1}{q^{2}}$ to $\Delta_{IR}=\frac{C_{IR}}{q^{2}}$ where
$C_{IR}(\frac{q^{2}}{\Lambda^{2}})$ is a profile with the properties
$C_{IR}(0)=0$ and $C_{IR}\rightarrow 
1$ as $q\rightarrow\infty$. We require that the two cutoffs are related by
\begin{equation}
C_{IR}\left(\frac{q^{2}}{\Lambda^{2}}\right) +
C_{UV}\left(\frac{q^{2}}{\Lambda^{2}}\right) = 1.
\end{equation}

No source terms have been included in the generating functionals, (2.1) and
(2.2). If a source term, normally denoted $J$, is included in the bare action,
then the renormalization procedure will result in additional $J$ dependent
terms in the renormalised action. In this case the correlation functions,
given by 
\begin{equation}
\frac{1}{Z[0]} \frac{\delta}{\delta J(x)}
\frac{\delta}{\delta J(y)} Z[J]
\end{equation}
equal to $<\!\!\phi (x)\phi (y)\!\!>$ at the bare level
will be completely unaltered by the renormalization procedure.

\section{The Polchinski flow equation}

The Polchinski flow equation [17] governs the flow of $S_{\Lambda}$ under
varying $\Lambda$. Here we present a constructive proof following the methods
found in ref. [18].
The partition function for $N$ component scalar
field theory with propagator $\Delta^{-1}$, arbitrary bare action 
$S_{\Lambda_{o}}[\phi]$ and source $J$, can be written
\begin{equation}
Z[J]=\int\!\! \mathcal{D}\phi\ e^{-\frac{1}{2}\phi_{a} .\Delta^{-1} .\phi_{a} -
S_{\Lambda_{o}}+J_{a}.\phi_{a}}
\end{equation}
where the field $\phi$ has a flavour index $a$ which runs from 1 to
$N$. Consider the following integral which we will show reduces to
(2.6). These fields $\phi_{IR}$ and $\phi_{UV}$ are integrated over freely.
\begin{equation}
Z[J]=\int\!\! \mathcal{D}\phi^{\phantom{a}}_{IR}\mathcal{D}
\phi^{\phantom{a}}_{UV}\  
e^{-\frac{1}{2}\phi^{\phantom{-1}}_{UV} .\Delta^{-1}_{UV} 
.\phi^{\phantom{-1}}_{UV} - \frac{1}{2}\phi^{\phantom{-1}}_{IR}
.\Delta^{-1}_{IR}  
.\phi^{\phantom{-1}}_{IR} - S^{\phantom{-1}}_{\Lambda_{o}}
[\phi^{\phantom{-1}}_{IR} +  
\phi^{\phantom{-1}}_{UV}] + J.(\phi^{\phantom{-1}}_{IR} +
\phi^{\phantom{-1}}_{UV})} 
\end{equation}
where
\begin{equation}
\Delta (p)=\Delta^{\phantom{a}}_{IR}(p)+\Delta^{\phantom{a}}_{UV}(p)
\end{equation}
and
\begin{equation}
\phi_{a}=\phi_{IR}+\phi_{UV}.
\end{equation}
For simplicity of notation we have suppressed the flavour index $a$ on the
right side of (2.9). Physically, we interpret the $\phi^{\phantom{a}}_{IR}$
fields as the momentum modes higher than
$\Lambda$\footnote{i.e. $q^{2}>\Lambda^{2}$,} and the
$\phi^{\phantom{a}}_{UV}$ fields as the modes that are lower than 
$\Lambda$. By changing variables in (2.7), $\phi^{\phantom{-1}}_{IR}=\phi -
\phi_{UV}$ and $\phi^{\phantom{'}}_{UV}=\phi^{'}_{UV} +
\Delta_{UV}\Delta^{-1}.\phi$, and  performing a Gaussian integral we
reproduce the partition function in (2.6) up to a constant of proportionality.

Now consider integrating only over the higher modes:
\begin{equation}
Z_{\Lambda}[J,\phi^{\phantom{a}}_{UV}]=\int\!\! \mathcal{D}
\phi^{\phantom{a}}_{IR}\  
e^{-\frac{1}{2}\phi^{\phantom{-1}}_{IR}.\Delta _{IR}^{-1}.
\phi^{\phantom{-1}}_{IR} - S^{\phantom{-1}}_{\Lambda_{o}}
[\phi^{\phantom{-1}}_{IR} + 
\phi^{\phantom{-1}}_{UV}] + J.(\phi^{\phantom{-1}}_{IR} +
\phi^{\phantom{-1}}_{UV})}. 
\end{equation}
Using (2.9) to eliminate $\phi^{\phantom{a}}_{IR}$ in favour of $\phi$, it is
straightforward to show that $Z_{\Lambda}$ does not depend on both $J$ and
$\phi^{\phantom{a}}_{UV}$ independently in the sense that
\begin{equation}
Z_{\Lambda}[J,\phi^{\phantom{a}}_{UV}]=e^{\frac{1}{2}J.
\Delta^{\phantom{a}}_{IR}.J + J.\phi^{\phantom{a}}_{UV}- 
S^{\phantom{1}}_{\Lambda}
[\Delta^{\phantom{a}}_{IR}.J+\phi^{\phantom{a}}_{UV}]} 
\end{equation}
for some functional $S_{\Lambda}$ which depends only on the sum $\Delta_{IR}.J
+ \phi_{UV}$. If we restrict the support of $J$ to low
energy modes only, \emph{i.e.} we set $\Delta^{\phantom{a}}_{IR}.J =0$, (2.11)
simplifies. However from (2.10) and (2.7),
\begin{equation}
Z[J]=\int\!\! \mathcal{D}\phi^{\phantom{a}}_{UV}\
Z_{\Lambda}[J,\phi^{\phantom{a}}_{UV}] 
e^{-\frac{1}{2}\phi^{\phantom{-1}}_{UV}.\Delta^{-1}_{UV}.
\phi^{\phantom{-1}}_{UV}},
\end{equation}
and we see that $S_{\Lambda}$ is nothing more than the interaction part of the
Wilsonian effective action, as defined in (2.3). The exact flow equation
follows readily from the fact that (2.10) depends on $\Lambda$ only through
the $\phi^{\phantom{-1}}_{IR}.\Delta_{IR}^{-1}.\phi^{\phantom{-1}}_{IR}$
term. Thus, differentiating $Z_{\Lambda}$ with respect to $\Lambda$ we 
immediately obtain the flow equation for $Z_{\Lambda}$:
\begin{equation}
\frac{\partial}{\partial\Lambda}Z_{\Lambda}[\phi^{\phantom{a}}_{UV},J] =
-\frac{1}{2} \left(\frac{\delta}{\delta J} -
\phi^{\phantom{a}}_{UV}\right). \left(\frac{\partial}{\partial 
\Lambda} \Delta^{-1}_{IR}\right).\left(\frac{\delta}{\delta
J}-\phi^{\phantom{a}}_{UV}\right) Z_{\Lambda} .
\end{equation}
Substituting (2.11), performing a change of variables $\Phi =
\Delta_{IR}.J+\phi_{UV}$ and relabelling $\phi = \Phi$, yields Polchinski's
version of the Wilson flow equation [17]:
\begin{equation}
\frac{\partial S_{\Lambda}}{\partial \Lambda} = \frac{1}{2} \frac{\delta
S_{\Lambda}}{\delta \phi}.\frac{\partial \Delta_{UV}}{\partial \Lambda}. 
\frac{\delta S_{\Lambda}}{\delta \phi}-\frac{1}{2}\mathrm{tr} \left(
\frac{\partial \Delta_{UV}}{\partial \Lambda}.\frac{\delta^{2}S_{\Lambda}}
{\delta\phi\delta\phi} \right)
\end{equation}
where the trace represents an integral over position space. In terms of
the cutoff function, $C=C_{UV}$, this is written
\begin{equation}
\frac{\partial S_{\Lambda}}{\partial \Lambda} = \frac{1}{\Lambda^{3}}\left(
\mathrm{tr}\left(C'.\frac{\delta^{2}S_{\Lambda}}{\delta\phi_{a}\delta\phi_{a}}
\right)- \frac{\delta S_{\Lambda}}{\delta\phi_{a}}.C'.\frac{\delta
S_{\Lambda}}{\delta\phi_{a}} \right).
\end{equation} 
where we have reintroduced the flavour index.

\section{The Legendre flow equation}

The Legendre flow equation will prove particularly useful for us as it can be
solved 
exactly in the large $N$ limit. However, we will extensively use the Legendre
flow equation for Fermions and thus omit a calculation of the large $N$ limit
for scalar field theory\footnote{we will however demonstrate that the LPA
provides exact exponents in this limit}. Consider the action with kinetic term
modified to include an infra-red (IR) cutoff, as described in section 2.1.
Then the generating functional can be written 
\begin{equation}
Z[J]=\int\!\! \mathcal{D}\phi
\
e^{-\frac{1}{2}\phi^{\phantom{1}}_{a}.\Delta^{-1}_{IR}.
\phi^{\phantom{-}}_{a} - 
S^{\phantom{1}}_{\Lambda}+J^{\phantom{1}}_{a}.\phi^{\phantom{1}}_{a}}
\end{equation}
where $a$ runs from 1 to $N$ as before, but here $\Delta_{IR}=\frac{C_{IR}}
{q^2}$ where $C_{IR}$ might take the form shown in figure 2.3. 
\begin{figure}
\vspace{-56mm}
\hspace{25mm}\scalebox{0.35}{\includegraphics*[0pt,0pt][1000pt,1000pt]{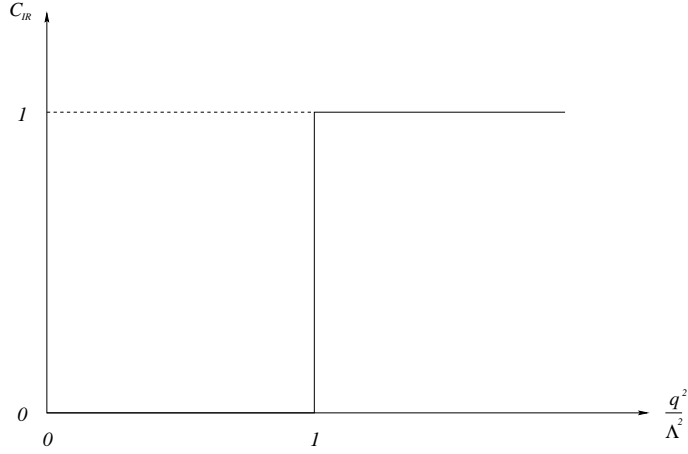}}
\caption{An infra-red cutoff function}
\end{figure}
We begin by defining the generating
functional of connected Greens functions by $W[J]=\ln(Z[J])$ and introducing
the Legendre effective action, $\Gamma_{\Lambda} [\phi^{c}]$, 
\begin{equation}
\Gamma_{\Lambda} [\phi^{c}] + \frac{1}{2}
\phi^{c}_{a}.\Delta^{-1}_{IR}.\phi^{c}_{a} = -W[J]+J_{a}.\phi^{c}_{a},
\end{equation}
where in the classical limit $h\rightarrow 0$, $\Gamma_{\Lambda}\rightarrow
S_{\Lambda}$. The classical field is defined as
\begin{equation}
\phi^{c}_{a}=\frac{\delta W}{\delta J_{a}},
\end{equation}
and corresponds to the expectation value of the field.
Differentiating (2.17) twice with respect to $\phi^{c}$, and then separately
twice with respect to $J$, we derive
\begin{equation}
\frac{\delta^{2}W}{\delta J_{a}\delta J_{b}} = 
\left(\frac{\delta^{2}\Gamma_{\Lambda}}
{\delta\phi^{c}_{a}\delta\phi^{c}_{b}} +
\delta_{ab}\Delta^{-1}_{IR}\right)^{-1}.
\end{equation}
We also differentiate (2.17) once with respect to $\Lambda$, to find
\begin{equation}
\left. \frac{\partial\Gamma_{\Lambda}}{\partial\Lambda}\right|_{\phi^{c}} +
\frac{1}{2}\phi^{c}_{a}.\frac{\partial \Delta^{-1}_{IR}}
{\partial\Lambda}.\phi^{c}_{a} = \left. -\frac{\partial W}
{\partial \Lambda}\right|_{J} .
\end{equation}
Finally we find an expression for the right side of (2.20) by differentiating 
the partition function with respect to $\Lambda$, 
\begin{equation}
\frac{\partial W}{\partial \Lambda} = -\frac{1}{2}\frac{\delta W}{\delta
J}.\frac{\partial \Delta^{-1}_{IR}}{\partial \Lambda}.\frac{\delta W}{\delta
J} - \frac{1}{2}\mathrm{tr}\left(\frac{\partial \Delta^{-1}_{IR}}{\partial
\Lambda}.\frac{\delta ^{2} W}{\delta J \delta J}\right)
\end{equation}
which then leads to the exact
Legendre flow equation upon substituting (2.19),
\begin{equation}
\frac{\partial\Gamma_{\Lambda}}{\partial\Lambda} = 
\frac{1}{2} \mathrm{tr} \left(\frac{\partial\Delta^{-1}_{IR}}
{\partial\Lambda}.(A^{-1})_{aa}\right)
\end{equation}
where
\begin{equation}
A_{ab}=\frac{\delta^{2}\Gamma_{\Lambda}}
{\delta\phi^{c}_{a}\delta\phi^{c}_{b}} + \delta_{ab}\Delta^{-1}_{IR}.
\end{equation}
The equations are best appreciated graphically. We view (2.21) as depicted in
figure 2.4, which shows how the $n$-point functions $W_{n}(p_{1},\cdots
,p_{n})$ of $W_{\Lambda}$ evolve. The vertices are drawn as open circles,
whereas the black dots represent the two point function, $\frac{\partial
\Delta^{-1}_{IR}}{\partial\Lambda}$. In the case of a sharp cutoff these
represent a delta function restricting momenta to $q=\Lambda$. The Polchinski
flow equation (2.14) can be viewed in a similar manner.  
\begin{figure}
\vspace{-90mm}
\hspace{12mm}\scalebox{0.45}{\includegraphics*[0pt,0pt][1000pt,1000pt]{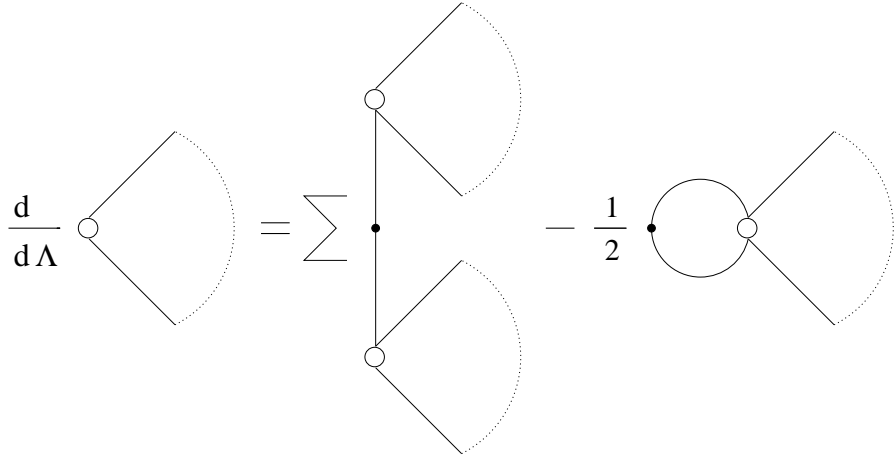}}
\caption{The flow equations for the vertices of the generating functional
$W_{\Lambda}$} 
\end{figure}
In fact the Polchinski and Legendre flow equations are intimately 
linked [18]. In particular the Wilsonian action behaves like the generator of
the connected diagrams and these effective actions are thus
found to be related by a generalised Legendre transform,
\begin{equation}
S_{\Lambda}[\phi_{a}] = \Gamma_{\Lambda}[\phi^{c}_{a}] +
\frac{1}{2}(\phi^{c}-\phi)_{a}.\Delta^{-1}_{IR}.(\phi^{c}-\phi)_{a}.
\end{equation}
Eliminating $\Gamma_{\Lambda}$ from (2.22) above, we retrieve the
Polchinski flow equation (2.14), if we impose (2.4). A more
detailed discussion of this relationship can be found in the literature [18].

\section{Application of the Local Potential Approximation}

The Local Potential Approximation (LPA) can be applied with considerable
success to either the Polchinski or Legendre flow equation. Here we choose to
focus our attention on the Polchinski flow equation [17]. Consider the flow
equation in 
terms of the cutoff function, $C$, given by (2.15). Here we 
impose the LPA by writing $S_{ren}$ as (1.45), thus
\begin{equation}
\frac{\delta S_{\Lambda}}{\delta \phi_{a}}.C'.\frac{\delta S_{\Lambda}}
{\delta \phi_{a}} = \int \!\! d^{D}x d^{D}y\ V'(\phi (x)) C'_{xy} V'(\phi (y)).
\end{equation}
Fourier transforming,
\begin{eqnarray}
C'_{xy} & = & \frac{1}{(2 \pi)^{D}}\! \int\!\! d^{D}q\  e^{iq(x-y)} C'
\left(\frac{q^{2}}{\Lambda^{2}}\right) \\ 
        & = &C'\left(-\frac{\partial^{2}}{\Lambda^{2}}\right)\delta(x-y). 
\end{eqnarray}
Considering $C'(-\frac{\partial^{2}}{\Lambda^{2}})$ as an expansion in
powers of $(-\frac{\partial^{2}}{\Lambda^{2}})$ we discard 
all but the leading term, $C'(0)$. Then we find that (2.25) can be written
\begin{equation}
\frac{\delta S_{\Lambda}}{\delta \phi_{a}}.C'.\frac{\delta S_{\Lambda}}
{\delta \phi_{a}} = C'(0)\! \int\!\! d^{D}x\ V'^{2}(\phi (x)).
\end{equation}
Later we will assume that $C'(0)<0$. In principle the $C'(0)$ could well be
taken to vanish however we take $C'(0)<0$ for the approximation to
proceed. This may be considered physically unreasonable however the flow
equation is ultimately independent of $C'(0)$ , suggesting that this
assumption may actually be sensible.
Similarly, we simplify the first term in (2.15) using (2.26) to find 
\begin{eqnarray}
\mathrm{tr} \left( C'.\frac{\delta^{2} S_{\Lambda}}{\delta \phi_{a}(x) \delta
\phi_{a}(y)} \right) & = & \int\!\! d^{D}x d^{D}y\ C'_{xy} 
\frac{\delta^{2} S_{\Lambda}}{\delta \phi_{a}(x) \delta \phi_{a}(y)} \\
                     & = & A\Lambda^{D}\! \int\!\! d^{D}x\ V''(\phi (x))
\end{eqnarray}
where we define the dimensionless $A$ through 
\begin{equation}
A\Lambda^{D} = \frac{1}{(2 \pi)^{D}}\! \int\!\! d^{D}q\ C'\left(\frac{q^{2}}
{\Lambda^{2}}\right) < 0.
\end{equation}
Then the `reduced Polchinski equation' reads
\begin{equation}
\frac{\partial V}{\partial \Lambda} = A\Lambda^{D-3} V'' -
\frac{C'(0)}{\Lambda^{3}} V'^{2}.
\end{equation}
Scaling to dimensionless quantities via $\phi \rightarrow \phi
\Lambda^{\frac{1}{2}(D-2)} \sqrt{|A|}$ (where as mentioned in section 1.12
$\eta=0$ in the LPA) and  
$V \rightarrow V \Lambda^{D} \frac{|A|}{|C'(0)|}$ and using renormalization
time as defined by (1.29) we finally arrive at a cutoff independent, and hence
scheme independent, flow equation, which will be the subject of our attention:
\begin{equation}
\frac{\partial V}{\partial t} = DV -\frac{1}{2}(D-2) \phi_{a} \frac{\partial V}
{\partial \phi_{a}} - \left( \frac{\partial V}{\partial \phi_{a}}\right)^{2}
+ \frac{\partial^{2} V}{\partial \phi_{a}^{2}}.
\end{equation}
The physical predictions (critical exponents) are independent of the
scaling\footnote{\emph{i.e.} under $\phi\rightarrow\mathcal{A}\phi$ and
$V\rightarrow\mathcal{B} V$ for arbitrary constants $\mathcal{A}$ and
$\mathcal{B}$} of $\phi$ and $V$, although the form of the potential will be
altered.

Given the definition of a fixed point as $V=V_{*}$ such that $\frac{\partial
V_{*}}{\partial t}=0$, it is immediately noticed that there are two trivial
solutions, the Gaussian and High Temperature fixed points which are given by
\begin{equation}
V^{G}_{*} = 0
\end{equation}
and
\begin{equation}
V^{HT}_{*} = \frac{1}{2} \phi^{2} -\frac{N}{D}
\end{equation}
respectively. We will shortly justify the name High Temperature for the latter.
In addition we will find at least one non-trivial fixed point
below four dimensions. From (2.33) we can deduce that with the exception of
the Gaussian solution, in the large $\phi$ regime, $V_{*}(\phi)\sim
\frac{1}{2}\phi^{2}$. 

To compute the critical exponents we perturb about the
fixed point solution as described in chapter one. We substitute $V(\phi,t)=
V_{*}(\phi)+\epsilon \delta V(\phi,t)$ into (2.33) and work to
$O(\epsilon)$. Then by separation of variables we find that the potential
takes the form of (1.31) as claimed and that the $f_{n}(\phi)$ satisfy
\begin{equation}
\frac{\partial^{2}f_{n}}{\partial \phi ^{2}_{a}} = (\lambda_{n} - D)f_{n}
+\frac{1}{2} (D-2) \phi_{a} \frac{\partial f_{n}}{\partial \phi_{a}} + 
2\frac{\partial V_{*}}{\partial \phi_{a}} \frac{\partial f_{n}}{\partial
\phi_{a}}. 
\end{equation}
For the trivial fixed points we can easily solve to
extract the $\lambda_{n}$. Restricting to $N=1$, we observe that the solutions
take the form of a polynomial,
$f_{n}(\phi)=\phi^{A}+B(A)\phi^{A-2}+C(A)\phi^{A-4}+\cdots$, with $A=2n$ for
integer $n$ such that $f_{n}(\phi )$ has no divergences. The $\lambda_{n}$ can
then be written in terms of $A$. In particular, for the trivial fixed points we
find
\begin{equation}
\lambda^{G}_{n} = D-n(D-2)
\end{equation}
and
\begin{equation}
\lambda^{HT}_{n} = D-n(D+2).
\end{equation}
Note that the eigenvalues for the Gaussian fixed point are the classical
dimension of the couplings. We observe, that for
the Gaussian fixed point in any 
dimension, the leading exponent defined by (1.34) is\footnote{this result is
in direct agreement with mean field theory 
[5]} $\nu=\frac{1}{2}$. However, for the High Temperature  
fixed point and for all $D>0$ we find that all the $\lambda_{n}<0$ implying
that there are just irrelevant 
directions around this fixed point. In addition, the exponents of
the High Temperature fixed point are independent of $N$ corresponding to
an infinitely massive theory with no propagation. 

\section{Flow and $\beta$-functions}

For illustration, we consider the simple case of the Gaussian fixed point
perturbed by the mass operator, for a single scalar field. Thus, using the
notation introduced in section 1.10, we set (for
$N=1$) $V(\phi,t)= g^{1}(t)+\frac{1}{2} g^{2}(t)\phi^{2}$. The
$\beta$-functions for $g^{1}$ and $g^{2}$ then follow easily by substitution in
(2.33): 
\begin{equation}
\beta^{1}(g)=\frac{\partial g^{1}}{\partial t}=Dg^{1}+g^{2}
\end{equation}
and
\begin{equation}
\beta^{2}(g)=\frac{\partial g^{2}}{\partial t}=2g^{2}(1-g^{2}).
\end{equation}
These can be solved for $g^{1}(t)$ and $g^{2}(t)$. As expected, at the
Gaussian ($g^{1}=g^{2}=0$) and the High Temperature ($g^{1}=-\frac{1}{D}$ and
$g^{2}=1$) fixed
points, the $\beta$-functions vanish. Now we can plot the $\beta$-function for
the mass (figure 2.5). We deduce that if we add a small mass
term to the Gaussian we will flow into the High Temperature fixed point. 
In other words the mass term corresponds to a relevant perturbation
for the Gaussian fixed point, but an irrelevant perturbation for the High 
Temperature fixed point.
\begin{figure}
\vspace{-17mm}
\hspace{27mm}\scalebox{0.345}{\includegraphics*[25pt,25pt][800pt,800pt]{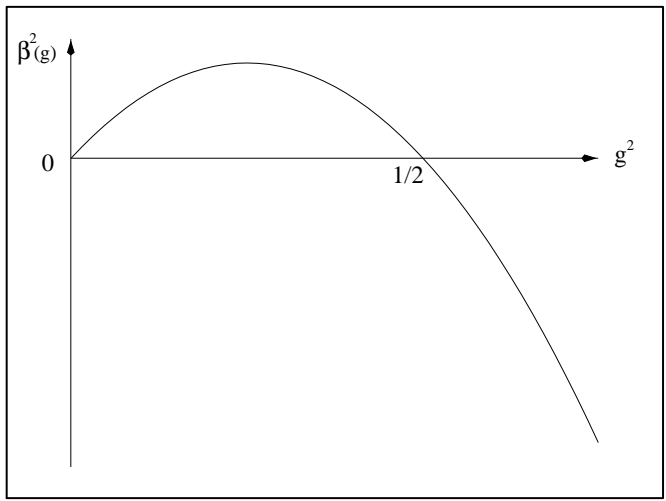}}
\caption{The $\beta$-function for the mass term of three dimensional one
component $Z_{2}$-invariant scalar field theory}
\end{figure}

\section{Truncations}

We are interested in finding non-trivial fixed points. Here, we outline
the method of truncations for one component theory. In this
approximation method (in addition to the LPA) we expand $V_{*}(\phi)$ in 
powers of
$\phi$ and truncate the series at some finite power, $2p$ ($p=\frac{1}{2}, 1,
\frac{3}{2},\cdots $). Begin by considering $V_{*}$ as an infinite series in
powers of $\phi$ 
\begin{equation}
V_{*} = \sum_{n=0}^{\infty} V_{n} \phi^{n}
\end{equation}
such that the fixed point version of (2.33) yields
\begin{equation}
\frac{n}{2} (D-2) V_{n} - D V_{n} = (n+1)(n+2)V_{n+2} -\!\!\!\! \sum_{u+v=n+2}
uvV_{u}V_{v}
\end{equation}
as the coefficients of a general power, $\phi^{n}$. In the last term $u\geq 1$
and $v\geq 1$ are integers. We will consider a $Z_{2}$-invariant theory (even
$V(\phi)$) which requires $V_{n}=0$ for odd $n$.

Similarly we write an eigenperturbation as a series\footnote{the $f_{n}$
defined here are distinct to the eigenfunctions used in the latter part of
section 2.4},
\begin{equation}
f = \sum_{n=0}^{\infty} f_{n} \phi^{n}
\end{equation}
such that the coefficients of $\phi^{n}$ in (2.36) yields
\begin{equation}
\frac{n}{2} (D-2) f_{n} - (D-\lambda) f_{n} = (n+1)(n+2)f_{n+2} - 
2\!\!\!\!\sum_{u+v=n+2} uvV_{u}f_{v}.
\end{equation}
Given that we truncate the series at a power $2p$ we find that the eigenvalues
are found by solving the determinant of a $(p+1) \times (p+1)$ matrix,
\begin{equation}
\det
\left( \begin{array}{cccc}
D-\lambda&2                &0                   &\cdots  \\
0        &2-8V_{2}-\lambda &12                  &\cdots  \\
0        &-16V_{4}         &4-D-16V_{2}-\lambda &\cdots  \\
0        &-24V_{6}         &-32V_{4}            &\cdots  \\
\vdots   &\vdots           &\vdots              &\ddots
\end{array} \right)
=0.
\end{equation}
For illustration consider the simplest case of $p=2$, which leads to three
simultaneous equations for the $V_{i}$:
\begin{equation}
V_{0}=-\frac{2}{D} V_{2},
\end{equation}
\begin{equation}
V_{2}=2V_{2}^{2}-6V_{4},
\end{equation}
and
\begin{equation}
(4-D)V_{4}=16V_{2}V_{4}.
\end{equation}
Solving, we reproduce both the Gaussian and High Temperature fixed points. 
Additionally we find a non-trivial fixed point,
\begin{equation}
V_{*}^{NT}=\frac{D-4}{8D}+\frac{4-D}{16}\phi^{2}+\frac{D^{2}-16}{768}\phi^{4}.
\end{equation}
Immediately we observe that it becomes degenerate with the Gaussian in $4$
dimensions and the High Temperature in $-4$ dimensions. The former is observed
generally. The first three eigenvalues for the Gaussian and High Temperature
fixed points,
given by (2.37) and (2.38), are reproduced exactly. In principle the 
non-trivial fixed point in three dimensions would have $\lambda = 3, 
\frac{3}{4}\pm\frac{1}{2} \sqrt{\frac{37}{4}}$, however (2.49) is actually
unbounded from below. This is unphysical and will be seen to be an artefact of
the truncation.

Generally, the method of truncations proves problematic. The
exponents calculated in this case are found to
initially converge but then fail to converge further with increasing $p$
[19]. Most notably spurious solutions are generated for $p>2$, and some 
procedure must be sought to distinguish the true fixed points. 
In some cases it has been found that some improvement can be made by expanding
around the minimum 
of the potential [20]. The truncation method has been found to
yield imaginary exponents ($\omega$ here) which would imply exotic flows near
the fixed point. However, later we will find that the exponents are expected
to be real, excluding the possibility of exotic flows.

\section{Shooting}

Exact non-trivial fixed points (within the LPA) are found by solving (2.33)
numerically. We use a method known as shooting [13]. We see that there is a
one parameter (denoted $s$) set of 
solutions, corresponding to a second order differential equation with a single
boundary condition. For a given value of $s$ we integrate out for $V(\phi)$
until it 
diverges, generally at finite $\phi =\phi_{c}$. Then we `scan' over $s$ and
look for non-divergent solutions at $s=s_{*}$, which correspond to our fixed
points\footnote{this is a requirement of Griffiths analyticity combined
with the required analyticity at the origin}.
\begin{figure}
\vspace{-17mm}
\hspace{23mm}\scalebox{0.35}{\includegraphics*[0pt,0pt][800pt,800pt]{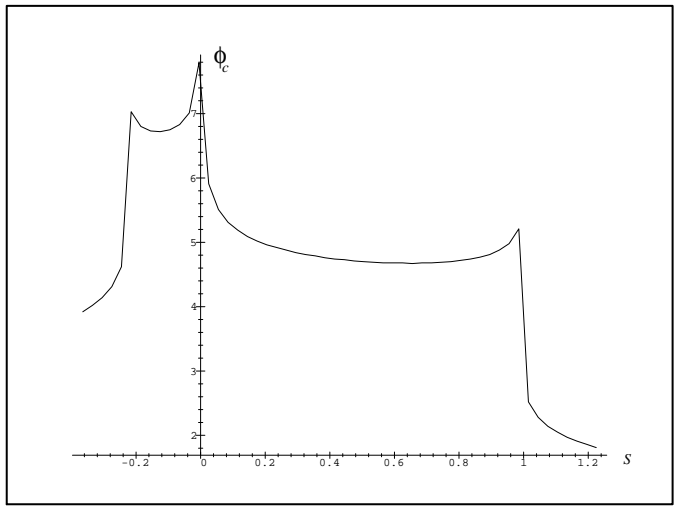}}
\caption{Shooting: $\phi_{c}$ v $s$}
\end{figure}
For consistency, we consider again a $Z_{2}$-invariant one component 
theory such that $V$ is even in $\phi$ and then set our boundary conditions as
\begin{equation}
V(0)=-\frac{s}{D}
\end{equation}
and
\begin{equation}
V'(0)=0
\end{equation}
where the $s=V''(0)$ and prime denotes differentiation with respect to
$\phi$. The first condition is a translation directly from (2.33) and the
second is due to the symmetry of the theory. 
A plot of $\phi_{c}$ against $s$ is shown
for three dimensions, in figure 2.6. The fixed points occur for diverging 
$\phi_{c}$; as expected from (2.34) and (2.35) the Gaussian occurs at
$s^{G}_{*}=0$ and the High Temperature occurs at  
$s^{HT}_{*}=1$. The non-trivial Wilson-Fisher fixed point [16] is observed at 
$s^{WF}_{*}=-0.229$. In figure 2.7, we plot $V_{*}^{WF}$.
\begin{figure}
\vspace{-34mm}
\hspace{23mm}\scalebox{0.41}{\includegraphics*[0pt,0pt][800pt,800pt]{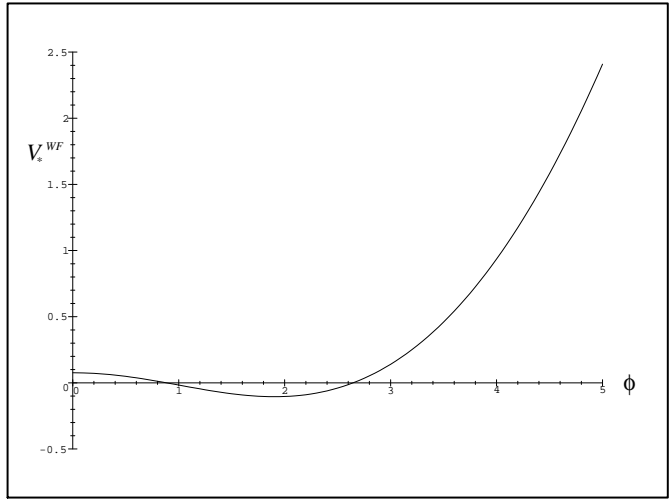}}
\caption{The Wilson-Fisher fixed point}
\end{figure}

Varying the dimension, we observe that there are no non-trivial fixed points
at or above four dimensions. In exactly four dimensions we find the
Wilson-Fisher fixed point is
degenerate with the Gaussian, where the Gaussian has a marginal eigenvalue,
$\lambda_{2}=0$. With decreasing dimension, the Wilson-Fisher fixed point
occurs for 
decreasing $s$, such that it moves away from the Gaussian (left in figure
2.6), and new non-trivial fixed points are created which are initially
degenerate with the Gaussian for dimensions where the $\lambda^{G}_{n}=0$
($n=2,3,\cdots$).  
These marginal eigenvalues allow the addition of the corresponding
perturbation to the Gaussian, creating a new fixed point as previously
discussed. The dimensions are given by
\begin{equation}
D=\frac{2(n+1)}{n}.
\end{equation}
Hence, below four dimensions we see at least one
non-trivial fixed point whereas below three dimensions we see at least
two (Appendix 2A). At and below two dimensions we observe a continuous line of
`oscillating' fixed points. The behaviour of $s_{*}$ with dimension $D$ is
illustrated for the Wilson-Fisher and two subleading fixed points in figure
2.8.  
\begin{figure}
\vspace{-17mm}
\hspace{23mm}\scalebox{0.35}{\includegraphics*[0pt,0pt][800pt,800pt]{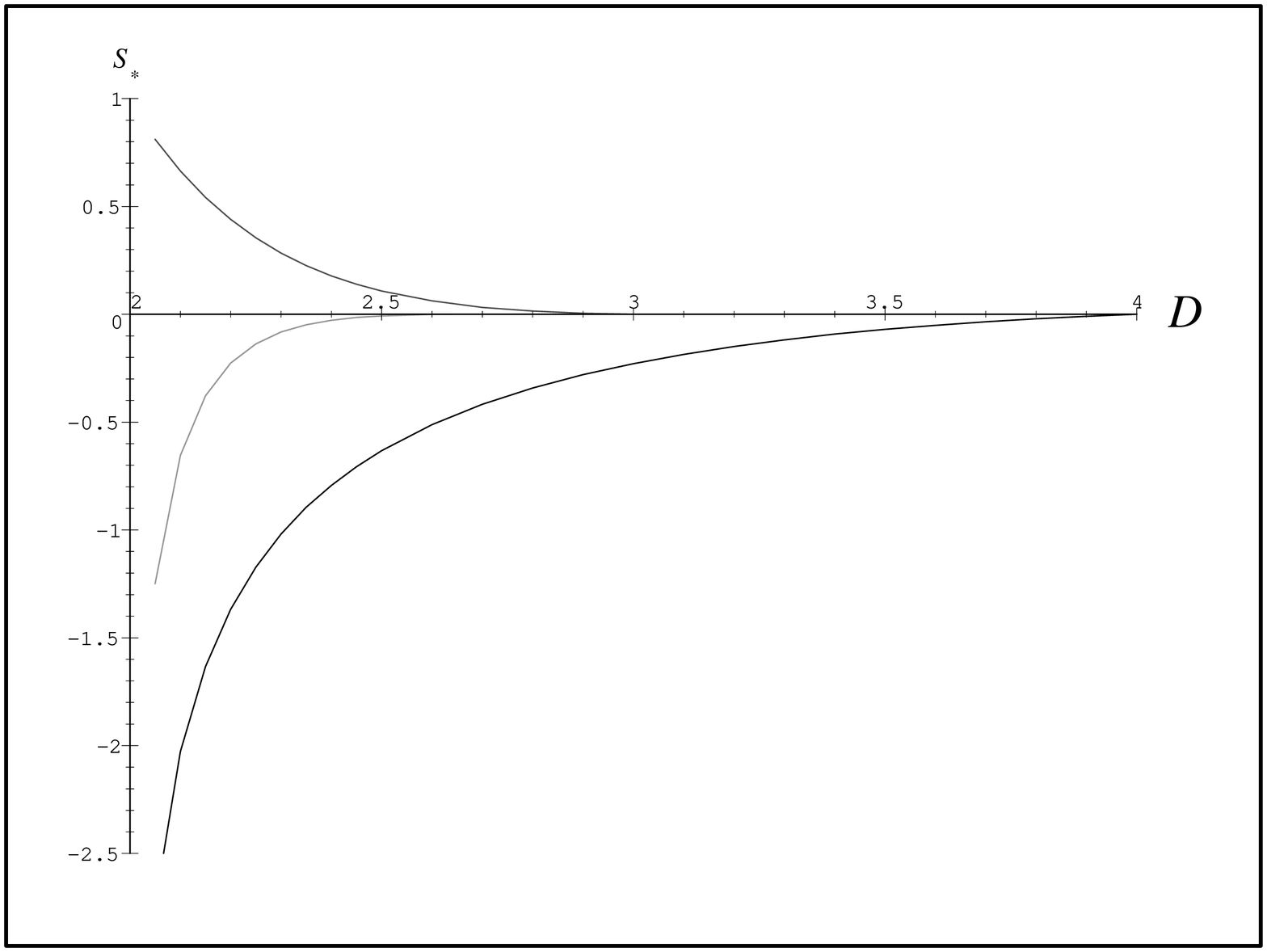}}
\caption{A plot of $s_{*}$ v $D$ for the Wilson-Fisher and two subleading
fixed points} 
\end{figure}

Finally (2.36) is solved for the exponents in a similar manner. Here we impose
two conditions,
\begin{equation}
f(0)=1
\end{equation}
and
\begin{equation}
f'(0)=0
\end{equation}
where the first is an arbitrary normalisation and the second is an
imposition of our choosing (we are restricting to even
perturbations). The solutions generally diverge and are found to depend on the
eigenvalue, $\lambda$. Thus, similar to before we sweep through a range of
$\lambda$ and look for acceptable (\emph{i.e.}
non-divergent) $f(\phi)$. For the Wilson-Fisher fixed point, the largest values
yield $\nu=0.6496$ and $\omega=0.6557$. Later, these will be compared to other
leading methods.

\section{Variation}

We begin by introducing $\rho(\phi,t)=e^{-V(\phi,t)}$ and
$G(\phi)=e^{-\frac{1}{4}(D-2)\phi_{a}^{2}}$ such that the Polchinski equation
(2.33) can be written [21,22]
\begin{equation}
a^{N}G\frac{\partial \rho}{\partial t} = - \frac{\delta \mathcal{F}}{\delta
\rho}
\end{equation}
where
\begin{equation}
\mathcal{F} [\rho ] = a^{N}\!\! \int\!\! d^{N}\phi\ G \left( \frac{1}{2}\left(
\frac{\partial \rho}{\partial \phi_{a}}\right)^{2}+\frac{D}{4}\rho^{2}\left(
1-2\ln\rho\right)\right)
\end{equation}
and $a>0$ is a normalisation factor for the $\phi$ measure to be determined
later. Then the fixed point is defined by
$\frac{\delta\mathcal{F}}{\delta\rho}=0$. An interesting reparameterisation of
(2.56) is discussed in Appendix 2B.

As discussed in chapter one we let $g^{i}(t)$ be a complete set of parameters
(the coupling constants) 
for $V$. Away from fixed points these will be infinite in number. Then using
(2.55), we write
\begin{eqnarray}
\frac{\partial \mathcal{F}}{\partial g^{i}} 
        & = & \int\!\! d^{N}\phi\ \frac{\delta\mathcal{F}}{\delta \rho} 
\frac{\partial \rho}{\partial g^{i}} \\
	& = & -a^{N}\!\! \int\!\! d^{N}\phi\ G \frac{\partial \rho}{\partial t}
	\frac{\partial \rho}{\partial g^{i}} \\ 
        & = & -M_{ij} \beta^{j}
\end{eqnarray}
where $\beta^{j}$ are the $\beta$-functions given by (1.41) and $M_{ij}$ is 
defined by
\begin{equation}
M_{ij} = a^{N}\!\! \int\!\! d^{N}\phi\ G \frac{\partial \rho}{\partial g^{i}}
\frac{\partial \rho}{\partial g^{j}}.
\end{equation}
Perturbing about the fixed point solution by writing $g^{j}=g^{j}_{*}+\delta
g^{j}=g^{j}_{*}+b^{j}e^{\lambda t}$, (2.59) yields 
\begin{eqnarray}
\frac{\partial^{2} \mathcal{F}}{\partial g^{i}\partial g^{j}} \delta g^{j} 
    & = & \frac{\partial \mathcal{F}}{\partial g^{i}} \\
    & = & -\lambda M_{ij} b^{j} e^{\lambda t}
\end{eqnarray}
such that (assuming that $b^{j}$ is not a null vector) we calculate the
eigenvalues, $\lambda$, by solving
\begin{equation}
\det \left( \lambda M_{ij} + \frac{\partial^{2} \mathcal{F}}{\partial
g^{i}\partial g^{j}}\right) = 0.
\end{equation}

The fixed point equation, $\frac{\delta \mathcal{F}}{\delta \rho} = 0$,
suggests the possibility of approximating $\rho$ by a variational ansatz,
$\rho=f(\phi,g^{i})$ where $i$ runs from $1$ to a finite $M$ and $f$ is of our
choosing. Then interpreting the functional derivative in (2.55) to include
only variations in this restricted set, we arrive again at (2.59) where
however, indices only run from $1$ to $M$. Thus geometrically, we restrict the
flows to a submanifold $\mathcal{M}$ parametrised by
$g^{1},\cdots ,g^{M}$. 

It is important that these methods reproduce at least the trivial fixed points
accurately. Here we will rederive the Gaussian and High Temperature fixed
points with the leading exponent using the methods described above. We take
$V=g^{1}(t)+\frac{1}{2} g^{2}(t)\phi^{2}$ as an ansatz and minimise the
functional in (2.56) 
\begin{eqnarray}
\frac{\partial \mathcal{F}}{\partial g^{i}}
	& = & a^{N}\!\!\int\!\! d^{N}\phi\ G \left(\rho ' \frac{\partial \rho
'}{\partial g^{i}}-D\rho\ln \rho \frac{\partial \rho}{\partial g^{i}}\right) \\
	& = & 0
\end{eqnarray}
to arrive at a pair of simultaneous equations for $g^{1}$ and
$g^{2}$. Performing the resulting Gaussian integrals (Appendix 2C), we arrive
at the following conditions
\begin{equation}
g^{2}(D+2g^{2})+2g^{1}D[\frac{1}{2}(D-2)+2g^{2}]=0
\end{equation}
and
\begin{equation}
3g^{2}(D+2g^{2})+2(g^{1}D-2g^{2})[\frac{1}{2}(D-2)+2g^{2}]=0.
\end{equation}
Solving these leads directly to the Gaussian ($g^{1}=g^{2}=0$) and High
Temperature ($g^{1}=-\frac{1}{D}$ and $g^{2}=1$) solutions. Then
differentiating (2.64) 
with respect to $g^{i}$ and substituting (2.60) into (2.63) and again
performing the resultant Gaussian integrals we reproduce the leading exponents
for the trivial fixed points. For the Gaussian (2.63) leads to $(\lambda - D)
(\lambda -2)=0$ and for the High Temperature we find $(\lambda-D)(\lambda+2)
=0$.

\section{The Wilson-Fisher fixed point by Variation}

The true test of the variation method lies in the accuracy with which it can
reproduce the Wilson-Fisher fixed point [22]. Thus, working again with a 
$Z_{2}$-invariant one component theory and using the simplest non-trivial
ansatz, $V(\phi,t)=g^{1}(t)+\frac{1}{2} g^{2}(t)\phi^{2}+g^{3}(t)\phi^{4}$, we
find an approximate solution to the Wilson-Fisher fixed  
point, corresponding to $g^{1}_{*}=0.05479$, $g^{2}_{*}=-0.13488$ and
$g^{3}_{*}=0.00773$. In figure 2.9, we plot the resulting form of $\rho_{*}$
and compare it to the exact non-trivial fixed point solution\footnote{exact
within the LPA}, solved by
shooting. Solving (2.63), we obtain, apart from $\lambda=3$, $\nu=0.6347$ from
the positive eigenvalue, and $\omega=0.6093$ from the remaining eigenvalue. 
These are $2\%$ and $8\%$ away from the exact$^{8}$ values respectively.

Clearly, it may be possible to find `spurious fixed points' that do not well
approximate the exact solutions, but these will not be stable under changing
or improving the ansatz manifold $\mathcal{M}$. In fact as $\mathcal{M}$
improves, such fixed points (if indeed there are any) will disappear. These
characteristics are in marked contrast to the general situation for
truncations, as discussed earlier. To improve $\mathcal{M}$ we may simply 
extend the series to include a $\phi^{6}$ term, or alternatively appeal to 
(2.33) to build a more sophisticated ansatz. In practice such
parameterisations are increasingly difficult to solve and the results presented
here are of sufficient accuracy. As applied here, the variational method 
provides no real advantage over shooting. The true potential of this method
lies in the relative ease in which  approximations for global flows may be
solved and approximate solutions found for more than one invariant. 

\begin{figure}
\hspace{20mm}\scalebox{0.55}{\includegraphics*[10pt,10pt][600pt,500pt]{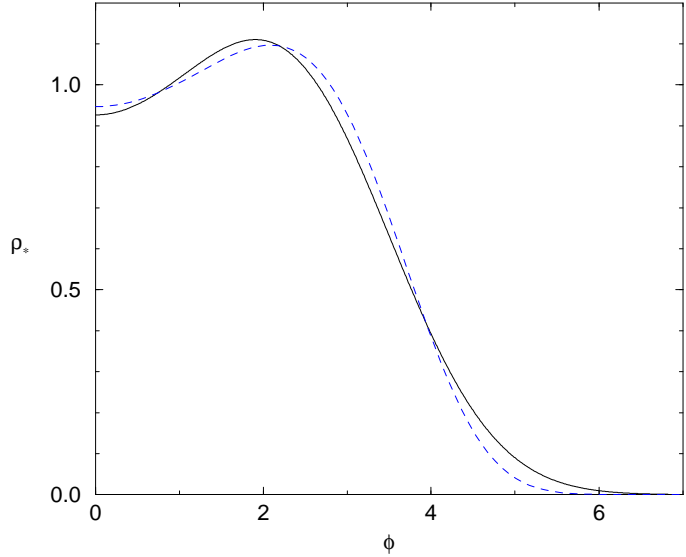}}
\caption{The simplest polynomial variational approximation to the Wilson-Fisher
fixed point (dashed line) compared to the exact solution (full line)}
\end{figure}

\section{The large $N$ limit of the LPA}

The large $N$ limit of the LPA to the Polchinski equation can be solved
analytically and compared with known exact results. It is thus a good testing
ground for the LPA. In fact, we will find that the LPA yields exact results in
this limit giving us added confidence in the reliability of this
approximation. Using other methods [1] various authors have found that, for
large $N$, there exists a fixed point in 
dimensions $2<D<4$ with $\nu = \frac{1}{D-2}$. Here we reproduce these
results, using the methods of Morris \emph{et. al.} [23]. 

We begin by defining $z(x)=\phi_{a}^{2}(x)>0$ such that
within the LPA $V$ can be
written as a function of $z$ and $t$ only, corresponding to an $O(N)$
invariant theory. Then the Polchinski equation (2.33) can be written
\begin{equation}
\frac{\partial V}{\partial t} = DV-((D-2)z+2N)\frac{\partial V}{\partial z} +
4z\frac{\partial ^{2} V}{\partial z^{2}} - 4z\left(\frac{\partial V}{\partial
z}\right)^{2}.
\end{equation}
We observe $V=0$ and $V=\frac{1}{2}z-\frac{N}{D}$ as the trivial fixed points,
however concentrate on non-trivial solutions. By considering figure 2.10
(due to $g\phi^{4}$) and requiring a non-trivial finite
limit, $g\sim Ng^{2}$ which implies that $g\sim\frac{1}{N}$. Then
$m\phi^{2}\sim\frac{1}{N}\phi^{4}$ which implies $\phi^{2}\sim N$ and thus
$V\sim N$ where the mass is $N$ independent.
\begin{figure}
\vspace{-70mm}
\hspace{26mm}\scalebox{0.4}{\includegraphics*[0pt,0pt][800pt,800pt]{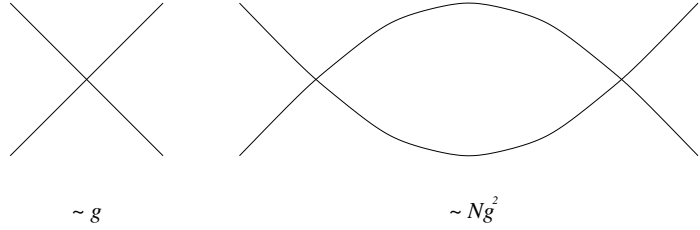}}
\caption{$N$-dependence of the two leading contributions to the four point
function} 
\end{figure}
Thus changing variables so as to scale out the $N$ dependences, and taking the
infinite $N$ limit the 
second order derivative in (2.68) vanishes. Then, differentiating with respect
to $z$ and defining $W=\frac{\partial V}{\partial z}$ we arrive at the
following flow equation,
\begin{equation}
\frac{\partial W}{\partial t}+((D-2)z+8zW-2)\frac{\partial W}{\partial z} = 2W
- 4W^{2}.
\end{equation}
Alternatively, shifting to a stationary point in $V$, $z_{o}$, (thus excluding
the High Temperature solution) and defining $z=z_{o}+y$, we arrive at [23]
\begin{equation}
\dot{W}-\dot{z_{o}}W'+(D-2)(z_{o}+y)W'-2W=2W'-8(z_{o}+y)WW'-4W^{2}.
\end{equation}
Here the dot denotes differentiation with respect to $t$ and the prime denotes
differentiation with respect to $y$.

However, because we have shifted to a minimum we require $W(0,t)=0$ in (2.70),
which yields $W'(0,t)=0$ and thus $W(y, t)=0$ (by expanding in powers of $y$),
unless
\begin{equation}
\dot{z_{o}}=(D-2)z_{o}-2.
\end{equation}
Substituting this non-Gaussian solution back into (2.70) yields 
\begin{equation}
\dot{W}+(D-2)yW'-2W=-8(z_{o}+y)WW'-4W^{2}.
\end{equation}
Then, expanding in powers of $y$,
\begin{equation}
W(y, t)=\sum^{\infty}_{n=1} W_{n}(t)y^{n},
\end{equation}
yields a series of equations,
\begin{eqnarray}
\dot{W_{1}}+((D-2)-2)W_{1}  & = & -8z_{o}W_{1}^{2},               \\
\dot{W_{2}}+(2(D-2)-2)W_{2} & = & -24z_{o}W_{1}W_{2}-8W_{1}^{2},  \\
\dot{W_{n}}+(n(D-2)-2)W_{n} & = & -8(n+1)z_{o}W_{1}W_{n}\cdots .
\end{eqnarray}
The final equation holds for $n>1$ and the dots stand for terms containing
products of $W_{m}$'s with $m<n$. Thus the $W_{n}(t)$ are all soluble in terms
of $W_{m}$, with $m<n$, and $z_{o}(t)$, and truncations to some finite $n$ are
exact.

Now consider a fixed point, $W(y, t)=W(y)$ and $z_{o}(t)=z_{o}$. Then by
(2.71)
\begin{equation}
z_{o}=\frac{2}{D-2},
\end{equation}
however as $z_{o}$ is the original field squared it must be positive,
requiring $D>2$. Note that since there is only one solution for $z_{o}$,
$W(y)$ can only cross the axis once. Substituting this into (2.74) yields
$W_{1}=0$ (the Gaussian) and 
\begin{equation}
W_{1}=\frac{(D-2)(4-D)}{16}.
\end{equation}
Since $W(y)$ crosses the $y$-axis only once, we must have
$W_{1}>0$ which implies we must have $D<4$ (as we have established $D>2$), so
that $W(y)>0$ for $y>0$. Otherwise the potential is unbounded
below. Substituting (2.78) into the series of equations (2.74) to (2.76) we
see that the $W_{n}$ exist and are unique (\emph{e.g.} $W_{2}=
\frac{6}{8-5D}W_{1}^{2}$).  

We deduce the leading exponent by perturbing the $z_{o}$ in (2.71) about the
fixed point solution, $z_{o}=z_{o*}+\delta z_{o}$, yielding $\dot{\delta
z_{o}} = (D-2)\delta z_{o}$, such that there is an eigenperturbation $\delta
z_{o}\propto e^{\lambda_{0}t}$, with $\lambda_{0}=D-2$ implying
\begin{equation}
\nu=\frac{1}{D-2}, \ \ \ \ \ \ \ 2<D<4.
\end{equation}
Now we can systematically generate the subleading exponents by perturbing
about fixed point solutions in (2.74) to (2.76). For example, by writing
$W_{1} = W_{1*}+\delta W_{1}$ in (2.74) (with $\delta z_{o}=0$), we arrive at
$\dot{\delta W_{1}} = 
(D-4)\delta W_{1}$, or $\delta W_{1} \propto e^{(D-4)t}$, yielding
$\lambda_{1}=D-4$. Then in (2.75) we find a new perturbation which takes the
form $\delta z_{o}(t)=0$, $\delta W_{1}(t)=0$ and $\delta W_{2}(t)\propto
e^{(D-6)t}$, yielding $\lambda_{2}=D-6$. In this way we
generate the set of eigenvalues in the large $N$ limit, $\lambda_{n}=D-2n$ for
$n=1,2,\cdots$.

\section{Discussion}

Referring to the numerical methods, the most outstanding feature is 
that they allow us to exhaustively search the entire
infinite dimensional space using a single parameter and to make physical
predictions for complex systems which can be measured experimentally. The
results for $\nu$ and $\omega$ found by shooting and the variational method
compare favourably with other leading methods (table 2.1) [1]. It is not yet
known how to apply the momentum expansion (beyond LPA) to the Polchinski flow
equation in a reparameterisation invariant [15] way. However the momentum
expansion at $O(p^{2})$ has been applied to the Legendre flow equation and
shows a modest improvement in the exponents, over the LPA [14]. Traditionally
it is difficult to estimate the errors introduced using these approximation
schemes, other than by comparison with the other methods, though recently
progress has been made [14]. The LPA is found to
be competitive, at least for small $N$ and away from $D=2$ where $\eta$ becomes
non-negligible.

The success of large $N$ limit of the LPA to the Polchinski equation is
somewhat 
fortunate. It should be noted that the flow equation (2.68) is not itself
exact, but belongs to a large class of equations which yield the correct
exponents [23]. Nevertheless, it is an important result and this limit will be
the focus of our attention for Fermionic field theory. It is hoped that
solving this limit will lead to an increased understanding of the problem and
thus help us to deal with finite $N$ theories. Similarly, truncations have
been found to be useful, albeit in a limited capacity. In fact from (2.62) we
can now see that provided the potential $V$ is real, we expect the eigenvalues,
$\lambda_{n}$, to be also real, as claimed. This contrasts with the imaginary
exponents which occur using the method of truncations.

\vspace{5mm}
\begin{table}[H]
\begin{center}
\begin{tabular}{|l|c|c|c|}                                     \hline
Method                                    &$\nu$  &$\omega$ \\ \hline \hline
Lattice calculation                       &0.6305 &         \\ \hline
$\epsilon$-expansion at $O(\epsilon^{5})$ &0.6310 &0.81     \\ \hline
Six loop perturbation series              &0.6300 &0.79     \\ \hline \hline
Local potential approximation \ (Pol)     &0.6496 &0.6557   \\ \hline
LPA Variation method \ (Pol)              &0.6347 &0.6093   \\ \hline
Local potential approximation \ (Leg)     &0.6604 &0.6285   \\ \hline \hline
Momentum expansion at $O(p^2)$\ (Leg)     &0.620  &0.898    \\ \hline
\end{tabular}
\caption{Exponents for three dimensional one component $Z_{2}$-invariant
scalar field theory}
\end{center}
\end{table}
\vspace{-5mm}

The above comparisons are made using results found in [1] and [14].

\newpage

\chapter*{Appendices}
\addcontentsline{toc}{chapter}
{\numberline{Appendices}}

\section*{Appendix 2A: Non-trivial fixed points}

We include the data for non-trivial fixed points below four dimensions,
including the first two subleading fixed points (table 2.2).

\vspace{5mm}
\begin{table}[H]
\begin{center}
\begin{tabular}{|c|c|c|c|c|}                                   \hline
\ \  D\ \ \ &$s_{*}$          &$s_{*}$          &$s_{*}$       \\ 
            &Wilson-Fisher    &1st subleading   &2nd subleading\\ \hline \hline
4.00        &0.000$^{\dagger}$&-                &-             \\ \hline
3.90        &-0.009           &-                &-             \\ \hline
3.80        &-0.021           &-                &-             \\ \hline 
3.70        &-0.035           &-                &-             \\ \hline
3.60        &-0.051           &-                &-             \\ \hline
3.50        &-0.070           &-                &-             \\ \hline 
3.40        &-0.092           &-                &-             \\ \hline
3.30        &-0.119           &-                &-             \\ \hline
3.20        &-0.149           &-                &-             \\ \hline
3.10        &-0.186           &-                &-             \\ \hline
3.00        &-0.229           &0.000$^{\dagger}$&-             \\ \hline
\end{tabular}
\end{center}
\end{table}
\begin{table}[H]
\begin{center}
\begin{tabular}{|c|c|c|c|c|}                                    \hline
2.90        &-0.280   &0.005 &-                                \\ \hline
\hspace{3pt}2.80\hspace{3pt} &\hspace{21pt}-0.342\hspace{21pt}
&\hspace{22pt}0.015\hspace{22pt} &-                            \\ \hline
2.70        &-0.417   &0.032 &-                                \\ \hline
2.60        &-0.512   &0.062 &\hspace{21pt}-0.001\hspace{21pt} \\ \hline
2.50        &-0.632   &0.108 &-0.008                           \\ \hline
2.45        &-0.707   &0.139 &-0.015                           \\ \hline
2.40        &-0.793   &0.178 &-0.028                           \\ \hline
2.35        &-0.895   &0.226 &-0.049                           \\ \hline
2.30        &-1.019   &0.284 &-0.082                           \\ \hline
2.25        &-1.172   &0.355 &-0.137                           \\ \hline
2.20        &-1.368   &0.440 &-0.226                           \\ \hline
2.15        &-1.633   &0.542 &-0.378                           \\ \hline
2.10        &-2.027   &0.664 &-0.654                           \\ \hline
2.05        &-2.738   &0.811 &-1.249                           \\ \hline
2.00        &-$\infty$ &1.000$^{\ddagger}$ &-$\infty$          \\ \hline
\end{tabular}
\caption{Wilson-Fisher and subleading fixed points.}
\end{center}
\end{table}

$\dagger$ Fixed points degenerate with the Gaussian. 
\newline$\ddagger$ Fixed points degenerate with the High Temperature.

\section*{Appendix 2B: Functional reparameterisation}

We can reparameterise the functional, $\mathcal{F}$, in such a manner, that
the first term in (2.56) looks like a kinetic term. As a consequence the
second term can be interpreted as a time varying potential, unless we deal with
two dimensions, when it is static. This is done by making the following change
of variables,
\begin{equation}
d\phi = d\tau \ e^{\frac{1}{4} (2-D)\phi^{2}}.
\end{equation}
Then, replacing $\mathcal{F}$ with $-S$ we arrive at the following, with the
desired kinetic term and time varying potential:
\begin{equation}
S=\int\!\! d\tau \left(\frac{1}{2} \dot{\rho} ^{2}-f(\tau)\rho
^{2}\left(\ln\rho- \frac{1}{2}\right)\right)
\end{equation}
where
\begin{equation}
f(\tau)=\frac{D}{2}\ e^{\frac{1}{2}(2-D)\phi ^2}
\end{equation}
and the dot denotes differentiation with respect to $\tau$.

Now we explicitly view (2.81) as an action, and think of a particle at position
$\rho$, at time\footnote{not to be confused with the renormalization time
defined by (1.29)} $\tau$. From (2.82) it is easy to see that
in dimension greater than two, $f(\tau)$, and thus the potential, decreases
with  time, whereas for dimension less than two the reverse is true. The form
of the potential,
\begin{equation}
V(\rho)=\rho^{2}\left(\ln\rho -\frac{1}{2}\right)
\end{equation}
is shown in figure 2.11. 
\begin{figure}
\vspace{-20mm}
\hspace{25mm}\scalebox{0.35}{\includegraphics*[0pt,0pt][800pt,800pt]{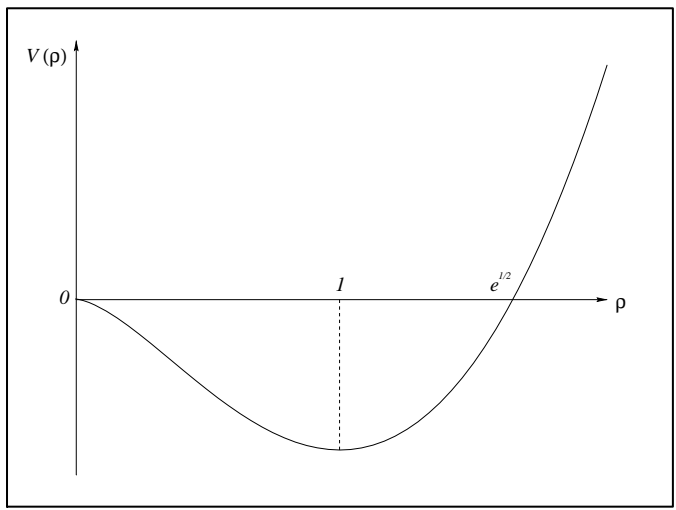}}
\caption{Time varying potential in reparameterised functional}
\end{figure}

We now view fixed points as motions in the 
potential well which are either at rest permanently, or, only come to rest at
$\rho=0$ in an infinite time (recall that in the large $\phi$ regime, 
$V_{*}(\phi)\sim\frac{1}{2}\phi^{2}$). For example, the Gaussian fixed point 
corresponds to a particle which simply remains 
static in the minimum of the well, whereas the High Temperature fixed point 
corresponds to a solution which lies at $\rho=\exp({\scriptstyle\frac{1}{3}})
\approx 1.40$ at $\tau=0$ and then moves towards $\rho=0$ as
$\tau\rightarrow\infty$. The 
solution for the  non-trivial fixed point has been found (figure 2.9), and
this corresponds to a particle 
which lies at $\rho\approx0.92$ at $\tau=0$. For this solution the particle 
oscillates once before coming to rest at $\rho=0$, again as
$\tau\rightarrow\infty$. These motions can be deduced directly by observing the
behaviour of $\rho$ with $\phi$.

\section*{Appendix 2C: Gaussian integration}

For completeness, we list the set of Gaussian integrals used:
\begin{equation}
I_{0}=\int ^{\infty}_{-\infty}\!\!\! dx\ e^{-\frac{1}{2} ax^{2}} = 
\sqrt{\frac{2\pi}{a}}
\end{equation}
found by considering $I_{0}^{2}$ and changing variables, $(x,y)\rightarrow
(r,\theta)$. Then we define
\begin{eqnarray}
I_{m} & = & \int ^{\infty}_{-\infty}\!\!\! dx\
      x^{m} e^{-\frac{1}{2} ax^{2}} \\
      & = & I_{0} \left. \left( \frac{\partial}{\partial b} 
      \right) ^{m} e^{\frac{b^2}{2a}} \right|_{b=0}
\end{eqnarray}
and calculate
\begin{eqnarray}
I_{2} & = & \frac{1}{a} \sqrt{\frac{2\pi}{a}}, \\
I_{4} & = & \frac{3}{a^2} \sqrt{\frac{2\pi}{a}}, \\
I_{6} & = & \frac{15}{a^3} \sqrt{\frac{2\pi}{a}}.
\end{eqnarray}
 
\chapter{Zamalodchikov's $C$-function}

In the preceding chapter we argued that the Local Potential Approximation
provides competitive estimates of the critical exponents, at least for scalar
field theory. The theoretical analysis presented here 
further vindicates and increases our confidence in the approximation. We begin
by introducing conformal field theory, leaving a more detailed analysis to
reference [4]. We review Zamalodchikov's $C$-theorem [24] using
the introductory material. The $C$-theorem, as presented here, only makes
sense in two dimensions, although a number of groups have sought to generalise
these ideas to other
dimensions [25-36]. However, within the Local Potential Approximation we
construct a function, $C$, which has an appropriate generalisation of the
Zamalodchikov properties, in general dimension, $D$. Our $C$-function `counts'
degrees of freedom at fixed points. We normalise such that it counts one
per Gaussian scalar and zero at the High Temperature fixed point
(corresponding to an infinitely massive theory). 

\newpage

\section{The conformal group}

As discussed in chapter one, we are interested in non-trivial fixed points which
correspond to an interacting scale free theory, known as a conformal field
theory. Here we will work mostly in two dimensions, however recent work
[22,25] has extended some of the aspects presented to four dimensions. We
begin by considering a $D$-dimensional Riemannian manifold upon which a
metric, $g_{\mu \nu}$, is defined. The interval between two points on the
manifold is given by 
\begin{equation}
ds^2=g_{\mu\nu}dx^{\mu}dx^{\nu}
\end{equation}
and a coordinate transformation, $x \rightarrow x'$, generates a change in the
metric tensor given by
\begin{equation}
g_{\mu \nu}(x)\rightarrow g_{\mu \nu}'(x') = \frac{\partial
x^{\alpha}}{\partial x^{'\mu}} \frac{\partial x^{\beta}}{\partial x^{'\nu}}
g_{\alpha \beta}(x).
\end{equation}
The conformal group is defined to be the subgroup of the transformations that
leaves the metric unchanged apart from a non-zero differential scaling
function $\Omega(x)$ such that
\begin{equation}
g_{\mu \nu}(x)\rightarrow g_{\mu \nu}'(x') = \Omega(x)g_{\mu \nu}(x).
\end{equation}
These metrics are said to be conformally related; the angle between vectors
and the ratios of magnitudes of vectors are the same on both manifolds.
However, since the length of the vectors differ, this
transformation corresponds to a rescaling. The Poincar\'{e} group leaves
the metric totally invariant and is thus a subgroup of the conformal group with
$\Omega =1$.

\section{Special properties in two dimensions}

The infinitesimal generators of the group can be determined by the analysis of
an infinitesimal coordinate transformation, $x^{\mu}\rightarrow x^{' \mu} =
x^{\mu} - \epsilon^{\mu}(x)$, which causes a line element change. In flat
space ($g_{\nu\mu}=\delta_{\nu\mu}$)\footnote{working in Euclidean space}
\begin{eqnarray}
ds^{2}\rightarrow ds^{'2} &=& dx^{'\nu}dx^{'}_{\nu}\\
                          &=& ds^{2} - (\partial_{\mu}\epsilon_{\nu}(x) +
                          \partial_{\nu}\epsilon_{\mu}(x))dx^{\mu}dx^{\nu},
\end{eqnarray}
where we have used (3.2) and defined $\partial_{\mu}=\frac{\partial}{\partial
x^{\mu}}$.  Note that if $\epsilon^{\mu}(x)=\epsilon^{\mu}$, \emph{i.e.}
a translation, then the line element is invariant under such a
transformation. For the transformation to be conformal we require by
(3.3) and (3.5) that
\begin{equation}
\Omega(x)ds^{2}=ds^{2}-(\partial_{\mu}\epsilon_{\nu} +
\partial_{\nu}\epsilon_{\nu})dx^{\mu}dx^{\nu}.
\end{equation}
Then using (3.1) we arrive at
\begin{equation}
\delta_{\mu\nu}(1-\Omega(x))dx^{\mu}dx^{\nu}=(\partial_{\mu}\epsilon_{\nu} +
\partial_{\nu}\epsilon_{\mu})dx^{\mu}dx^{\nu}
\end{equation}
and for general $dx^{\mu}$
\begin{equation}
\delta_{\mu\nu}(1-\Omega(x)) = \partial_{\mu}\epsilon_{\nu} +
\partial_{\nu}\epsilon_{\mu}. 
\end{equation}
Then, using $\delta^{\mu\nu}\delta_{\mu\nu} = \delta^{\nu}_{\nu} = D$, we find
that the scaling function is given by 
\begin{equation}
\Omega(x)=1-\frac{2}{D}(\partial.\epsilon).
\end{equation}
where $\partial.\epsilon=\partial^{\mu}\epsilon_{\mu}$. 

In two dimensions, substituting (3.9) into (3.8) yields the Cauchy-Riemann
conditions for the analyticity of $\epsilon_{0}+i\epsilon_{1}$,
\begin{equation}
\partial_{0}\epsilon_{0}=\partial_{1}\epsilon_{1}\ \ \mathrm{and}\ \ 
\partial_{0}\epsilon_{1}=-\partial_{1}\epsilon_{0}.
\end{equation} 
This motivates us to use complex coordinates defined by
\begin{equation}
z=x^{0}+ix^{1}\ \ \mathrm{and}\ \ \bar{z}=x^{0}-ix^{1},
\end{equation}
and write the infinitesimal transformation as  
\begin{equation}
\epsilon(z)=\epsilon_{0}+i\epsilon_{1}\ \ \mathrm{and}\ \
\bar{\epsilon}(\bar{z}) = \epsilon_{0}-i\epsilon_{1}.
\end{equation} 
Then the conformal transformations coincide with the coordinate
transformations
\begin{equation}
z\rightarrow f(z)\ \ \mathrm{and}\ \ \bar{z} \rightarrow \bar{f}(\bar{z})
\end{equation}
which have an infinite dimensional local algebra, as $f(z)$ and
$\bar{f}(\bar{z})$ can be any analytic functions. These are the generators of
the local conformal algebra with Euclidean line element $ds^{2}=(dx^{0})^{2} +
(dx^{1})^{2}=dzd\bar{z}$ transforming as
\begin{equation}
ds^{2}\rightarrow\left| \frac{\partial f}{\partial z}\right|^{2}dzd\bar{z} =
\Omega dzd\bar{z}.
\end{equation}
However, if we take the transformed coordinates
\begin{equation}
z\rightarrow z'=(1-\epsilon_{n} z^{n})z
\end{equation}
and
\begin{equation}
\bar{z}\rightarrow\bar{z}'=(1-\bar{\epsilon}_{n} \bar{z}^{n})\bar{z}
\end{equation}
where $n$ ranges over the integers, then it is easily verified that the
infinitesimal generators are given by
\begin{equation}
L_{n}=-z^{n+1}\frac{\partial}{\partial z}=-z^{n+1}\partial_{z}
\end{equation}
and
\begin{equation}
\bar{L}_{n}=-\bar{z}^{n+1}\partial_{\bar{z}}. 
\end{equation}
Now we can construct the algebra:
\begin{equation}
\left[L_{m},L_{n}\right]=(m-n)L_{m+n},
\end{equation}
\begin{equation}
\left[\bar{L}_{m},\bar{L}_{n}\right]=(m-n)\bar{L}_{m+n}
\end{equation}
and
\begin{equation}
\left[L_{m},\bar{L}_{n}\right]=0.
\end{equation}
The latter condition, (3.21), shows that the local conformal
algebra is a direct sum of two independent subalgebras. The action of the
$D=2$ conformal group thus factorises into independent actions on $z$ and
$\bar{z}$  and the Greens functions of the theory are such that $z$ and
$\bar{z}$ can be treated as independent variables.

\section{The stress tensor as the generator of scaling}

Let us now consider a Euclidean action $S(\phi)$ depending only on the field
and its first partial derivatives, invariant under translation (associated
with momentum conservation), rotation (angular momentum conservation) and
dilation. Let us perform an infinitesimal change of variables:
\begin{equation}
x_{\mu}\rightarrow x'_{\mu}=x_{\mu}-\epsilon_{\mu}(x).
\end{equation}
As discussed in Appendix 3A, due to translational invariance, the variation of
the action involves only the partial derivatives of $\epsilon_{\mu}(x)$:
\begin{equation}
\delta S(\phi)=\int\!\! d^{D}x\ T_{\mu\nu}(x)\partial_{\mu}\epsilon_{\nu}(x)
\end{equation}
where $T_{\mu\nu}(x)$ is the stress tensor. Imposing other conformal
symmetries, we deduce some required properties of the stress tensor [1]. For
example rotation invariance implies $\delta S$ vanishes for
\begin{equation}
\epsilon_{\mu}=\Lambda_{\mu\nu}x^{\nu}
\end{equation}
in which $\Lambda_{\mu\nu}$ is an arbitrary  antisymmetric matrix. Therefore
the stress tensor must be symmetric:
\begin{equation}
T_{\mu\nu}=T_{\nu\mu}.
\end{equation}
Similarly, dilation invariance corresponds to
\begin{equation}
\epsilon_{\mu}=\lambda x_{\mu}
\end{equation}
and implies the vanishing of the trace of the stress tensor:
\begin{equation}
T_{\mu\mu}\equiv\Theta=0.
\end{equation}

In complex coordinates we have $\partial_{\bar{z}}z=\partial_{z}\bar{z}=0$ as
$z$ and $\bar{z}$ are treated independent. We refer the metric to the complex
coordinate frame using (3.2),
\begin{equation}
g=
\left( \begin{array}{cc}
0            & \frac{1}{2}   \\
\frac{1}{2}  & 0
\end{array} \right),
\end{equation}
and similarly calculate the stress tensor,
\begin{equation}
T_{zz}=\frac{1}{4}(T_{00}-2iT_{10}-T_{11}),
\end{equation}
\begin{equation}
T_{\bar{z}\bar{z}}=\frac{1}{4}(T_{00}+2iT_{10}-T_{11})
\end{equation}
and
\begin{equation}
T_{z\bar{z}}=T_{\bar{z}z}=\frac{1}{4}T_{\mu\mu}.
\end{equation}
Furthermore, we can calculate the derivatives of (3.29) and (3.30) with
respect to $\bar{z}$ and $z$ respectively, such that by conservation of the
stress tensor (see Appendix 3A), $\partial_{\mu}T_{\mu\nu}=0$, we deduce
\begin{equation}
\partial_{\bar{z}}T_{zz}+\partial_{z}T_{\bar{z}z} =
\partial_{\bar{z}}T+\frac{1}{4}\partial_{z}\Theta = 0 
\end{equation}
and similarly
\begin{equation}
\partial_{z}T_{\bar{z}\bar{z}}+\partial_{\bar{z}}T_{z\bar{z}} =
\partial_{z}\bar{T}+\frac{1}{4}\partial_{\bar{z}}\Theta = 0.
\end{equation}
Here we have defined $T(z,\bar{z})\equiv T_{zz}(z,\bar{z})$ and
$\bar{T}(z,\bar{z})\equiv T_{\bar{z}\bar{z}}(z,\bar{z})$, $\Theta$ is
given by (3.27) and we have used (3.31).
Finally we note that for a scale invariant field theory\footnote{corresponding
to a fixed point} (3.31) vanishes by (3.27) and we find 
$\partial_{\bar{z}}T=\partial_{z}\bar{T}=0$ which implies that the stress
tensor factorises into independent pieces, $T(z,\bar{z})=T(z)$ and
$\bar{T}(z,\bar{z})=\bar{T}(\bar{z})$.

\section{Conformal weights}

In general, the fields in a conformal theory obey the two dimensional
equivalent of the tensor transformation law,  
\begin{equation}
\Phi(z,\bar{z})\rightarrow\left(\frac{\partial f}{\partial z}\right)^{h}\left(
\frac{\partial \bar{f}}{\partial \bar{z}}\right)^{\bar{h}}\Phi \left(f(z),
\bar{f}(\bar{z})\right)
\end{equation}
where ($h,\bar{h}$) are the conformal weights of $\Phi$. We observe that
$T=T_{zz}$ and $\bar{T}=T_{\bar{z}\bar{z}}$ have conformal weights (2,0) and
(0,2) respectively. Similarly the trace of the stress tensor,
$\Theta=4T_{z\bar{z}}$, has conformal weight (1,1). Now consider two fields
$\Phi_{1}(z,\bar{z})$ and $\Phi_{2}(z,\bar{z})$ with conformal weights
($h_{1},\bar{h}_{1}$) and ($h_{2},\bar{h}_{2}$) respectively. Then by (3.34)
$<\Phi_{1}(z,\bar{z})\Phi_{2}(0,0)>\rightarrow a^{(h_{1}+h_{2})}
\bar{a}^{(\bar{h}_{1}+\bar{h}_{2})}<\Phi_{1}(f(z),\bar{f}(\bar{z}))
\Phi_{2}(0,0)>$ for $f(z)=az$. For a rotation $a=e^{i\theta}$, and thus
we require 
\begin{equation}
<\Phi_{1}(z,\bar{z})\Phi_{2}(0,0)> = \frac{E(z\bar{z})}{z^{h_{1}+h_{2}}
\bar{z}^{\bar{h}_{1}+\bar{h}_{2}}}
\end{equation}
for the transformation to have no effect. In (3.35) $E$ is an arbitrary scalar
function. If, in addition we have scale invariance (corresponding to a fixed
point), \emph{i.e.} $a=re^{i\theta}$, we have a stronger condition for
invariance,
\begin{equation}
<\Phi_{1}(z,\bar{z})\Phi_{2}(0,0)> = \frac{E}{z^{h_{1}+h_{2}}
\bar{z}^{\bar{h}_{1}+\bar{h}_{2}}}
\end{equation}
where now $E$ must be a constant.
 
\section{The operator product expansion}

To generate the operator product expansion we construct a conserved charge on
the conformal $z$-plane,
\begin{equation}
Q=\frac{1}{2\pi i}\int\! \left(dzT(z)\epsilon(z)+d\bar{z}\bar{T}
(\bar{z})\bar{\epsilon}(\bar{z})\right),
\end{equation}
associated with the conserved stress tensor discussed above. For the full
conformal group there are an infinite set of conserved currents (associated
with scale invariance at a fixed point), $\epsilon(z)T(z)$ and their
antiholomorphic partners. These charges can then be used to generate an
infinitesimal symmetry variation in the field, $\Phi$, given by
\begin{eqnarray}
\delta_{\epsilon,\bar{\epsilon}}\Phi(w,\bar{w})&=&\frac{1}{2\pi i}\int [dzT(z)
\epsilon(z), \Phi (w,\bar{w})] \nonumber \\
&+&\frac{1}{2\pi i}\int [d\bar{z} \bar{T}(\bar{z})
\bar{\epsilon}(\bar{z}), \Phi(w,\bar{w})]  \\
&=&\frac{1}{2\pi i} \oint_{w} dz \epsilon(z) T(z) \Phi(w,\bar{w}) \nonumber \\
&+&\frac{1}{2\pi i} \oint_{\bar{w}} d\bar{z} \bar{\epsilon}(\bar{z})\bar{T}
(\bar{z})\Phi(w,\bar{w})
\end{eqnarray}
where we have combined the two $z$ integrations in the commutator into a
single contour integral around $w$ [4] (figure 3.1) and
\begin{figure}
\vspace{-188mm}
\hspace{7mm}\scalebox{0.6}{\includegraphics*[0pt,0pt][1000pt,1000pt]{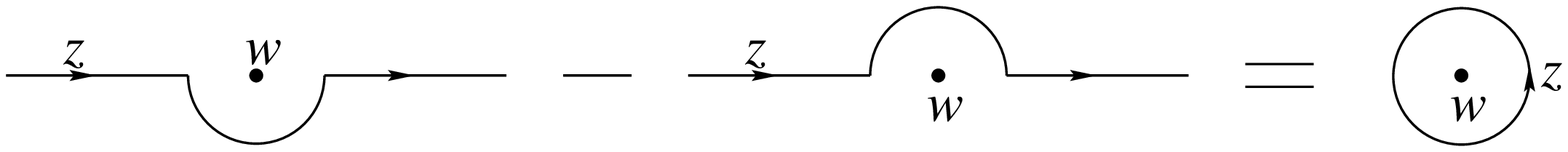}} 
\caption{Evaluation of the commutator (3.38) on the conformal plane}
\end{figure}
normalised with a factor $\frac{1}{2\pi i}$. However,
taking\footnote{consistent with the transformation used in section 3.2 for
small $\epsilon (z)=-z^{n+1}$}
$f(z)=z+\epsilon(z)$ and using (3.34) we can expand for small $\epsilon$ to
find
\begin{eqnarray}
\Phi (w,\bar{w}) &\rightarrow& (1+h\partial_{z}\epsilon)
(1+\bar{h}\partial_{\bar{z}}\bar{\epsilon}) (\Phi + \epsilon\partial_{z}\Phi +
\bar{\epsilon}\partial_{\bar{z}}\Phi)  \\
                 &=& \Phi + h(\partial_{z}\epsilon)\Phi +
                 \bar{h}(\partial_{\bar{z}}\bar{\epsilon})\Phi +
                 \epsilon\partial_{z}\Phi +
                 \bar{\epsilon}\partial_{\bar{z}}\Phi.
\end{eqnarray}
Then, for (3.39) to be consistent with (3.41) we deduce that we must have
\begin{equation}
T(z)\Phi(w,\bar{w})=\frac{h}{(z-w)^{2}}\Phi(w,\bar{w}) +
\frac{1}{z-w}\partial_{w}\Phi(w,\bar{w}) + \cdots
\end{equation}
and similarly
\begin{equation}
\bar{T}(\bar{z})\Phi(w,\bar{w})=\frac{\bar{h}}{(\bar{z}-\bar{w})^{2}}
\Phi(w,\bar{w}) + \frac{1}{\bar{z}-\bar{w}}\partial_{\bar{w}}\Phi(w,\bar{w}) +
\cdots
\end{equation}
where the dots contain contributions analytic in $z$. We may verify (3.42) and
(3.43) by substituting into (3.39) and performing the contour integration,
reproducing the variation in (3.41). The expressions (3.42) and (3.43) are
known as the short distance operator expansion and apply generally to a tensor
of weight ($h,\bar{h}$).

Alternatively, in the corresponding quantum theory, $T(z)$ is written as a
normal 
ordered product of two fields and Wicks rules can be used to reproduce the
operator expansions above. Then, it is found that in the 
case of the expansion of $T$ with itself an additional term arises due to
double Wick contractions, which can be written [4]
\begin{equation}
T(z)T(w)=\frac{c}{2(z-w)^{4}}+\frac{2}{(z-w)^{2}}T(w)+\frac{1}{(z-w)^{2}}
\partial_{w}T(w)+\cdots
\end{equation}
where $c$ is known as the conformal anomaly\footnote{also known as the
Virasoro central charge} [37], and the coefficient of $T(w)$
reflects the fact that $T$ has a conformal weight $h=2$. A similar
expression for the expansion of $\bar{T}$ with itself can be found, however we
will now concentrate purely on the unbarred case. Observe that as the stress
tensor written as a normal ordered product annihilates the vacuum, the
two-point function is given by 
\begin{equation}
<T(z)T(0)>=\frac{c}{2z^{4}},
\end{equation}
and thus it is expected that $c\geq0$ for a unitary theory. Note that this is
precisely the form required at a fixed point by (3.36).

\section{The Virasoro algebra}  

To generate the Virasoro algebra [4] it is useful to define a Laurent expansion
of the stress tensor in terms of operator modes,
\begin{equation}
T(z)=\sum_{n\epsilon Z} z^{-n-2} L_{n}
\end{equation}
with
\begin{equation}
L_{n}=-\frac{1}{2\pi i}\oint dz z^{n+1} T(z).
\end{equation}
Classically these are simply Fourier modes but become operators modes in the
context of quantum field theory. From (3.15) we substitute $\epsilon (z)=
-z^{n+1}$ into (3.39) to verify that (3.47) is indeed the infinitesimal
generator. The exponent, $z^{-n-2}$, is chosen to give the operators, $L_{n}$,
a scaling dimension of $n$ under a scaling $z\rightarrow a z$, consistent
with (3.18), since then $T(z)\rightarrow\frac{1}{a^{2}}T(a z)$ by
(3.44) and thus $L_{n}\rightarrow a^{n}L_{n}$ in (3.47).

To derive the algebra satisfying the above modes we must commute the operators
$L_{n}$ to give a double contour integral,
\begin{equation}
[L_{n},L_{m}]=\left(\oint\frac{dz}{2\pi i}\oint\frac{dw}{2 \pi i} -
\oint\frac{dw}{2\pi i}\oint\frac{dz}{2 \pi i}\right) z^{n+1}T(z)w^{m+1}T(w).
\end{equation}
The $z$ contour integration can then be performed by initially fixing $w$ and
deforming the difference between the two $z$ integrations in (3.48) into a
single contour around $w$ (figure 3.2). If we then use the stress tensor self
product, (3.44), and finally integrate over $w$ we can write the commutator as,
\begin{eqnarray}
[L_{n},L_{m}] & = & \oint\frac{dw}{2\pi i} w^{m+1}\oint_{w}\frac{dz}{2\pi i}
z^{n+1}\left(\frac{c}{2(z-w)^{4}}\right. \nonumber \\
              & + & \left. \frac{2}{(z-w)^{2}}T(w)+\frac{1}{(z-w)^{2}}
\partial_{w}T(w)+\cdots\right)  \\ 
              & = & \oint\frac{dw}{2\pi i}w^{m+1} 
\left(\phantom{\frac{1}{1}}\!\!\!\!
\frac{c}{12}(n+1)n(n-1)w^{n-2}\right. \nonumber \\  
              & + & \left. 2(n+1)w^{n}T(w)+w^{n+1}\partial_{w}T(w)+\cdots
\phantom{\frac{1}{1}}\!\!\!\!\! \right).  
\end{eqnarray}
where we have explicitly performed the $z$ integration. Re-expressing the
third integral in (3.50) by parts and combining it with the second term, we
finally arrive at 
\begin{equation}
[L_{n},L_{m}]=\frac{c}{12}(n^{3}-n)\oint\frac{dw}{2\pi i}w^{n+m-1} +
(n-m)\oint\frac{dw}{2\pi i}w^{n+m+1}T(w).
\end{equation}
Referring to (3.47) we identify
\begin{figure}
\vspace{-170mm}
\hspace{7mm}\scalebox{0.6}{\includegraphics*[0pt,0pt][1000pt,1000pt]{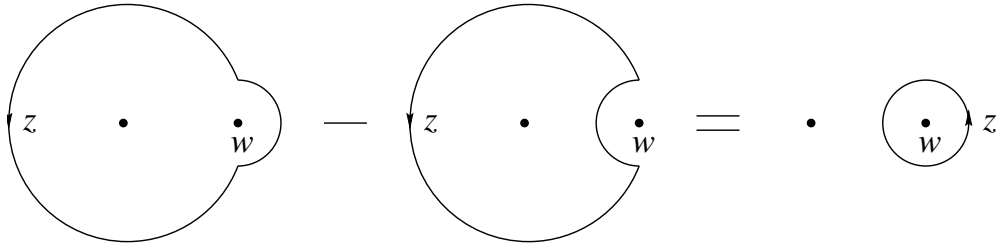}}
\caption{Evaluation of the commutator (3.48) on the conformal plane}
\end{figure}
the second integral as $L_{m+n}$ and then performing the
first integral we derive the Virasoro algebra,
\begin{equation}
[L_{n},L_{m}]=\frac{c}{12}(n^{3}-n)\delta_{n+m,0}+(n-m)L_{n+m}
\end{equation}
and similarly
\begin{equation}
[\bar{L}_{n},\bar{L}_{m}]=\frac{\bar{c}}{12}(n^{3}-n)\delta_{n+m,0} +
(n-m)\bar{L}_{n+m}.
\end{equation}
Every conformally invariant quantum field theory has a particular
representation of this algebra with some value of $c$ or $\bar{c}$. We observe
that the Virasoro algebra reduces to the classical case for $c=\bar{c}=0$,
(3.19) and (3.20).

\section{Counting}

In the following sections of this chapter we motivate, review, and discuss the
implications of Zamalodchikov's celebrated $C$-theorem [24] for two dimensional
quantum field theory, and establish three
important properties for his $C$-function. The third property depends on the
conformal anomaly and thus only makes sense in two dimensions. A number of
groups have sought to 
generalise these ideas to higher dimensions [25]. Then, within the
LPA, we will display a $C$-function which has the first two of these properties
and an appropriate generalisation of the third property in any dimension
$D$. At fixed points, our $C$-function is
extensive, videlicet additive in 
mutually non-interacting degrees of freedom, as is also true of the conformal
anomaly. 

Suppose that the fields of the theory form two non-interacting sets. Let us
write
$\phi_{a}=\phi_{a}^{(1)}$ when the field belongs to the first set, and
$\phi_{a}=\phi_{a}^{(2)}$ when it belongs to the second set. Similarly, as they
are non-interacting, the couplings $g$ split into two sets denoted $g^{(1)}$
and $g^{(2)}$ and the action can be written
\begin{equation}
S[\phi,t]=S^{(1)}[\phi^{(1)},t]+S^{(2)}[\phi^{(2)},t].
\end{equation}
Then, the variation in the action and thus the stress tensor, can be written
as a sum of two separate parts as can be seen from (3.23). Furthermore, the
generators given by (3.47) separate into two independent pieces and thus the
Virasoro algebra in (3.52) leads to the conclusion that $c$ is extensive at
fixed points, \emph{i.e.}
\begin{equation}
c(g_{*})=c(g_{*}^{(1)})+c(g_{*}^{(2)}). 
\end{equation}
This counting will be the third property of our $C$-function
derived below. 

The interpretation given to this is one of counting degrees of freedom. At the
Gaussian fixed point, $V^{G}_{*}=0$, the conformal anomaly counts one degree of
freedom per scalar. At the High Temperature fixed point,
$V_{*}^{HT}=\frac{1}{2}\phi^{2}-\frac{N}{D}$, the conformal anomaly vanishes,
corresponding to an infinitely massive theory\footnote{in units of
$\Lambda$} with no degrees of freedom. Thus, we will normalise such that our 
$C$ counts one for each Gaussian scalar and zero at the High Temperature fixed
point.

\section{Zamalodchikov's $C$-theorem}

Zamalodchikov's $C$-thereom provides an explicit geometric framework for the
space of two dimensional quantum field theories [21,22] and demonstrates the
irreversibility of renormalization group flows, thus proving that exotic flows
are missing. Zamalodchikov established three properties for his $C$-function,

I) There exists a function $C(g)\geq 0$ of such a nature that
\begin{equation}
\frac{d C}{d t}\equiv\beta^{i}(g) \frac{\partial C(g)}{\partial g^{i}} \leq 0
\end{equation}
where $g^{i}$ form an infinite set of parameters (the coupling constants
discussed in section 1.10) 
and the beta functions are defined by (1.41).

II) $C(g)$ is stationary only at the fixed points of the renormalization
group, \emph{i.e.} when $g(t)=g_{*}$.

III) The value of $C(g)$ at the fixed point $g_{*}$ is the same as the
corresponding conformal anomaly [37]. (This property thus only makes sense in
two dimensions.)

The proof relies on rotational invariance, reflection positivity and the
conservation of the stress tensor (a property which is a general consequence
of translational invariance and is also valid away from the critical
point). Consider some point on a renormalization group trajectory (refer to
figure 1.4) specified by 
a set of couplings, $g^{i}$. For the time being we suppress the dependence on
$g^{i}$. As discussed, away from the fixed point the Hamiltonian is no longer
invariant under dilations and in addition to the components $T$ and $\bar{T}$
the stress tensor has non-zero trace, $\Theta$. These components have
conformal weights (2,0), (0,2) and (1,1) respectively, so that by (3.35) we
can write their two-point functions as
\begin{eqnarray}
<T(z,\bar{z})T(0,0)> &=& \frac{F(z\bar{z})}{z^{4}}, \\
<\Theta(z,\bar{z})T(0,0)>\ \  =\ \  <T(z,\bar{z})\Theta(0,0)> &=&
\frac{G(z\bar{z})}{z^{3}\bar{z}}, \\
<\Theta(z,\bar{z})\Theta(0,0)> &=& \frac{H(z\bar{z})}{z^{2}\bar{z}^{2}}.
\end{eqnarray} 
where $F, G$ and $H$ are non-trivial scalar functions. Taking the correlation
function of (3.32) with $T(0,0)$ and $\Theta(0,0)$ yields two equations
\begin{eqnarray}
G &=& \frac{1}{3}(\dot{G}+4\dot{F}) \\
G &=& \dot{G}+\frac{1}{4}(\dot{H}-2H)
\end{eqnarray}
where we define $\dot{F}=z\bar{z}F'(z\bar{z})$. Then, defining
\begin{equation}
C=2F-G-\frac{3}{8}H
\end{equation}
and eliminating $G$ from (3.60) and (3.61) yields
\begin{equation}
\dot{C}=-\frac{3}{4}H.
\end{equation}
Reflection positivity requires that $<\!\Theta\Theta\!>\ \geq 0$, so that by
(3.59) 
$H\geq 0$. Then from (3.63) $\dot{C}\leq 0$ which implies $C'\leq 0$ as
$z\bar{z}\geq 0$. This in turn implies that $C$ is a non-increasing function of
$r=(z\bar{z})^{\frac{1}{2}}=((x^{0})^{2}+(x^{1})^{2})^{\frac{1}{2}}\geq 0$ as
$C'=2r\frac{\partial C}{\partial r}$. 

Now imagine making a renormalization group transformation such that the
lattice spacing $a\rightarrow a+\delta a$. Since our $C$-function is
dimensionless, this is equivalent to sending $r\rightarrow r-\delta r$, and
the coupling constants $g^{i}$ will flow according to the renormalization
group equation,
\begin{equation}
\left(\beta_{i}(g)\frac{\partial}{\partial g^{i}}-r\frac{\partial}{\partial r}
\right)C(r,g^{i})=0.
\end{equation}
If we now define $C(g)\equiv C(r_{o},g^{i})$, where $r_{o}$ is some arbitrary
but fixed length scale, then given that $\frac{\partial C}{\partial r}\leq 0$,
we deduce the first Zamalodchikov property for $C$. Moreover, $C$ is
stationary if and only if $H=0$, which by reflection positivity, implies
$\Theta=0$, so 
that the theory is scale invariant (condition for (3.27)) and thus corresponds
to a fixed point, which implies property two. Finally at such a
fixed point, the vanishing of the trace of 
the stress tensor ($\Theta$) by (3.27) implies that $G=H=0$ by (3.58) and
(3.59). Thus (3.62) becomes $F=\frac{1}{2}C$, so that
comparing (3.45) with (3.57) shows that $C=c$ verifying the final
Zamalodchikov property.

\section{$M_{ij}$ as a metric}

We begin by noticing that $M_{ij}$, defined by (2.60), can be interpreted as a
positive definite metric [22] in the space of quantum field
theories. Initially we consider the interval,
\begin{eqnarray}
ds^{2} & = & M_{ij}(g)dg^{i}dg^{j}  \\
       & = & a^{N}\!\!\int\!\! d^{N}\phi\ G\left(dg^{i}\frac{\partial
       \rho}{\partial g^{i}}\right) ^{2}  \\
       & \geq & 0.
\end{eqnarray}
The final observation (positive definiteness) holds providing $V$ is real
(corresponding to a unitary theory in Minkowski space). The equality in (3.67)
is only reached if $dg^{i}=0$. Perhaps more importantly we observe that the
interval is indeed invariant under a change of coordinates,
\begin{eqnarray}
ds^{2} & = & M'_{ij}(g')dg'^{i}dg'^{j} \\
       & = & M'_{ij}(g')\frac{\partial g'^{i}}{\partial g^{k}}\frac{\partial
       g'^{j}}{\partial g^{l}}dg^{k}dg^{l} \\
       & = & M_{kl}dg^{k}dg^{l}.
\end{eqnarray}

\section{A $C$-function representation of the LPA}

We define our $C(g)$-function through
\begin{equation}
\mathcal{F}=\frac{Db^{C}}{4}
\end{equation}
where $b>1$ and $\mathcal{F}$ is defined by (2.56). This $C$-function
satisfies our appropriate generalisation of Zamalodchikov's properties in
general dimension, $D$. Generally
\begin{equation}
\frac{\partial b^{C}}{\partial g^{i}} = \frac{\partial C}{\partial g^{i}}
b^{C} \ln(b).
\end{equation}
However, by differentiating (3.71) with respect to $g^{i}$ and using (2.59) we
find 
\begin{equation}  
\frac{\partial b^{C}}{\partial g^{i}} = -\frac{4}{D} {M}_{ij} \beta^{j}
\end{equation} 
and therefore 
\begin{equation}
\frac{\partial C}{\partial g^{i}}=-\hat{M}_{ij}\beta^{j}
\end{equation}
where we have defined $\hat{M}_{ij}$ through
$M_{ij}=(\mathcal{F}\mathrm{ln}b)\hat{M}_{ij}$. Thus, we find that
\begin{eqnarray}
\frac{d C}{d t} & = & \beta^{i} \frac{\partial C}{\partial g^{i}} \\
                & = & -\beta^{i}\hat{M}_{ij}\beta^{j}.
\end{eqnarray}
Using an argument similar to that used to derive (3.67) we can now write
\begin{eqnarray}
\frac{d C}{d t} & = & a^{N}\!\!\int\!\! d^{N}\phi\
G\left(\beta^{i}\frac{\partial \rho}{\partial g^{i}}\right)^{2} \\
                & \leq & 0
\end{eqnarray}
confirming the first Zamalodchikov property. Given that the $\beta$-functions
vanish at a fixed point, the second property follows
immediately from (3.74).

The appropriate generalisation of the third property is found by supposing
that the fields form two mutually non-interacting sets. Using our previous
notation we write
$\phi_{a}=\phi_{a}^{(1)}$ when the field belongs to the first set, and
$\phi_{a}=\phi_{a}^{(2)}$ when it belongs to the second set, such that
the couplings $g$ also split into two sets denoted $g^{(1)}$
and $g^{(2)}$ and by (3.54) the potential can be written
\begin{equation}
V(\phi,t)=V^{(1)}(\phi^{(1)},t)+V^{(2)}(\phi^{(2)},t)
\end{equation}
and thus $\rho$ factorises: $\rho=\rho^{(1)}\rho^{(2)}$. However, substituting
the fixed point equation $\frac{\delta \mathcal{F}}{\delta \rho}=0$ into
(2.56), we derive that fixed points $\rho (\phi,t)=\rho_{*}(\phi)$ satisfy
\begin{equation}
\mathcal{F}[\rho_{*}]=\frac{D}{4}a^{N}\!\!\int\!\! d^{N}\phi\ G \rho_{*}^{2}
\end{equation}
and we immediately notice that
$\mathcal{F}[\rho_{*}]=\frac{4}{D}\mathcal{F}[\rho_{*}^{(1)}] 
\mathcal{F}[\rho_{*}^{(2)}]$. Thus from (3.71) we observe that at fixed points
our $C$ is extensive, as required by (3.55).

In two dimensions, at the Gaussian fixed point, the conformal anomaly counts
one degree of freedom per scalar, \emph{i.e.} is equal to $N$, and at the High
Temperature fixed point, it vanishes. We will take this to be true in general
dimension $D$. Thus the normalisation
factors $a$ and $b$ can be uniquely determined by requiring that our $C$ agree
with this counting in any dimension $D$. Substituting $C=0$ in (3.71), and
$\rho_{*}=\exp (-V_{*}^{HT})$ in (3.80) and performing the Gaussian integral
yields
\begin{equation}
a=e^{2/D}\sqrt{\frac{D+2}{4\pi}}.
\end{equation}
Similarly, for the Gaussian fixed point we substitute $C=N$ in (3.71) and
$\rho_{*}=\exp (-V_{*}^{G})=1$ in (3.80) and find
\begin{equation}
b=e^{2/D}\sqrt{\frac{D+2}{D-2}}.
\end{equation}
Note that as required, $b>1$, at least for $D\geq2$.

\section{Examples}

Consider the simple example of the Gaussian fixed point perturbed by the mass
operator, for a single scalar field. Thus we set $V(\phi,
t)=g^{1}(t)+\frac{1}{2}g^{2}(t)\phi^{2}$. Then the $\beta$-functions are given
by (2.39) and (2.40). Integrating (2.40) yields
\begin{equation}
g^{2}(t)=\frac{1}{1+ke^{-2t}}
\end{equation}
and then (2.39) yields
\begin{eqnarray}
g^{1}(t) &=& g^{1}_{0}e^{Dt}+e^{Dt}\!\int\!\!\frac{e^{-D\tau}}{1+ke^{-2\tau}}
     d\tau \\ 
     &=& g^{1}_{0}e^{Dt}-\frac{1}{2}\int^{1}_{0}\!\!
     \frac{u^{D/2-1}}{1+uke^{-2t}}du, 
\end{eqnarray}
where $u=e^{2t-2\tau}$ and  $k$ and $g^{1}_{0}=g^{1}(0)$ are 
constants of integration. Considering the limit 
$t\rightarrow -\infty$, we see that this solution indeed emanates from the
Gaussian fixed point. We normalised the special solution in (3.85) so that
with $g^{1}_{0}=0$, it tends to $V_{*}^{HT}$ as $t\rightarrow +\infty$.

Since $\rho$ is Gaussian, $\mathcal{F}$ is readily determined using (2.56),
\begin{equation}
\mathcal{F}=\frac{a}{2}e^{-2g^{1}}\left(2g^{1}D +
\frac{4(g^{2})^{2}+6g^{2}D-2D+D^{2}}{D-2+4g^{2}}\right)
\sqrt{\frac{\pi}{D-2+4g^{2}}}. 
\end{equation}
In particular, using (3.81) we verify
\begin{equation}
\mathcal{F}^{G}=\frac{D}{4}e^{2/D}\sqrt{\frac{D+2}{D-2}}
\end{equation}
and
\begin{equation}
\mathcal{F}^{HT}=\frac{D}{4}
\end{equation}
for the Gaussian ($g^{1}=g^{2}=0$) and High Temperature ($g^{1}=-\frac{1}{D}$
and $g^{2}=1$) respectively. Then using (3.71) and (3.82) we deduce that
$C^{G}=1$ and $C^{HT}=0$ as expected. Indeed we may further verify that when
$g^{1}_{0}=0$, $C(t)$ flows 
from 1 to 0 as $t$ runs from $-\infty$ to $+\infty$. Note that if
$g^{1}_{0}\neq0$, then $C(t)\rightarrow\pm\infty$ as $t\rightarrow\infty$,
depending on the sign of $g^{1}_{0}$. This seems in contradiction with the idea
that $C$ counts degrees of freedom, since evidently the vacuum energy should
not figure \emph{per se} in this counting. However, we recall from (3.55) that
$C$ is extensive (\emph{i.e.} counts) only at fixed points, and with
$g^{1}_{0}\neq0$ the system never reaches another fixed point as
$t\rightarrow\infty$. 

From the Gaussian form of $G$, we recognise that at both the Gaussian and High
Temperature fixed points, the metric (2.60) is diagonalised by choosing the
operators $\mathcal{O}_{i}=\frac{\partial V}{\partial g^{i}}$ to be products
of Hermite polynomials $H_{n}$ in the $\phi_{a}$. Since these also turn out to
diagonalise $\frac{\partial^{2}\mathcal{F}}{\partial g^{i}\partial
g^{j}}(g_{*})$, they are the eigenperturbations, and the corresponding
eigenvalues follow straightforwardly. Choosing the Gaussian fixed point
$\rho_{*}=1$ for example, and again specialising to the case of a single
scalar field for simplicity, we thus take
$\mathcal{O}_{n}=H_{n}(\frac{\phi}{2}\sqrt{D-2})$, $n=0,1,\cdots$. From (2.56)
we then obtain 
\begin{equation}
\frac{\partial^{2} \mathcal{F}}{\partial g^{i} \partial g^{j}} =
-DM_{ij}+a\int\!\! d\phi\ G  \frac{\partial \mathcal{O}_{i}}{\partial \phi} 
\frac{\partial \mathcal{O}_{j}}{\partial \phi}.
\end{equation}
However, using $H_{n}'=2nH_{n-1}$ where prime denotes differentiation with
respect to the argument, we deduce $\frac{\partial \mathcal{O}_{n}}{\partial
\phi}=n\sqrt{D-2}\mathcal{O}_{n-1}$ so that (3.89) yields
\begin{equation}
\frac{\partial^{2} \mathcal{F}}{\partial g^{i} \partial g^{j}} =
-DM_{ij}+ij(D-2)M_{i-1\ j-1}.
\end{equation}
Then, given that the metric has  non-zero components [38]
$M_{nn}=2^{n+1}n!a\sqrt{\frac{\pi}{D-2}}$, from (2.62) we recover the Gaussian 
spectrum of eigenvalues, 
\begin{eqnarray}
\lambda_{n} &=& -\frac{1}{M_{nn}}
\frac{\partial^{2} \mathcal{F}}{\partial g^{n} \partial g^{n}} \\
        &=& D-\frac{1}{2}(D-2)n
\end{eqnarray}
as expected. The High Temperature fixed point can be solved in a similar
manner.  

\section{Discussion}

The $C$-theorem has the interpretation that renormalization group flows go
`downhill'. As described it rules out the existence of exotic flows and in
particular restricts the possible fixed points to which unstable directions at
a given fixed point may flow. Various authors [39] have reported on the
appealing interpretation of the $C$-function as a kind of entropy of
information about the critical system. Under renormalization, information is
lost about the short distance behaviour of the correlation functions,
corresponding to a decreasing $C$.

The value of the $C$-function at a fixed point is interpreted as the number of
degrees of freedom, which may initially seem counter intuitive given that the
$C$-function we have defined (3.71) is non-integer at the Wilson-Fisher fixed
point\footnote{the exact LPA solution yields $c=0.9867$ whereas variation
yields $c=0.9896$}. However, the following example may convince us that
this is in fact not as unusual as we might imagine. Consider a ball attached to
a solid wall on a spring. Initially the ball might have one degree of freedom
(position in one dimension). Then imagine increasing the spring
stiffness in a smooth manner until eventually the ball is permanently
stationary. The degrees of freedom of the ball have reduced smoothly from one
to zero. 

It is probably not possible within the Local Potential Approximation to
establish a more concrete link between our $C$-function  and Zamalodchikov's
$C$. Of course, it would be very interesting to understand if these
observations generalise to higher orders in the derivative expansion [40]
(which likely would allow a direct comparison with Zamalodchikov's $C$), or
indeed generalise to an exact expression along the present lines.

\newpage

\chapter*{Appendices}
\addcontentsline{toc}{chapter}
{\numberline{Appendices}}

\section*{Appendix 3A: The stress tensor}

Generally, if the Lagrangian density $\mathcal{L}$ depends only on the field
$\phi (x)$ and its derivative $\partial_{\mu}\phi (x)$, the variation in the
Lagrangian can be written
\begin{eqnarray}
\delta\mathcal{L}&=&\frac{\delta\mathcal{L}}{\delta \phi(x)}\delta\phi(x) +
\frac{\delta\mathcal{L}}{\delta(\partial_{\mu}\phi(x))}
\delta\partial_{\mu}\phi (x) \\
                 &=&\left(\frac{\delta\mathcal{L}}{\delta \phi(x)} -
\partial_{\mu}\frac{\delta\mathcal{L}}{\delta(\partial_{\mu}\phi(x))}
\right)\delta\phi (x) +
\partial_{\mu}\left(\frac{\delta\mathcal{L}}{\delta(\partial_{\mu}\phi(x))} 
\delta\phi (x)\right)
\end{eqnarray}
where we have used
\begin{equation}
\delta\partial_{\mu}\phi(x)=\partial_{\mu}\delta\phi(x).
\end{equation}
To derive the equations of motion we assume that the arbitrary variations,
$\delta\phi(x)$, vanish at infinity so that the last term in (3.94) integrates
to zero in the variation of the action. However we suppose that $\phi(x)$
satisfies the equations of motion and take $\delta\phi(x)$ not as an arbitrary
variation, but a translation:
\begin{equation}
\delta\phi(x)=-\epsilon^{\nu}\partial_{\nu}\phi(x).
\end{equation}
Then the first bracket in (3.94) vanishes and substituting (3.96) we find
\begin{equation}
\delta\mathcal{L}=-\epsilon^{\nu}\partial_{\mu}
\left(\frac{\delta\mathcal{L}}{\delta(\partial_{\mu}\phi(x))}
\partial_{\nu}\phi(x)\right).
\end{equation}
However, $\mathcal{L}$ is not invariant under a global transformation, rather
we have 
\begin{equation}
\delta\mathcal{L}=-\epsilon^{\mu}\partial_{\mu}\mathcal{L}.
\end{equation}
Thus, from (3.97) and (3.98)
\begin{equation}
\epsilon^{\nu}\partial_{\mu}\left(\frac{\delta\mathcal{L}}
{\delta(\partial_{\mu}\phi)}\partial_{\nu}\phi - 
\delta^{\mu}_{\nu}\mathcal{L}\right)=0
\end{equation}
which, since it is true for all $\epsilon^{\nu}$, implies
\begin{equation}
\partial^{\mu}T_{\mu\nu}=0
\end{equation}
where we have defined
\begin{equation}
T_{\mu\nu}=\frac{\delta\mathcal{L}}
{\delta(\partial_{\mu}\phi)}\partial_{\nu}\phi - 
\delta^{\mu}_{\nu}\mathcal{L}.
\end{equation}
The stress tensor, $T_{\mu\nu}$, is the conserved quantity corresponding to
translational invariance.

Now consider the variation of the action with a local translation
$\epsilon^{\nu}(x)$,
\begin{eqnarray}
\delta S &=& \!\int\!\!d^{D}x\delta\mathcal{L} \\
         &=& \! -\!\int\!\!d^{D}x\left(\frac{\delta\mathcal{L}}{\delta\phi(x)}
         \epsilon^{\nu}(x)\partial_{\nu}\phi(x) +
         \frac{\delta\mathcal{L}}{\delta(\partial_{\mu}\phi(x))}
         \partial_{\mu}(\epsilon^{\nu}(x)\partial_{\nu}\phi(x))\!\right) \\
         &=& \! -\!\int\!\!d^{D}x\left(\epsilon^{\nu}(x)
         \left(\frac{\delta\mathcal{L}}{\delta\phi(x)}\partial_{\nu}\phi(x) +
         \frac{\delta\mathcal{L}}{\delta(\partial_{\mu}\phi(x))}
         \partial_{\mu}\partial_{\nu}\phi(x)\right)\right. \\
         &+&\left. \partial_{\mu}\epsilon^{\nu}(x)\frac{\delta\mathcal{L}}
         {\delta(\partial_{\mu}\phi(x))}\partial_{\nu}\phi(x)\right) \nonumber
\end{eqnarray} 
where we have used (3.95) and (3.96). Replacing the inner brackets of the
first term in (3.104) with $\partial_{\nu}\mathcal{L}$ and integrating the
last term by parts we finally arrive at
\begin{eqnarray}
\delta S &=& \!\int\!\!d^{D}x\epsilon^{\nu}(x)\partial_{\mu}\left(
\frac{\delta\mathcal{L}}{\delta(\partial_{\mu}\phi(x))}\partial_{\nu}\phi(x) -
\delta^{\mu}_{\nu}\mathcal{L}\right) \\
         &=& \!\int\!\!d^{D}x\epsilon^{\nu}(x)\partial^{\mu}T_{\mu\nu}
\end{eqnarray}
by the definition of the stress tensor (3.101).
\newcommand{\sla}[1]{#1\!\!\!/}

\chapter{Fermionic field theory}

In the previous two chapters we have assembled a considerable portfolio of
evidence in support of the Local Potential Approximation. However, it must be
admitted that (here and in general) most success has arisen from focusing on
purely Bosonic field theories, and ultimately we must extend these
constructions to include Fermions in a reliable manner. Here we will be
specifically interested in the Fermionic Polchinski and Legendre flow
equations. For a finite number of components, the LPA to the Legendre flow
equation is significantly more difficult to implement than for the scalars and
the Polchinski flow equation will form the focus of our attention. However, a
technical problem involving chiral symmetry emerges and we find that the
Polchinski LPA is trivial. These results are 
compared with exact solutions obtained in the large $N$ limit using the
Fermionic
Legendre flow equation\footnote{only the Legendre flow equation is exact in
the large $N$ limit [23]}. As a result we develop an understanding of the
Polchinski LPA for a finite number of components, which are presented towards
the end of this chapter.

\section{Introduction}

In this chapter we will be interested in Fermionic field theory, and in
particular spinors, which we label $\psi_{\alpha a}$, where here $\alpha$ runs
over internal spinor indices from 1 to $N_{i}$ and $a$ is the flavour index
running  from 1 to $N_{f}$. These satisfy the well known anti-commutation
relations including 
\begin{equation}
\{\psi_{\alpha a},\psi_{\beta b}\}=0\ \ \mathrm{and}\ \ 
\{\psi_{\alpha
a},\psi^{\dag\beta}_{\phantom{\dag\beta}b}\}=\hbar\delta_{\alpha}
^{\phantom{\alpha}\beta}\delta_{ab}\delta (x-y).
\end{equation}
This so called
Grassmann nature implies that $\psi_{i}\psi_{i}=0$ where $i$ runs over all
indices, and thus that a function of a
finite number of Fermion fields can be expanded as an exact finite
series. It is useful to use the traceless gamma matrices [1]
$\gamma^{\mu}$ ($\mu=0,\cdots,3$) and $\gamma^{5}$=$\gamma^{0}\gamma^{1}
\gamma^{2}\gamma^{3}$, which satisfy
$\{\gamma^{5},\gamma^{\mu}\}=0$ and $(\gamma^{0})^{2}=I$ where $I$ is the
unit matrix. The adjoint spinor is then defined by
$\bar{\psi}_{a}^{\alpha}=\psi^{\dag\beta}_{a}
\gamma^{0\phantom{\beta}\alpha}_{\phantom{0}\beta}$. We will be using
derivatives $\frac{\delta}{\delta \bar{\psi}}$ and $\frac{\delta}{\delta
\psi}$ which both act on the left in the conventional manner.

A theory is said to be chirally symmetric if it is unchanged under
\begin{equation}
\psi\rightarrow\gamma_{5}\psi
\end{equation}
and
\begin{equation} 
\bar{\psi}\rightarrow\bar{\psi}\gamma_{5}.
\end{equation}
In particular we observe that a chirally invariant theory has no mass term,
$\bar{\psi}\psi$, or indeed any odd power of $\bar{\psi}\psi$. However the
kinetic term $\bar{\psi}\sla{q}\psi$ is acceptable, where we define
$\sla{q}=\gamma^{\mu} q_{\mu}$. Ultimately we are interested in the
Gross-Neveu model [41] which can be defined by the bare Lagrangian
\begin{equation}
\mathcal{L}=\bar{\psi}_{a}^{\alpha}(x)\sla{\partial}_{\alpha}
^{\phantom{\alpha}\beta} 
\psi_{\beta a}(x) +  g(\bar{\psi}^{\alpha}_{a}(x)\psi_{\alpha a}(x))^{2} 
\end{equation}
which is seen to be chirally invariant.

\section{Conditions on Fermionic potentials}

To derive the condition that must be satisfied by the Fermionic potential we
impose that the Minkowski Hamiltonian density,
\begin{equation}
\mathcal{H}[\psi (x), \bar{\psi}(x)]=\dot{\psi}(x)^{\phantom{\alpha}}_{\alpha
a} \Pi^{\alpha}_{a} (x) - \mathcal{L}_{M}
[\psi (x), \bar{\psi}(x)]
\end{equation}
where the Lagrangian is given by
\begin{equation}
\mathcal{L}_{M}[\psi (x), \bar{\psi}(x)]=i\bar{\psi}^{\alpha}_{a}(x)
\sla{\partial}^{\phantom{\alpha}\beta}_{\alpha} \psi^{\phantom{\beta}}_{\beta
a}(x) - V(\psi^{\dag}(x)\gamma_{0}\psi (x)) 
\end{equation}
and
\begin{equation}
\Pi_{a}^{\alpha} (x)=\frac{\partial\mathcal{L}_{M}}{\partial\dot{\psi}_{\alpha
a}(x)}=-i\psi^{\dag\alpha}_{a}(x),
\end{equation}
must be positive definite. For convenience we suppress the indices on the
arguments of functions. The dots in (4.5) and (4.7) denote differentiation
with respect to $t=x^{0}$. By substituting (4.6) and (4.7) into (4.5) we find
an alternative expression for the Hamiltonian density
\begin{equation}
\mathcal{H}[\psi (x),\bar{\psi} (x)]=\psi^{\dag\alpha}_{a}(x)\ \!
(\mathbf{a}.\mathbf{p})^{\phantom{\alpha}\beta}_{\alpha}\ \!
\psi(x)^{\phantom{\beta}}_{\beta a} + 
V(\psi^{\dag}(x)\gamma_{0}\psi(x))
\end{equation}
where $a^{i} = \gamma^{0}\gamma^{i}$ and $p^{i}=-i\partial^{i}$. 

We begin by substituting this Hamiltonian into the equation of motion
\begin{equation}
i\dot{\psi}_{\alpha a}(x)=\left[\psi_{\alpha a}(x),\int\!\!d^{D}y \mathcal{H}[\psi (y),
\bar{\psi}(y)]\ \right]
\end{equation}
such that using (4.1) we can perform the integrals over delta functions,
yielding
\begin{equation}
i\dot{\psi}^{\phantom{\beta}}_{\alpha a}(x)=(\mathbf{a}.\mathbf{p})_{\alpha}
^{\phantom{\alpha}\beta}\ \psi^{\phantom{\beta}}_{\beta a} (x) +
\left[\psi_{\alpha a} (x), \int\!\! d^{D}yV(\psi^{\dag}(y)\gamma^{0}\psi
(y))\right]. 
\end{equation}
Comparing this with (4.8) we observe that
\begin{eqnarray}
\mathcal{H}[\psi(x),\bar{\psi}(x)]
= i\psi^{\dag\alpha}_{a}(x)\dot{\psi}^{\phantom{\alpha}}_{\alpha a} (x) -
\psi^{\dag\alpha}_{a}(x) 
\left[\psi_{\alpha a} (x), \int\!\! d^{D}yV(\psi^{\dag}(y)\gamma^{0}\psi
(y))\right] \nonumber \\
+\ V(\psi^{\dag}(x)\gamma^{0}\psi(x)). \hspace{50mm}
\end{eqnarray}
Now consider a term in the potential which takes the form
$(\psi^{\dag}(x)\gamma^{0}\psi(x))^{n}$. For this term the commutator in
(4.11) can be expanded as
\begin{eqnarray}
\left[\psi_{\alpha a} (x), \int\!\! d^{D}y(\psi^{\dag}(y)\gamma^{0}\psi
(y))^{n}\right] 
= n \int\!\! d^{D}y \left[\psi_{\alpha a} (x), \psi^{\dag}(y)\gamma^{0}\psi
(y)\right] \nonumber \\
\times\ (\psi^{\dag}(y)\gamma^{0}\psi (y))^{n-1} + O(\hbar^{2}).\hspace{33mm}
\end{eqnarray}
Using (4.1) we compute the commutator on the right side of (4.12) to find
\begin{eqnarray}
\psi^{\dag\alpha}_{a}(x)\left[\psi^{\phantom{\alpha}}_{\alpha a} (x), \int\!\!
d^{D}y(\psi^{\dag}(y)\gamma^{0}\psi 
(y))^{n}\right] = n\psi^{\dag\alpha}_{a}(x)\gamma^{0\phantom{\alpha}\beta}
_{\phantom{0}\alpha}\psi^{\phantom{\beta}}_{\beta a}(x) \nonumber \\
\times\ (\psi^{\dag}(x)\gamma^{0}\psi(x))^{n-1} + O(\hbar^{2})\hspace{30mm}
\end{eqnarray}
and thus deduce that the Hamiltonian (4.11) can be written
\begin{eqnarray}
\mathcal{H}[\psi (x),\bar{\psi}(x)] =
i\psi^{\dag\alpha}_{a}(x)\dot{\psi}^{\phantom{\alpha}}_{\alpha a}(x)-
\psi^{\dag\alpha}_{a} (x)\gamma^{0\phantom{\alpha}\beta}_{\phantom{0}\alpha}
\psi^{\phantom{\beta}}_{\beta a}(x) 
V'(\psi^{\dag}(x)\gamma^{0}\psi(x)) \nonumber \\
+\ V(\psi^{\dag}(x)\gamma^{0}\psi(x))
+ O(\hbar^{2}).\hspace{38mm}
\end{eqnarray}
Defining $z(x)=\bar{\psi}_{a}^{\alpha}(x)\psi^{\phantom{\alpha}}_{\alpha a}
(x) =\psi^{\dag\alpha}_{a}(x)\gamma^{0\phantom{\alpha}\beta}
_{\phantom{0}\alpha} \psi(x)^{\phantom{\alpha}}_{\beta a}$ we thus 
require that the quantity 
\begin{equation}
V(z) - zV'(z)
\end{equation}
must be bounded from below in mean field situations, where $\hbar^{2}$ can be
neglected. More generally we see from (4.12) that it corresponds to neglecting
terms involving higher derivatives of the potential, \emph{i.e.} those of the
form 
\begin{equation}
z\frac{\partial^{n} V}{\partial z^{n}}(z)
\end{equation}
with $n>1$. We will find that these terms may be neglected in the large $N$
limit. Furthermore, we will learn that the correct interpretation (at large
$N$) is to treat the $z(x)$ as a real variable. At finite $N$ we find that
treating $z(x)$ as even Grassmann (and thus a potential, $V(z)$, as a finite
series) yields nonsense results typical of truncations [19,20]. Thus,
throughout we treat $z(x)$ as a real number and not even Grassmann as one
might naively expect.
Finally, the condition that (4.15) is bounded from below implies that any
potential growing larger than linear for large $z(x)$, must be bounded from
below.

\section{The Legendre flow equation}

The structure of Fermionic field theory leads to a Legendre flow equation that
looks somewhat different to the analogous equation for the scalars
(2.22). However, in common with scalar theory it can be solved exactly and
analytically in the large $N$ limit. The application of the LPA to Legendre
flow equation is non-trivial and thus at finite $N$ we concentrate on the
Polchinski formalism. Ultimately we will use the large $N$ limit of the
Legendre equation to develop an understanding of the LPA to Polchinski's
equation. Begin by consider the generating functional regulated by an
infra-red cutoff, similar to scalar field theory, 
\begin{equation}
Z(\bar{\zeta},\zeta)=\int(d\bar{\psi})(d\psi)e^{-\ \bar{\psi}^{\alpha}_{a}\
\Delta^{-1\phantom{\alpha}\beta}_{IR\alpha}\
\psi_{\beta a}^{\phantom{\rho}}-\
S_{\Lambda}^{\phantom{\rho}}[\bar{\psi},\psi]\ +\ \bar{\psi}^{\alpha}_{a}\
\zeta_{\alpha a}^{\phantom{\rho}}\ +\ \bar{\zeta}^{\alpha}_{a}\ \psi_{\alpha
a}^{\phantom{\rho}}} 
\end{equation}
where $\bar{\zeta}$ and $\zeta$ are the sources.
We begin by deriving the necessary preliminaries to finding
the flow equation. Define $Z(\bar{\zeta},\zeta)=e^{W(\bar{\zeta},\zeta)}$ and
introduce the Legendre effective action, $\Gamma_{\Lambda}$, by
\begin{equation}
\Gamma_{\Lambda}(\bar{\psi}^{c}_{\phantom{j}},\psi^{c}_{\phantom{j}})\ +\
\bar{\psi}^{c\alpha}_{\phantom{c}a}\ \Delta^{-1\phantom{\alpha}\beta}_{IR
\alpha}\ 
\psi^{c}_{\phantom{c}\beta a}\ =\ -W(\bar{\zeta},\zeta)\ +\
\bar{\psi}^{c\alpha}_{\phantom{c}a}\ \zeta^{\phantom{\rho}}_{\alpha a}\ +\
\bar{\zeta}^{\alpha}_{a}\ \psi^{c}_{\phantom{c}\alpha a}
\end{equation}
so that taking the classical limit, $h\rightarrow 0$, $S_{\Lambda}\rightarrow
\Gamma_{\Lambda}$. Then, if we differentiate (4.18) with respect to
$\psi^{c}$ and $\bar{\psi}^{c}$ separately, we find
\begin{equation}
\frac{\delta\Gamma_{\Lambda}}{\delta\psi^{c}_{\phantom{c}\alpha a}}\ -\
\bar{\psi}^{c\beta}_{\phantom{c}a}\
\Delta^{-1\phantom{\beta}\alpha}_{IR\beta}\ =\ 
-\ \bar{\zeta}^{\alpha}_{a}
\end{equation}
and
\begin{equation}
\frac{\delta\Gamma_{\Lambda}}{\delta\bar{\psi}^{c\alpha}_{\phantom{c}a}}\ +\
\Delta^{-1\phantom{\alpha}\beta}_{IR\alpha}\psi^{c}_{\phantom{c}\beta a}\ =\ 
\zeta^{\phantom{\rho}}_{\alpha a}
\end{equation}
where the classical fields are defined by 
\begin{equation}
\bar{\psi}^{c\alpha}_{\phantom{c}a}\ =\ -\ \frac{\delta W}{\delta
\zeta^{\phantom{\rho}}_{\alpha a}}
\end{equation}
and
\begin{equation}
\psi^{c}_{\phantom{c}\alpha a}\ =\ \frac{\delta W}{\delta
\bar{\zeta}^{\alpha}_{a}} 
\end{equation}
similar to the scalar case (2.18). Then, by differentiating (4.19) and (4.20)
with respect to $\bar{\psi}^{c}$ or $\psi^{c}$ and differentiating (4.21)
and (4.22) once 
with respect to $\bar{\zeta}$ or $\zeta$ we can rewrite the following matrix
equation  
\begin{equation}
\left( \begin{array}{cc}
\frac{\delta\psi^{c}}{\delta\bar{\zeta}}
  &
\frac{\delta\psi^{c}}{\delta\zeta}
 \\
\frac{\delta\bar{\psi}^{c}}{\delta\bar{\zeta}}
  & 
\frac{\delta\bar{\psi}^{c}}{\delta\zeta}
\end{array} \right)_{ab\ \alpha}^{\phantom{ab\ \alpha}\beta}
\left( \begin{array}{cc}
\frac{\delta\bar{\zeta}}{\delta\psi^{c}}
  &
\frac{\delta\bar{\zeta}}{\delta\bar{\psi}^{c}}
 \\
\frac{\delta\zeta} {\delta\psi^{c}} 
  & 
\frac{\delta\zeta}{\delta\bar{\psi}^{c}}
\end{array} \right)_{bd\ \beta}^{\phantom{bd\ \beta}\gamma}
=
\left( \begin{array}{cc}
I & 0 \\
0 & I 
\end{array} \right)_{ad\ \alpha}^{\phantom{ad\ \alpha}\gamma}
\end{equation}
as
\begin{eqnarray}
\left( \begin{array}{cc}
\frac{\delta^{2} W}{\delta\bar{\zeta}\delta\bar{\zeta}}
  &
\frac{\delta^{2} W}{\delta\zeta\delta\bar{\zeta}}
 \\
-\ \frac{\delta^{2} W}{\delta\bar{\zeta}\delta\zeta}
  & 
-\ \frac{\delta^{2} W}{\delta\zeta\delta\zeta}
\end{array} \right)_{ab\ \alpha}^{\phantom{ab\ \alpha}\beta}
\left( \begin{array}{cc}
-\ \frac{\delta^{2}\Gamma_{\Lambda}}{\delta\psi^{c}\delta\psi^{c}}
  &
-\ \frac{\delta^{2}\Gamma_{\Lambda}}{\delta\bar{\psi}^{c}\delta\psi^{c}}
+ \Delta^{-1}_{IR}
 \\
\frac{\delta^{2}\Gamma_{\Lambda}}{\delta\psi^{c}\delta\bar{\psi}^{c}}
+ \Delta^{-1}_{IR}
  & 
\frac{\delta^{2}\Gamma_{\Lambda}}{\delta\bar{\psi}^{c}\delta\bar{\psi}^{c}}
\end{array} \right)_{ad\ \alpha}^{\phantom{ad\ \alpha}\gamma} \nonumber \\
=
\left( \begin{array}{cc}
I & 0 \\
0 & I 
\end{array} \right)_{ad\ \alpha}^{\phantom{ad\ \alpha}\gamma}\ \ \ \ \ \ \ \ \ \ \ \ \ \ \ \ \ \ \ \ \ \ \ \ 
\end{eqnarray}
where $I$ is the unit matrix

With the preliminaries complete we begin the derivation by differentiating the
generating functional, (4.17), with respect to the cutoff, $\Lambda$,
\begin{equation}
\frac{\partial Z}{\partial \Lambda}=\frac{\delta}{\delta
\zeta^{\phantom{\rho}}_{\alpha a}}\left(\frac{\partial
\Delta^{-1\phantom{\alpha}\beta}_{IR\alpha}}{\partial 
\Lambda} \frac{\delta Z}{\delta \bar{\zeta}^{\beta}_{a}}\right)
\end{equation}
or alternatively, substituting $Z=e^{W}$,
\begin{equation}
\frac{\partial W}{\partial \Lambda}=\frac{\delta W}{\delta
\zeta^{\phantom{\rho}}_{\alpha a}}\frac{\partial
\Delta^{-1\phantom{\alpha}\beta}_{IR\alpha}}{\partial \Lambda}\frac{\delta
W}{\delta \bar{\zeta}^{\beta}_{a}}\ +\ 
\mathrm{tr}\left(\frac{\partial
\Delta^{-1\phantom{\alpha}\beta}_{IR\alpha}}{\partial \Lambda} 
\frac{\delta}{\delta \zeta^{\phantom{\rho}}_{\alpha a}}\frac{\delta W}{\delta
\bar{\zeta}^{\beta}_{a}}\right).
\end{equation}
Then we differentiate (4.18) with respect to $\Lambda$ and use (4.21) and
(4.22) to find
\begin{equation}
\left. \frac{\partial W}{\partial \Lambda}\right|_{\zeta ,\bar{\zeta}} =-\ \!\!
\left. \frac{\partial \Gamma_{\Lambda}}{\partial \Lambda}\ \right|_{\psi^{c} ,
\bar{\psi}^{c}}\!\! -\ \bar{\psi}^{c\alpha}_{\phantom{c}a}
\frac{\partial\Delta^{-1\phantom{\alpha}\beta} 
_{IR\alpha}}{\partial \Lambda}\psi^{c}_{\phantom{c}\beta a}.
\end{equation}
Equating (4.26) and (4.27) and again using (4.21) and (4.22) we finally arrive
at our flow equation
\begin{equation}
\frac{\partial\Gamma_{\Lambda}}{\partial\Lambda}=-\frac{1}{2}\mathrm{tr}\left(
\left( \begin{array}{cc}
0 & \frac{\partial \Delta^{-1}_{IR}}{\partial \Lambda} \\
\frac{\partial \Delta^{-1}_{IR}}{\partial \Lambda} & 0 
\end{array} \right)_{\alpha}^{\phantom{\alpha}\beta}
\left( \begin{array}{cc}
\frac{\delta^{2} W}{\delta\bar{\zeta}\delta\bar{\zeta}}
  &
\frac{\delta^{2} W}{\delta\zeta\delta\bar{\zeta}}
 \\
-\frac{\delta^{2} W}{\delta\bar{\zeta}\delta\zeta}
  & 
-\frac{\delta^{2} W}{\delta\zeta\delta\zeta}
\end{array} \right)_{aa\ \beta}^{\phantom{aa\ \beta} \alpha}
\right)
\end{equation}
which can be rewritten using (4.24),
\begin{eqnarray}
\frac{\partial\Gamma_{\Lambda}}{\partial\Lambda}=-\frac{1}{2}\mathrm{tr}\left(
\left( \begin{array}{cc}
0 & \frac{\partial \Delta^{-1}_{IR}}{\partial \Lambda} \\
\frac{\partial \Delta^{-1}_{IR}}{\partial \Lambda} & 0 
\end{array} \right)_{\alpha}^{\phantom{\alpha}\beta} \right. \ \ \ \ \ \ \ \ \
\ \ \ \nonumber \\ 
\times \left. \left( \begin{array}{cc}
-\ \frac{\delta^{2}\Gamma_{\Lambda}}{\delta\psi^{c}\delta\psi^{c}} 
  &
-\ \frac{\delta^{2}\Gamma_{\Lambda}}{\delta\bar{\psi}^{c}\delta\psi^{c}}
+ \Delta^{-1}_{IR}
 \\
\frac{\delta^{2}\Gamma_{\Lambda}}{\delta\psi^{c}\delta\bar{\psi}^{c}}
+ \Delta^{-1}_{IR}
  & 
\frac{\delta^{2}\Gamma_{\Lambda}}{\delta\bar{\psi}^{c}\delta\bar{\psi}^{c}}
\end{array} \right)^{-1\phantom{aa\ \beta}\alpha}_{\phantom{-1}aa\ \beta}
\right)
\end{eqnarray}
which will be our starting point for solving the large $N$ limit
exactly. Compare and contrast (4.29) with (2.22).

\section{The exact large $N$ limit}

As discussed, the Legendre flow equation is exactly solvable in the large $N$
limit and thus we will use this limit to establish some exact results with
which we will later compare the LPA (as applied to the Polchinski formalism).
We begin by assuming that we can write
$\Gamma_{\Lambda}=\Gamma_{\Lambda}[z(x)]$ for some $\Lambda$ [23], where we
have defined the scalar
$z(x)=\bar{\psi}^{\alpha}_{a}(x)\psi^{\phantom{\rho}}_{\alpha a}(x)$. Then
differentiating $\Gamma_{\Lambda}$ with respect to the fields,
we observe that the diagonal terms of (4.29) become zero and the off diagonal
terms take the form 
\begin{equation}
\delta_{ab} \Delta^{-1\phantom{\beta}\alpha}_{IR\beta} +
\delta_{ab}\delta^{\phantom{\beta}\alpha}_{\beta}\frac{\delta
\Gamma_{\Lambda}}{\delta z} 
+ \bar{\psi}^{\alpha}_{a}\psi^{\phantom{\rho}}_{\beta b}
\frac{\delta^{2}\Gamma_{\Lambda}}{\delta z\delta z}.
\end{equation}
Thus the inverse in (4.29) is of the form $(aI+b)^{-1} =
a^{-1}I-a^{-1}ba^{-1} + a^{-1}ba^{-1}ba^{-1}-\cdots$ where $a$ is a flavour
singlet, $I$ is the unit matrix in flavour space and $b$ is an arbitrary
matrix in flavour space. With the trace over flavours in (4.29) this implies
that the third term in (4.30) contributes negligibly for large
$N_{f}=\delta_{aa}$. Taking this limit, (4.29) becomes 
\begin{equation}
\frac{\partial\Gamma_{\Lambda}}{\partial\Lambda}=-N_{f}\mathrm{tr}\left(
\frac{\partial 
\Delta^{-1}_{IR}}{\partial \Lambda}^{\phantom{\alpha}\beta}_{\alpha} 
\left(\frac{\delta\Gamma_{\Lambda}}{\delta z}I  + 
\Delta^{-1}_{IR}\right)^{-1\phantom{\beta}\alpha}_{\phantom{-1}\beta}\right).
\end{equation}
If for some value of $\Lambda$, $\Gamma_{\Lambda}$ is a functional of $z$
only (as assumed) then $\frac{\delta\Gamma_{\Lambda}}{\delta z}I
+ \Delta^{-1}_{IR}$ is also a 
functional of $z$ only, and thus, given that (4.31) is a first order
differential 
equation in $\Lambda$, we have that this form is preserved by the flow,
\emph{i.e.} $\Gamma_{\Lambda}=\Gamma_{\Lambda}[z]$ for all
$\Lambda$\footnote{this has resulted in a restriction to a submanifold in the
coupling constant space [23]}. Furthermore, the fact that the flow equation
(4.31) now contains 
only zero or one point functions allows for a considerable simplification
[23]: Set $z(x)=z$ a constant. Then, since the effective action may be written
as a derivative expansion (to all orders), we see that only a potential term
of the form 
\begin{equation}
\Gamma_{\Lambda}=\int d^{D}x V(z,\Lambda)
\end{equation}
survives. Thus, substituting (4.32) into (4.31) and transforming into momentum
space 
\begin{equation}
\frac{\partial V}{\partial \Lambda}=N_{f}\Omega\int_{0}^{\infty}dq q^{D-1}
\frac{\partial\Delta^{\phantom{-1}}_{IR}}{\partial \Lambda}_{\alpha}^{\phantom{\alpha}\beta}
\Delta^{-1\phantom{\beta}\gamma}_{IR\beta} \left(I+\Delta^{\phantom{-1}}_{IR}
\frac{\partial  
V}{\partial z} \right)^{-1\phantom{\gamma}\alpha}_{\phantom{-1}\gamma}
\end{equation}
where we have used $\frac{\partial \Delta^{-1}_{IR}}{\partial \Lambda} =
-\Delta^{-1}_{IR}\frac{\partial \Delta_{IR}}{\partial \Lambda}
\Delta^{-1}_{IR}$ and $\Omega 
(2\pi)^{D}$ is the solid angle of a $(D-1)$-sphere. Now we introduce an
infra-red cutoff function, $C^{\phantom{-1}}_{IR}(i\frac{\sla{q}}{\Lambda})$,
such that
\begin{equation}
\Delta^{\phantom{IR\alpha}\beta}_{IR\alpha} = 
(\alpha+i\sla{q})^{-1\phantom{\alpha}\gamma}_{\phantom{-1}\alpha}
C^{\phantom{IR\gamma}\beta}_{IR\gamma},  
\end{equation}
where $\alpha$ is an unspecified constant\footnote{not to be
confused with the indices} and
$\Delta_{IR\alpha}^{\phantom{IR\alpha}\beta}\sim
-i\sla{q}_{\phantom{-1}\alpha}^{-1\phantom{\alpha}\beta}$ for large momentum.
Note that this cutoff generally breaks the chiral symmetry (later we will
choose to restrict the cutoff function and the purpose of the seemingly
redundant $\alpha$ will become clear).
Here, as with the scalars, $C_{IR}\rightarrow 1$ as $q\rightarrow\infty$ such
that the physics is independent of the infra-red cutoff at scales much larger
than $\Lambda$ and the theory is well regulated.
In terms of the cutoff function (4.33) can now be written
\begin{eqnarray}
\frac{\partial V}{\partial \Lambda} &=& -iN_{f}\Omega\int_{0}^{\infty}dq
q^{D-1} \left( \frac{1}{\Lambda^{2}} (\alpha\sla{q}^{-1}_{\phantom{\rho}} +
i)^{-1\phantom{\alpha}\beta}_{\phantom{-1}\alpha} C'^{\phantom{IR\beta}\gamma}
_{\phantom{'}IR\beta} C^{-1\phantom{\beta}\delta}_{IR\gamma} (\alpha + i 
\sla{q})^{\phantom{\delta}\epsilon}_{\delta} \right. \nonumber \\
& \times & \left. \left(I+(\alpha +
i\sla{q})^{-1} C^{\phantom{-1}}_{IR} \frac{\partial V}{\partial
z} \right)^{-1\phantom{\epsilon}\alpha}_{\phantom{-1}\epsilon}\right).
\end{eqnarray}
After scaling into dimensionless quantities, $V\rightarrow\Lambda^{D}V$, 
$z\rightarrow\Lambda^{D-1}z$, $q\rightarrow\Lambda q$ and
$\alpha\rightarrow\Lambda\alpha$, and using renormalization time defined by
(1.29) this becomes,
\begin{eqnarray}
\frac{\partial V}{\partial t} &=& DV+(1-D)z\frac{\partial V}{\partial z}+
iN_{f}\Omega\int_{0}^{\infty}dq q^{D-1}
\left((\alpha \sla{q}^{-1}+i)^{-1\phantom{\alpha}\beta}_{\phantom{-1}\alpha}
C'^{\phantom{IR\beta}\gamma}_{\phantom{'}IR\beta}
C^{-1\phantom{\gamma}\delta}_{IR\gamma}  
\right. \nonumber \\
& \times & \left. (\alpha+i\sla{q})^{\phantom{\delta}\epsilon}_{\delta}
\left(I +
(\alpha+i\sla{q})^{-1} C^{\phantom{-1}}_{IR}
\frac{\partial V}{\partial
z}\right)^{-1\phantom{\epsilon}\alpha}_{\phantom{-1}\epsilon}\right). 
\end{eqnarray}
The cutoff function takes the general form,
\begin{equation}
C^{\phantom{IR\alpha}\beta}_{IR\alpha}=g(q^{2})
\delta^{\phantom{\alpha}\beta}_{\alpha} + 
ih(q^{2})\sla{q}^{\phantom{\alpha}\beta}_{\alpha}
\end{equation}
and
\begin{equation}
C^{-1\phantom{\alpha}\beta}_{IR\alpha}=b(q^{2})
\delta^{\phantom{\alpha}\beta}_{\alpha} + 
ic(q^{2})\sla{q}^{\phantom{\alpha}\beta}_{\alpha}
\end{equation}
with
\begin{equation}
b=\frac{g}{g^{2}+h^{2}q^{2}}
\end{equation}
and
\begin{equation}
c=-\frac{h}{g^{2}+h^{2}q^{2}}
\end{equation}
where $b,c,g$ and $h$ are arbitrary functions of the now dimensionless $q^{2}$.
After some manipulation (4.36) can be written
\begin{eqnarray} 
\frac{\partial V}{\partial t} &=& DV+(1-D)z\frac{\partial V}{\partial z}+
iN_{f}\Omega\int_{0}^{\infty}dq q^{D-1}
\left(\frac{\alpha\sla{q} - iq^{2}}
{\alpha^{2} +
q^{2}}\delta_{\alpha}^{\phantom{\alpha}\beta}\right. \nonumber \\ 
& \times & \left((2q^{2}h'+h) -
2ig'\sla{q}\right)_{\beta}^{\phantom{\beta}\gamma}  
(b + ic\sla{q})_{\gamma}^{\phantom{\gamma}\delta} \nonumber \\
& \times & \left.
(\alpha + i\sla{q})_{\delta}^{\phantom{\delta}\epsilon}
\frac{H - G\sla{q}}{H^{2}-q^{2}G^{2}}
\delta_{\epsilon}^{\phantom{\delta}\alpha}\right)  
\end{eqnarray}
where prime denotes differentiation with respect to the argument, and
\begin{equation}
H(q^{2})=1+\frac{\alpha g - h q^{2}}{\alpha^{2} + q^{2}}
\frac{\partial V}{\partial z}
\end{equation}
and
\begin{equation}
G(q^{2})=i\frac{\alpha h - g}{\alpha^{2} + q^{2}}
\frac{\partial V}{\partial z}.
\end{equation}
We choose specifically $h=0$ and $g=\theta_{\epsilon}(q,1)$, where
$\theta_{\epsilon}(q,1)$ is a smooth cutoff function, satisfying
$\theta_{\epsilon}(q,1)\approx 0$ for $q<1-\epsilon$,
$\theta_{\epsilon}(q,1)\approx 1$ for $q>1+\epsilon$ and
$0<\theta_{\epsilon}<1$. Ultimately we take the sharp cutoff limit,
$\theta_{\epsilon}(q,1)\rightarrow\theta(q-1)$ as $\epsilon\rightarrow 0$, and
the flow equation will simplify enormously. Note that the cutoff breaks the
chiral symmetry unless $\alpha =0$. Using
$\sla{q}_{\alpha}^{\phantom{\alpha}\alpha } =
q_{\mu}\mathrm{tr}(\gamma^{\mu})=0$ and writing $N=N_{f}N_{i} =
N_{f}\delta^{\phantom{\alpha}\alpha}_{\alpha}$, we arrive at
\begin{equation}
\frac{\partial V}{\partial t} = DV+(1-D)z\frac{\partial V}{\partial z} +
2N\Omega\int_{0}^{\infty}dq q^{D+1}\frac{Hg'}{g(H^{2}-q^{2}G^{2})}.
\end{equation}
Then using (4.42) and (4.43) and the lemma presented in Appendix 4A [18] we
take the $\epsilon\rightarrow 0$ limit,
\begin{equation}
\frac{\partial V}{\partial t} = DV+(1-D)z\frac{\partial V}{\partial z}+
N\Omega\left[ \mathrm{ln}\left(\frac{t^{2}}{t^{2}\left(\frac{\partial
V}{\partial z}\right)^{2} +
2\alpha t\frac{\partial V}{\partial z} + \alpha^{2} + 1}\right)
\right]^{t=1}_{t=0}
\end{equation}
so that we finally reach the desired flow equation by a change of variables,
\begin{equation}
DV\rightarrow DV-N\Omega\left[ \mathrm{ln}(t^{2})\right]^{t=1}_{t=0}
\end{equation}
yielding
\begin{equation}
\frac{\partial V}{\partial t} = DV+(1-D)z\frac{\partial V}{\partial z} - 
\mathrm{ln}\left(1+\frac{\left(\frac{\partial V}{\partial z}\right)^{2} +
2\alpha\frac{\partial V}{\partial z}}{\alpha^{2}+1}\right).
\end{equation}
In the last step we scaled out the $N$ and $\Omega$ dependencies via
$V\rightarrow\Omega NV$ and $z\rightarrow\Omega Nz$ (note that these scalings
justify the neglect of terms of the form (4.16) in (4.15)). Differentiating
(4.47) with respect to $z$ and defining $W=\frac{\partial V}{\partial z}$ we
find 
\begin{equation}
\frac{\partial W}{\partial t} = W + \left((1-D)z -
\frac{2(W+\alpha)}{\alpha^{2} + 1 + W^{2} + 2\alpha W}\right)
\frac{\partial W}{\partial z}.
\end{equation}
The flow equation in its forms (4.47) and (4.48) will form the focus of our
attention. We realise the Gaussian solution, $W=V=0$, but in the following
will concentrate on non-trivial solutions.

\section{Solutions in the large $N$ limit}

We now solve the large $N$ limit [42,43] of the Legendre flow equation ((4.47)
and (4.48)) for the fixed points and eigenvalue spectrum. Initially we choose
$\alpha =0$ to analyse specifically the solutions with a cutoff introduced in
a chirally invariant manner, however ultimately we will allow $\alpha$ to be
non-zero to make meaningful comparisons with the Polchinski approach to be
developed later. 
The eigenvalues are found by perturbing the potential in (4.47). We
substitute $V=V_{*}+\epsilon\sum_{n} U_{n} e^{\lambda_{n} t}$, use the fixed
point version of (4.47) and expand the logarithmic term, yielding to
$O(\epsilon)$ 
\begin{equation}
(\lambda_{n}-D)U_{n} = \left((D-1)z+\frac{2(W_{*}+\alpha)}{\alpha^{2} +
1 + W_{*}^{2} +2\alpha W_{*}}\right)\frac{\partial U_{n}}{\partial z}.
\end{equation}
Using the fixed point version of (4.48) it is straightforward to see that
\begin{eqnarray}
\frac{\partial U_{n}}{\partial W_{*}}&=&\frac{\partial U_{n}}{\partial z}\left/
\frac{\partial W_{*}}{\partial z}\right. \nonumber \\
&=& (\lambda_{n} -D)\frac{U_{n}}{W_{*}}
\end{eqnarray}
and it is easily verified that
\begin{equation}
U_{n}=K_{1} W_{*}^{D-\lambda_{n}}.
\end{equation}
where $K_{1}$ is an arbitrary constant.
Insisting that the eigenfunctions, $U_{n}$, are analytic and providing that
$W_{*}$ crosses the $z$-axis we thus deduce that the power in (4.51) must be a
positive integer, $n$, and therefore that
\begin{equation}
\lambda_{n} = D-n
\end{equation}
regardless of $\alpha$. (Note, that these eigenvalues are degenerate with the
Gaussian in two dimensions.) 

Specialising to three dimensions the exact flow equation, in its form
(4.48), can be solved analytically using the method of 
characteristics. The solution is found to be
\begin{eqnarray}
z&=&\frac{1}{\alpha^{2}+1}\left(\alpha(\alpha^{2}-3)
\mathrm{ln}\left(\frac{W^{2}}{W^{2}+2\alpha
W+\alpha^{2}+1}\right)\right. \nonumber \\ 
&+& \left. 2(3\alpha^{2}-1)
\mathrm{tan}^{-1}(W+\alpha)\phantom{\frac{1}{1}}\!\!\!\!\!\right)W^{2}
+\frac{4\alpha^{2}W}{(\alpha^{2}+1)^{2}} \nonumber \\
&-&\frac{2W+\alpha}{\alpha^{2} + 1}+ f_{1}(We^{-t})W^{2}.
\end{eqnarray}
where the arbitrary function, $f_{1}(We^{-t})$, is set constant (denoted by
$f_{1}$) for a fixed point solution. However, we find we need to impose
conditions on the range of $f_{1}$ for the solution to be considered
physical. We require that the solution $W$ is defined for all $z$ and that
(4.15), and thus the negative potential, is bounded from below. After a little
analysis of (4.53) we deduce the large $W$ behaviour as
\begin{equation}
z\sim\frac{(3\alpha^{2}-1)\pi}{\alpha^{2}+1}|W|W +
f_{1}W^{2},
\end{equation}
which leads to a restriction in the range of $f_{1}$:
\begin{equation}
\frac{(3\alpha^{2}-1)\pi}{\alpha^{2}+1} < f_{1} < 
\frac{(1-3\alpha^{2})\pi}{\alpha^{2}+1}.
\end{equation}
Here we have imposed that $W(z)>0$ for large negative $z$ and that $W(z)<0$
for large positive $z$. 
Furthermore we demand that $W$ is not double valued in $z$. This is done by
simply differentiating (4.53) and studying its turning points. By experiment
we find that (4.55) provides a good approximation to this restriction.
\begin{figure}
\vspace{-122mm}
\hspace{7mm}
\scalebox{0.5}{\includegraphics*[0pt,0pt][1000pt,1000pt]{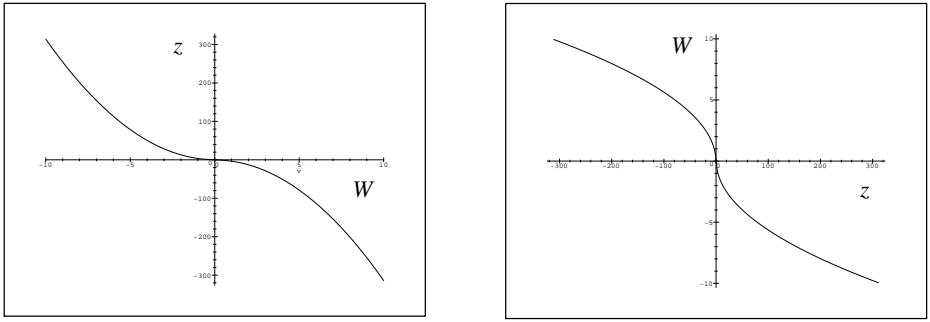}} 
\caption{Solution to the large $N$ limit of the Legendre flow equation in
three dimensions with cutoff introduced in a chirally symmetric manner}
\end{figure}
\vspace{-6mm}

Choosing $\alpha=0$ yields a flow equation
for the cutoff introduced in such a manner as to maintain chiral symmetry
which has the following solution:
\begin{equation}
z=-2W-2W^{2}\mathrm{tan}^{-1}(W)+f_{1}W^{2}.
\end{equation}
A chirally symmetric potential must be even in $z$ and thus we
observe that (4.56) actually breaks this symmetry unless we set
$f_{1}=0$. This solution is plotted in figure 4.1.
It is noticed that independent of $f_{1}$, the solution will always satisfy
$W(0)=0$ and thus the eigenvalue spectrum 
is indeed given by (4.52), $\lambda_{n}=3-n$. These results are in direct
agreement with those of Zinn-Justin [43]. The marginal eigenvalue,
$\lambda_{3}=0$, is associated with a line of fixed points parameterised by
$f_{1}$. 
We find that the double valued condition discussed above restricts the
acceptable solutions to the range $-3.0708<f_{1}<3.0708$. 

Allowing $\alpha$ to be non-zero, we can study the symmetry breaking
scenario. For example we could start by fixing $f_{1}$ and altering $\alpha$,
such that for the two boundedness conditions, (4.55), to be consistent we
require $\alpha^{2}<\frac{1}{3}$. Thus the region of `parameter space' allowed
is constrained (figure 4.2).
However, generally the double valued condition restricts the region allowed
even further. For example for $\alpha=0.25$ (4.55) implies
$-2.4024<f_{1}<2.4024$ whereas in actual fact we find $-2.2489<f_{1}<2.1325$.
We recognise that (4.53) is unchanged if we simultaneously let
$\alpha\rightarrow -\alpha$, $z\rightarrow -z$ and $W\rightarrow
-W$ (this is discrete chiral symmetry) and thus 
focus on just $\alpha >0$. The imposition of (4.55) and the observation that
(4.53) has no divergences has the implication that
all the acceptable solutions satisfy $W(z)=0$ for some $z$ and thus for
all these fixed points the eigenvalue spectrum is again $\lambda_{n}=3-n$.
\begin{figure}
\vspace{-75mm}
\hspace{40mm}
\scalebox{0.4}{\includegraphics*[0pt,0pt][1000pt,1000pt]{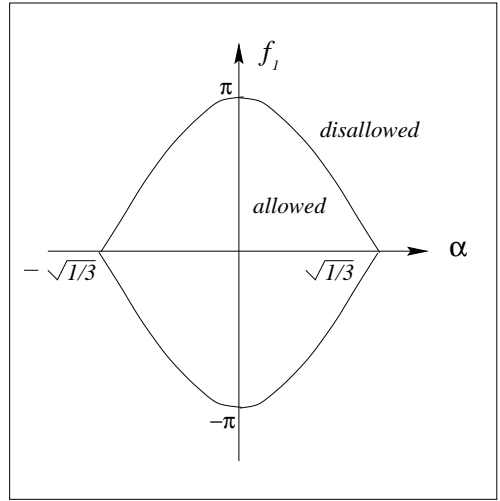}} 
\caption{Region of ($\alpha$, $f_{1}$) space allowed by the boundedness
condition (4.55)}
\end{figure}

\section{The Polchinski flow equation}

Here we present a constructive proof of the Polchinski flow equation for a
purely Fermionic theory. This will form the focus of our attention when we
apply the LPA. Begin by considering the generating functional
regulated by an overall momentum cutoff $\Lambda_{o}$, made explicit later. 
For a
Fermion field $\psi(x)$ with propagator $\Delta(p)$ and arbitrary bare action
$S_{\Lambda_{o}}[\bar{\psi},\psi]$, we write
\begin{equation}
Z(\bar{\zeta},\zeta)=\int(d\bar{\psi})(d\psi)e^{-\ \bar{\psi}^{\alpha}_{a}\
\Delta^{-1\phantom{\alpha}\beta}_{\phantom{-1}\alpha}\
\psi_{\beta a}^{\phantom{\rho}}-\
S_{\Lambda_{o}}^{\phantom{\rho}}[\bar{\psi},\psi]\ +\ \bar{\psi}^{\alpha}_{a}\
\zeta_{\alpha a}^{\phantom{\rho}}\ +\ \bar{\zeta}^{\alpha}_{a}\ \psi_{\alpha
a}^{\phantom{\rho}}} 
\end{equation}
where, as before $\bar{\zeta}$ and $\zeta$ are sources, $\alpha$ and $\beta$
run over internal 
spinor indices from 1 to $N_{i}$ and $a$ is the flavour index which runs
from 1 to $N_{f}$. Following the methods of [18] presented in chapter two we
split the propagator and field into high and low momentum parts,
\begin{equation}
\Delta(p)=\Delta^{\phantom{a}}_{IR}(p)+\Delta^{\phantom{a}}_{UV}(p)
\end{equation}
and
\begin{equation}
\psi=\psi_{IR}+\psi_{UV}
\end{equation}
respectively. Then we can write (4.57) as
\begin{equation}
Z(\bar{\zeta},\zeta) = \int\! (d\bar{\psi}_{UV}^{\phantom{\rho}})
(d\psi_{UV}^{\phantom{\rho}}) 
e^{-\ \bar{\psi}_{UV a}^{\phantom{UV} \alpha}\
\Delta^{-1\phantom{\alpha}\beta}_{UV\alpha}  
\ \psi_{UV \beta a}^{\phantom{\rho}}}\ Z_{\Lambda}
[\bar{\psi}_{UV}^{\phantom{\rho}}, \psi_{UV}^{\phantom{\rho}}, \bar{\zeta},
\zeta]  
\end{equation}
where
\begin{eqnarray}
Z_{\Lambda} [\bar{\psi}_{UV}^{\phantom{\rho}}, \psi_{UV}^{\phantom{\rho}},
\bar{\zeta}, \zeta] = \int\! (d\bar{\psi}_{IR}^{\phantom{\rho}})
(d\psi_{IR}^{\phantom{\rho}})
e^{-\ \bar{\psi}_{IR a}^{\phantom{IR}\alpha}\
\Delta^{-1\phantom{\alpha}\beta}_{IR\alpha}\ 
\psi_{IR \beta a}^{\phantom{\rho}}} \ \ \nonumber \\
\times \ e^{-\ S_{\Lambda_{o}}^{\phantom{\rho}}
[\bar{\psi}_{IR}^{\phantom{\rho}}+ 
\bar{\psi}_{UV}^{\phantom{\rho}},\psi_{IR}^{\phantom{\rho}}+
\psi_{UV}^{\phantom{\rho}}]\ +\ (\bar{\psi}_{IR}^{\phantom{\rho}}+
\bar{\psi}_{UV}^{\phantom{\rho}})^{\alpha}_{a}\ \zeta_{\alpha
a}^{\phantom{\rho}}\ +\ \bar{\zeta}^{\alpha}_{a}\
(\psi_{IR}^{\phantom{\rho}}+\psi_{UV}^{\phantom{\rho}})_{\alpha
a}^{\phantom{\rho}}}. 
\end{eqnarray}   
We verify (4.60) by eliminating $\psi_{IR}^{\phantom{\rho}}$ and
$\bar{\psi}_{IR}^{\phantom{\rho}}$ using (4.59) and substituting
$\psi_{UV}^{\phantom{\rho}}=\psi_{UV}^{\phantom{\rho}'} +
\Delta_{UV}^{\phantom{\rho}}\ \Delta^{-1}\ \psi$ to retrieve (4.57) up to a
multiplicative factor which we ignore. We now make the momentum cutoffs
explicit by defining $\Delta_{IR}=[\theta_{\epsilon}(p,\Lambda) -
\theta_{\epsilon}(p,\Lambda_{o})]\Delta$ and $\Delta_{UV} =
[1-\theta_{\epsilon}(p,\Lambda)]\Delta$ where similar to our previous analysis
$\theta_{\epsilon}$ is a smooth 
function satisfying $\theta_{\epsilon}(p,\Lambda)\approx 0$ for $p<\Lambda -
\epsilon$, $\theta_{\epsilon}(p,\Lambda)\approx 1$ for $p>\Lambda -
\epsilon$ and $0\leq\theta_{\epsilon}(p,\Lambda)\leq 1$ (figure 4.3).
\begin{figure}
\vspace{-135mm}
\scalebox{0.5}{\includegraphics*[0pt,0pt][1000pt,1000pt]{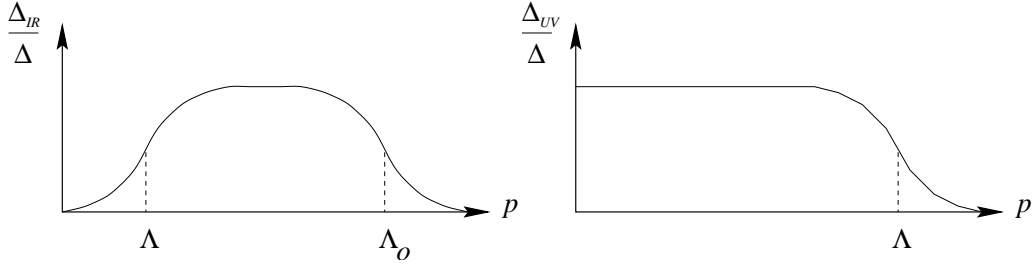}} 
\caption{Ultra-violet and Infra-red cutoff functions}
\end{figure}
The derivation of the flow equation starts by substituting
$\psi_{IR}=\psi-\psi_{UV}$ into (4.61), yielding
\begin{eqnarray}
Z_{\Lambda}=e^{-\ \bar{\psi}^{\phantom{UV}\alpha}_{UV a}\
\Delta^{-1\phantom{\alpha}\beta}_{IR\alpha}\ \psi_{UV\beta
a}^{\phantom{\rho}}\ -\ 
S_{\Lambda_{o}}^{\phantom{\rho}}[\frac{\delta}{\delta \bar{\zeta}},
\frac{\delta}{\delta \zeta}]} \ \ \ \ \ \ \ \ \ \ \ \ \ \ \ \ \ \ \ \ \
\nonumber \\ 
\times \int\!(d\bar{\psi})(d\psi)
e^{-\ \bar{\psi}^{\alpha}_{a}\ \Delta^{-1\phantom{\alpha}\beta}_{IR\alpha}\
\psi_{\beta a}^{\phantom{\rho}}\ 
+\ \bar{\psi}^{\alpha}_{a}\ [\Delta^{-1}_{IR}\ \psi_{UV}^{\phantom{\rho}} 
+ \zeta]^{\phantom{a}}_{\alpha a}\ +\ [\bar{\psi}^{\phantom{UV}}_{UV}\
\Delta^{-1}_{IR} + \bar{\zeta}]^{\alpha}_{a}\ \psi_{\alpha
a}^{\phantom{\rho}}}. 
\end{eqnarray}
Performing the Gaussian integral in (4.62)
\begin{eqnarray}
\!\!\!\!\!Z_{\Lambda}\!\!\! &=&\!\!\! e^{-\
\bar{\psi}^{\phantom{UV}\alpha}_{UV a}\ 
\Delta^{-1\phantom{\alpha}\beta}_{IR\alpha}\ \psi_{UV\beta
a}^{\phantom{\rho}}\ -\ 
S_{\Lambda_{o}}^{\phantom{\rho}}[\frac{\delta}{\delta \bar{\zeta}},
\frac{\delta}{\delta \zeta}]
\ +\ [\bar{\psi}^{\phantom{UV}}_{UV}\
\Delta^{-1}_{IR} + \bar{\zeta}]_{a}^{\alpha}\
\Delta_{IR\alpha}^{\phantom{IR\alpha}\beta}\ [\Delta^{-1}_{IR}\
\psi_{UV}^{\phantom{\rho}} + \zeta]^{\phantom{a}}_{\beta a}} \\
   &=&\!\!\! e^{\ \bar{\zeta}^{\alpha}_{a}\
   \Delta^{\phantom{IR\alpha}\beta}_{IR\alpha}\ 
\zeta_{\beta a}^{\phantom{\rho}}\ +\ \bar{\psi}^{\phantom{UV}\alpha}_{UV a}\ 
\zeta^{\phantom{\rho}}_{\alpha a}\ +\ \bar{\zeta}^{\alpha}_{a}\
\psi^{\phantom{\rho}}_{UV \alpha a}\ 
-\ S_{\Lambda}^{\phantom{\rho}}[\bar{\psi}_{UV}^{\phantom{\rho}} +
\bar{\zeta}\Delta_{IR}^{\phantom{\rho}},\psi_{UV}^{\phantom{\rho}} +
\Delta_{IR}^{\phantom{\rho}}\zeta]} 
\end{eqnarray}
so that (4.60) becomes
\begin{eqnarray}
Z(\bar{\zeta},\zeta)\!\! &=&\!\! \int\! (d\bar{\psi}_{UV}^{\phantom{\rho}})
(d\psi_{UV}^{\phantom{\rho}}) 
e^{-\ \bar{\psi}_{UV a}^{\phantom{UV} \alpha}\
\Delta^{-1\phantom{\alpha}\beta}_{UV 
a} \ \psi_{UV \beta a}^{\phantom{\rho}}\ 
+\ \bar{\zeta}^{\alpha}_{a}\ \Delta^{\phantom{IR\alpha}\beta}_{IR\alpha}\
\zeta_{\beta a}^{\phantom{\rho}}} \nonumber \\ 
& \times & e^{\ \bar{\psi}^{\phantom{UV}\alpha}_{UV a}\ 
\zeta^{\phantom{\rho}}_{\alpha a}\ +\ \bar{\zeta}^{\alpha}_{a}\
\psi^{\phantom{\rho}}_{UV \alpha a}\ 
-\ S_{\Lambda}^{\phantom{\rho}}[\bar{\psi}_{UV}^{\phantom{\rho}} +
\bar{\zeta}\Delta_{IR}^{\phantom{\rho}},\psi_{UV}^{\phantom{\rho}} +
\Delta_{IR}^{\phantom{\rho}}\zeta]}
\end{eqnarray}
for some functional, $S_{\Lambda}$.

Now consider that $\bar{\zeta}$ and $\zeta$ only couple to the low momentum
modes, \emph{i.e.} such that
$\bar{\zeta}\Delta_{IR}=\Delta_{IR}\zeta=0$. Then, in this limit (4.65) becomes
\begin{eqnarray}
Z(\bar{\zeta},\zeta)   &=&   \int\! (d\bar{\psi}_{UV}^{\phantom{\rho}})
(d\psi_{UV}^{\phantom{\rho}}) 
e^{-\ \bar{\psi}_{UV a}^{\phantom{UV} \alpha}\
\Delta^{-1\phantom{\alpha}\beta}_{UV a} 
\ \psi_{UV\beta a}^{\phantom{\rho}}} \nonumber \\
    &\times& e^{\ \bar{\psi}^{\phantom{UV}\alpha}_{UV a}\ 
\zeta^{\phantom{\rho}}_{\alpha a}\ +\ \bar{\zeta}^{\alpha}_{a}\
\psi^{\phantom{\rho}}_{UV\alpha a}\ 
-\ S_{\Lambda}^{\phantom{\rho}}[\bar{\psi}_{UV}^{\phantom{\rho}},
\psi_{UV}^{\phantom{\rho}}]}\ \ \ \ \ \ \ \ 
\end{eqnarray}
and thus $S_{\Lambda}$ coincides with the Wilsonian effective
action. Differentiating (4.61) with respect to the cutoff, $\Lambda$, gives
\begin{equation}
\frac{\partial Z_{\Lambda}}{\partial \Lambda}=\left(\frac{\delta}{\delta\zeta}
+ \bar{\psi}_{UV}\right)^{\alpha}_{a}\ \frac{\partial \Delta^{-1}_{IR
}}{\partial \Lambda}^{\phantom{\alpha}\beta}_{\alpha}\
\left(\frac{\delta}{\delta \bar{\zeta}} 
- \psi_{UV}\right)^{\phantom{\rho}}_{\beta a}\ Z_{\Lambda}.
\end{equation}
Defining $\underline{\psi}_{\alpha a}^{\phantom{\rho}} = \psi_{UV
\alpha a}^{\phantom{\rho}}\ +\ \Delta_{IR\alpha}^{\phantom{IR\alpha}\beta}   
\ \zeta_{\beta a}^{\phantom{\rho}}$ and
$\underline{\bar{\psi}}^{\alpha}_{a}=\bar{\psi}^{\phantom{UV}\alpha}_{UVa}\ +\ 
\bar{\zeta}^{\beta}_{a}\ \Delta_{IR\beta}^{\phantom{IR\beta}\alpha}$ and
substituting (4.64) into (4.67) we arrive at our flow equation, 
\begin{equation}
\frac{\partial S_{\Lambda}}{\partial \Lambda}= \frac{\delta
S_{\Lambda}}{\delta \underline{\psi}^{\phantom{\rho}}_{\alpha a}}\ K_{\Lambda
\alpha}^{\phantom{\Lambda\alpha}\beta}\ \frac{\delta
S_{\Lambda}}{\delta\underline{\bar{\psi}}^{\beta}_{a}}\ -\
\mathrm{tr} \left(K_{\Lambda\alpha}^{\phantom{\Lambda\alpha}\beta}\
\frac{\delta^{2} 
S_{\Lambda}}{\delta \underline{\psi}^{\phantom{\rho}}_{\alpha a} \delta
\underline{\bar{\psi}}^{\beta}_{\alpha}}\right)
\end{equation}
where
\begin{equation}
K_{\Lambda}=\frac{\partial \Delta_{IR}}{\partial \Lambda}=-\frac{\partial
\Delta_{UV}}{\partial \Lambda}.
\end{equation}
It is possible to confirm this solution by eliminating $S_{\Lambda}$ in favour
of $S_{ren}\ =\ \underline{\bar{\psi}}^{\alpha}_{a}\
\Delta^{-1\phantom{\alpha}\beta}_{UV\alpha}\ 
\underline{\psi}^{\phantom{\rho}}_{\beta a}\ +\ S_{\Lambda}$ and then showing
that $\frac{\partial Z}{\partial \Lambda}=0$ as required. 
Note the similarity between (2.14) and (4.68).

\section{Application of the Local Potential Approximation}

Previous attempts to solve the flow equation [44] (4.68) have involved
introducing an ultra-violet cutoff function
$C_{UV}(\frac{q^{2}}{\Lambda^{2}})$, as described for the scalars. However, typically various other
approximations have been made in addition to the LPA\footnote{most notably
truncations}, leading to results of dubious reliability. Here we only
require that the cutoff function be dimensionless and
Lorentz invariant, and so for the Fermions we allow $C=C_{UV}$ to be a
function of $\sla{q}$, $C(i\frac{\sla{q}}{\Lambda})$, and thus write
\begin{equation}
\Delta^{\phantom{UV\alpha}\beta}_{UV\alpha} =
-iC^{\phantom{\alpha}\gamma}_{\alpha}\sla{q}^{-1\phantom{\gamma}\beta}
_{\phantom{-1}\gamma},
\end{equation}
where $C(0)=1$ and $C$ falls to zero sufficiently quickly as 
$q\rightarrow\infty$, such that the theory is well regulated.

Substituting (4.70) into (4.68) and replacing $\underline{\psi}$ with $\psi$,
\begin{equation}
\frac{\partial S_{\Lambda}}{\partial \Lambda}
=\frac{\delta S_{\Lambda}}{\delta \psi^{\phantom{a}}_{\alpha a}}
\frac{C^{\prime}}{\Lambda^{2}}\delta_{\alpha}^{\phantom{\alpha}\beta}
\frac{\delta S_{\Lambda}}{\delta \bar{\psi}^{\beta}_{a}}
-\mathrm{tr}\left(\frac{C^{\prime}}{\Lambda^{2}}\delta^{\phantom{\alpha}\beta}_{\alpha}
\frac{\delta^{2} S_{\Lambda}} 
{\delta \psi^{\phantom{a}}_{\alpha a} \delta \bar{\psi}_{a}^{\beta}}
\right)
\end{equation}
where the prime denotes differentiation with respect to its argument.
Now we introduce the LPA by writing
\begin{equation}
S_{\Lambda}=\int d^{D}x V(\bar{\psi},\psi;\Lambda)
\end{equation}
such that $V$ does not contain any derivative terms.
Similarly we restrict 
$C(\sla{\partial})$. Within this approximation we find that (4.71) reduces to
\begin{equation}
\frac{\partial V}{\partial \Lambda}=\frac{1}{\Lambda^{2}}
\frac{\partial V}{\partial\psi^{\phantom{a}}_{\alpha a}}C^{\prime}(0)
\frac{\partial V}{\partial\bar{\psi}_{a}^{\alpha}}
-\tau\Lambda^{D-2}\frac{\partial^{2}V}
{\partial\psi^{\phantom{a}}_{\alpha a}\partial\bar{\psi}_{a}^{\alpha}},
\end{equation}
where $I\tau=\Lambda^{-D}(2\pi)^{-D}\int d^{D}qC^{\prime}(\sla{q}/\Lambda)$ for
a dimensionless constant $\tau$ and the $C^{\prime}(0)$ has been transformed
into  
configuration space. By scaling $V\rightarrow V\Lambda^{D}$, $\psi\rightarrow
\psi\Lambda^{\frac{1}{2}(D-1)}$, $\bar{\psi}\rightarrow\bar{\psi}
\Lambda^{\frac{1}{2}(D-1)}$, defining the renormalization time by (1.29) 
and finally scaling out the remaining cutoff function dependencies 
we derive the following scheme independent flow equation:
\begin{equation}
\frac{\partial V}{\partial t}=DV-\frac{1}{2}(D-1)\left(
\psi^{\phantom{a}}_{\alpha a} \frac{\partial V}
{\partial\psi^{\phantom{a}}_{\alpha a}} +
\bar{\psi}_{a}^{\alpha}\frac{\partial V} 
{\partial \bar{\psi}_{a}^{\alpha}}\right)+\frac{\partial V}
{\partial\psi^{\phantom{a}}_{\alpha a}} 
\frac{\partial V}{\partial \bar{\psi}_{a}^{\alpha}}-\frac{\partial^{2} V}
{\partial\psi^{\phantom{a}}_{\alpha a}\partial\bar{\psi}_{a}^{\alpha}}
\end{equation}
where all quantities are now dimensionless.
Note the striking similarity between this and the scalar case, (2.33). The
Gaussian and High Temperature solutions are given by
\begin{equation}
V_{*}^{G}=0
\end{equation}
and
\begin{equation}
V_{*}^{HT}=\frac{N}{D}+\bar{\psi}^{\alpha}_{a}\psi^{\phantom{a}}_{\alpha a}
\end{equation}
respectively. 

By perturbing about a fixed point solution, $V(\bar{\psi},\psi,t)=
V_{*}(\bar{\psi},\psi)+\epsilon\sum_{n} f_{n}(\bar{\psi},\psi)e^{\lambda_{n}
t}$, where  $V_{*}$ is the fixed point under consideration and $\epsilon$ is a
small parameter, we 
easily find an equation for the perturbations around a fixed point (ignoring 
$O(\epsilon^{2})$),
\begin{eqnarray}
\lambda_{n} f_{n}=Df_{n}-\frac{1}{2}(D-1)\left(
\psi^{\phantom{\rho}}_{\alpha a} \frac{\partial f_{n}}
{\partial\psi^{\phantom{\rho}}_{\alpha a}} +
\bar{\psi}_{a}^{\alpha}\frac{\partial f_{n}} 
{\partial \bar{\psi}_{a}^{\alpha}}\right) \nonumber \\
+\frac{\partial f_{n}}{\partial\psi^{\phantom{\rho}}_{\alpha a}}   
\frac{\partial V_{*}}{\partial \bar{\psi}_{a}^{\alpha}}+\frac{\partial V_{*}}
{\partial\psi^{\phantom{\rho}}_{\alpha a}}\frac{\partial f_{n}} {\partial
\bar{\psi}_{a}^{\alpha}} 
-\frac{\partial^{2} f_{n}}{\partial\psi^{\phantom{\rho}}_{\alpha a} 
\partial\bar{\psi}_{i}^{a}}.\ \ \ \ \ \ 
\end{eqnarray}
This can then be used to find the exponents which are related to the
$\lambda_{n}$ in the familiar way (section 2.4). As for the scalars, the
eigenvalues for the Gaussian fixed point are given by the mass dimension of
the couplings:
\begin{equation}
\lambda^{G}_{n}=D-n(D-1)
\end{equation}
where $n$ is a positive integer. The
eigenvalues for the High Temperature fixed point are independent of $N$,
\begin{equation}
\lambda^{HT}_{n}=D-n(D+1),
\end{equation}
as expected for an infinitely massive theory (seen by calculating the $\beta$
function for the mass, similar to that described for scalar field theory). 

\section{The large $N$ limit of the LPA}

We now consider the large $N$ limit of the LPA Polchinski flow equation (4.74)
and compare it to the exact results presented in section 4.5. Similar to
before we define 
$z(x)=\bar{\psi}^{\alpha}_{a}(x)\psi^{\phantom{a}}_{\alpha a}(x)$, and
consider $V=V(z,t)$. Then the flow equation becomes 
\begin{equation}
\frac{\partial V}{\partial t} = DV - ((D-1)z+N)\frac{\partial
V}{\partial z} - z\left(\frac{\partial V}{\partial
z}\right)^{2} - z\frac{\partial^{2} V}{\partial z^{2}}.
\end{equation}
Scaling out the $N$ dependence via $V\rightarrow NV$ and $z\rightarrow
Nz$ and
taking the limit, the double derivative on the right of (4.80) vanishes. We
retrieve the trivial fixed points $V_{*}^{G}=0$ and
$V_{*}^{HT}=\frac{1}{D}+z$. Then differentiating (4.80) with respect to
$z$, 
\begin{equation}
\frac{\partial W}{\partial t} = W - W^{2} + \left((1-D)z
-2zW-1\right)\frac{\partial W}{\partial z}
\end{equation}
where $W=\frac{\partial V}{\partial z}$.

As before we expand the potential in (4.80) as
$V=V_{*}+\epsilon\sum_{n} U_{n}e^{\lambda_{n} t}$ for small $\epsilon$. The
usual manipulation leads to an equation for the eigenfunctions
\begin{equation}
\frac{1}{U_{n}}\frac{\partial U_{n}}{\partial z}=\frac{\lambda_{n}
-D}{(1-D)z-2zW_{*}-1}.  
\end{equation}
Substituting the trivial solutions into (4.82) we rapidly reproduce the
eigenvalue spectrum for the Gaussian and High Temperature fixed points.
We rewrite (4.82) as
\begin{eqnarray}
\frac{\partial \mathrm{ln}(U_{n})}{\partial W_{*}} &=& \frac{\lambda_{n}
-D}{\frac{\partial W_{*}} {\partial
z}\left((1-D)z-2zW_{*}-1\right)}  \\
&=&\frac{\lambda_{n}-D}{W^{\phantom{2}}_{*}-W_{*}^{2}}
\end{eqnarray}
where we have used the fixed point version of (4.81). Integrating (4.84) we
solve for $U_{n}$ in terms of $W_{*}$,
\begin{equation}
U_{n}=K_{2}\left(\frac{W_{*}-1}{W_{*}}\right)^{D-\lambda_{n}},
\end{equation}
where $K_{2}$ is a constant of integration. Thus we deduce that if
$W_{*}=1$ for some $z$ then for $U_{n}$ to be analytic at that $z$, the
exponent in (4.85) must be
a positive integer ($n$), implying $\lambda_{n} = D-n$, whereas if $W_{*}=0$
for some $z$ we would have $\lambda_{n} = D+n$.
\begin{figure}
\vspace{-122mm}
\hspace{7mm}
\scalebox{0.5}{\includegraphics*[0pt,0pt][1000pt,1000pt]{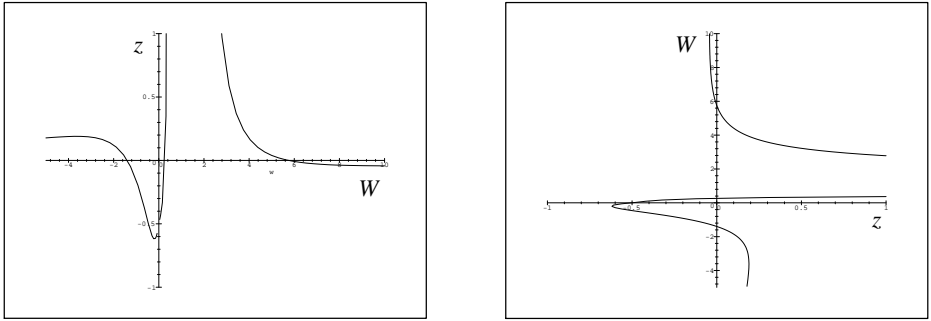}} 
\caption{Solution to the large $N$ limit of the LPA Polchinski flow equation
in three dimensions}
\end{figure}

Again, solving using the method of characteristics we find the solution in
three dimensions to be
\begin{equation}
z=\frac{1}{(W-1)^{4}}\left(\frac{3}{2}W^{2}\mathrm{ln}(W^{2})-W^{3}+3W
-\frac{1}{2} +W^{2}f_{2}\left(\frac{We^{-t}}{W-1}\right)\right)
\end{equation}
where $f_{2}(\frac{We^{-t}}{W-1})$ is an arbitrary function. For a fixed point
solution $f_{2}(\frac{We^{-t}}{W-1})$ must be set to a
constant ($f_{2}$). The $f_{2}=0$ solution is shown
in figure 4.4. However, from (4.86) we see that for large $z$, $z\sim 
-\frac{1}{W}$ and furthermore that we have exactly one divergence at $W=1$,
unless $f_{2}=-\frac{3}{2}$. Therefore by continuity, we deduce that none of
these fixed points can represent acceptable fixed points\footnote{The reader
can convince himself or herself of this by graphical analysis}. The remaining
possibility, $f_{2}=-\frac{3}{2}$ 
is plotted in figure 4.5 and is discarded for similar reasons.
Hence there are no acceptable non-trivial fixed points in three dimensions.
\begin{figure}
\vspace{-122mm}
\hspace{7mm}
\scalebox{0.5}{\includegraphics*[0pt,0pt][1000pt,1000pt]
{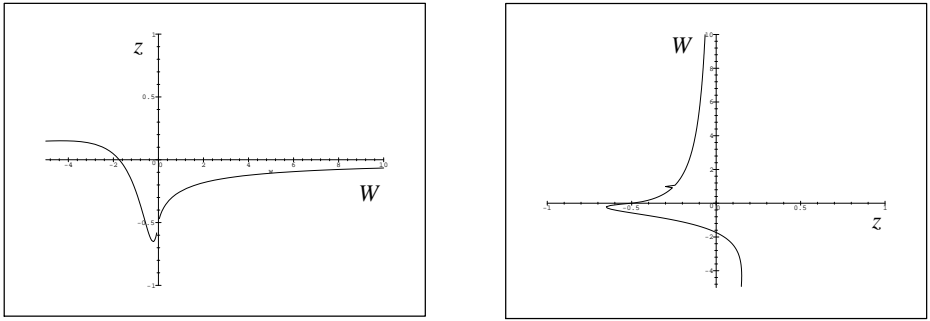}} 
\caption{Non-divergent solution to the large $N$ limit of the LPA Polchinski
flow equation in three dimensions}  
\end{figure}

\section{Finite $N$}

To study finite $N$ theories we use (4.80). We start by substituting a series
ansatz, truncated at some finite power of $z$. Generally we find
solutions 
degenerate with the Gaussian at $D=2$. However, similar to the case with
scalar field theory we quickly generate large numbers of spurious
solutions. The simplest ansatz yields the following solution, 
\begin{equation}
V_{*}^{NT}=\frac{2-D}{2D}+\left(\frac{1}{2}-\frac{D}{4}\right)z +
\left(\frac{1}{8}-\frac{D^{2}}{32}\right)z^{2},
\end{equation}
which also collapses to the High Temperature fixed point in $D=-2$, in a manner
reminiscent of scalar field theory. However, as discussed we treat $z(x)$ as a
real number and not as an even Grassmann and therefore we solve (4.80)
numerically. 

However, solving numerically,  we find that non-trivial fixed points entirely
disappear, unlike with scalar field
theory. As discussed in section 2.7 to solve for a fixed point we simply
set $\frac{\partial V}{\partial t}=0$ and integrate (4.80) out from a given
$V(0)$ until we encounter a divergence at $z=z_{c}$. We then invoke Griffiths
analyticity\footnote{and analyticity at the origin} and look for the finite
number of non-divergent solutions [13], \emph{i.e.} ones for which
$z_{c}\rightarrow\infty$. However, due to the fact
that the cutoff function is not chirally invariant, $V$ cannot be assumed even
in $z$. Thus we integrate out for both positive and negative
$z$. Figure 4.6 shows the results for three dimensional four component
theory. For $z>0$ and $V(0)<0$ we find that the
solution to (4.80) quickly diverges whereas all solutions with $V(0)>0$ are
non-divergent and seemingly acceptable fixed points. However, as discussed,
we have to integrate out (4.80) for negative $z$. In this case the
solutions diverge for all $V(0)$ with the exception of the Gaussian and High
Temperature fixed points indicated. 
These qualitative features do not change with $N$. Thus we conclude that
consistent with the large $N$ limit of the LPA to Polchinski's equation we
find no acceptable non-trivial fixed 
points in three dimensions. New divergences in $z_{c}$ are generated for $z<0$ 
at the Gaussian critical dimensions, those for which the $\lambda^{G}_{n}$,
given by (4.78), are marginal:
\begin{equation}
D=\frac{n+1}{n}
\end{equation}
where $n=1,2,\cdots$. Thus, for $D<2$ we
find a new divergence in $z_{c}$ as can be seen from figure 4.7. This
divergence is degenerate with the Gaussian in exactly two dimensions, but
moves away from the Gaussian as we lower the dimension. It is straightforward
to see that this does not correspond to a fixed point as this solution diverges
in the positive $z$ direction. However for $D<\frac{3}{2}$ the divergence
occurs for $V(0)>0$ and corresponds to an acceptable fixed point, labelled
NTFP
\newpage
\begin{figure}[H]
\vspace{-75mm}
\hspace{20mm}
\scalebox{0.36}{\includegraphics*[0pt,0pt][1000pt,1000pt]
{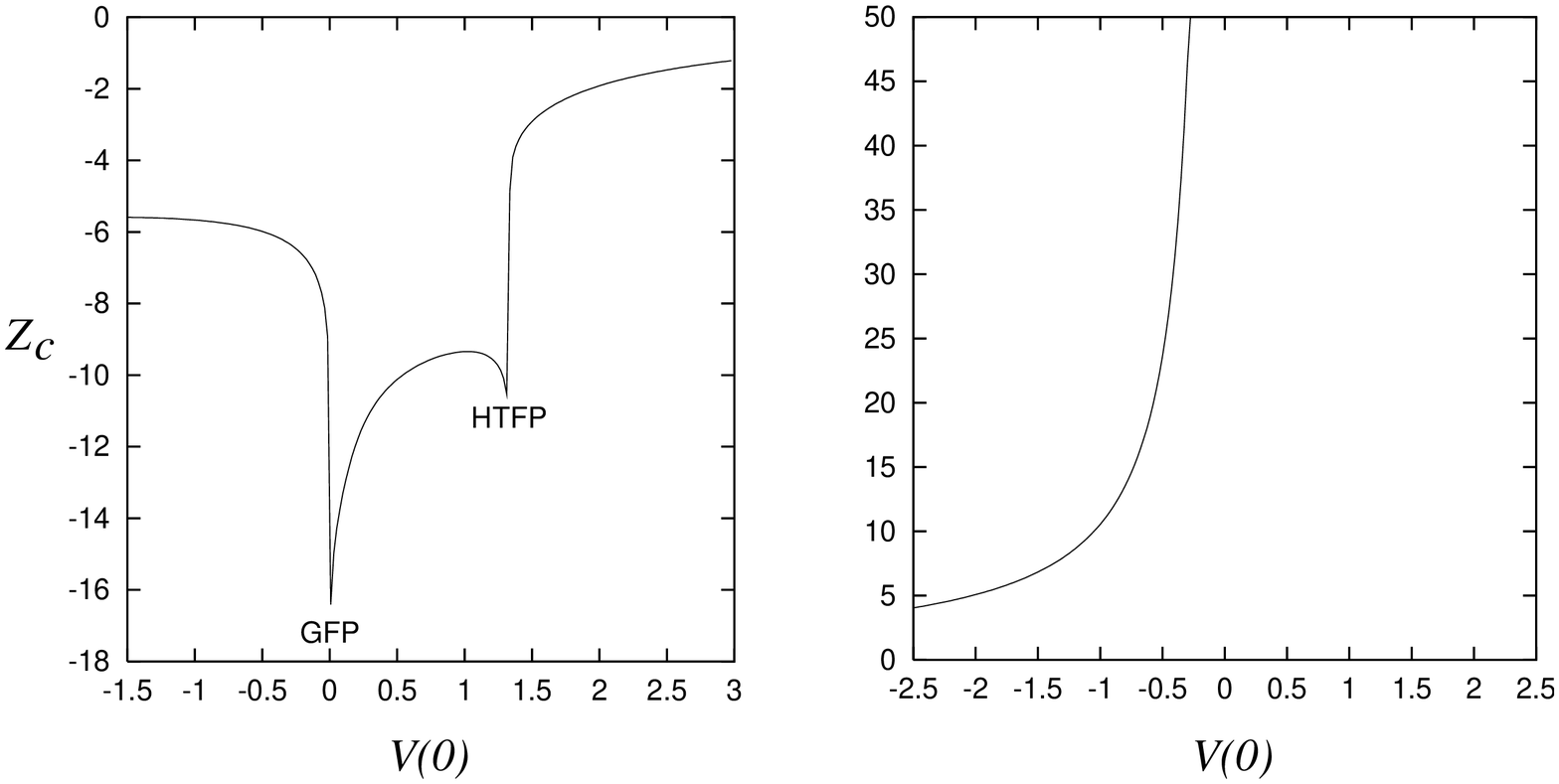}} 
\caption{$z_{c}$ versus $V(0)$ for the Fermionic Polchinski flow equation at
$N=4$ and $D=3$}  
\end{figure}
\begin{figure}[H]
\vspace{-85mm}
\hspace{20mm}
\scalebox{0.36}{\includegraphics*[0pt,0pt][1000pt,1000pt]
{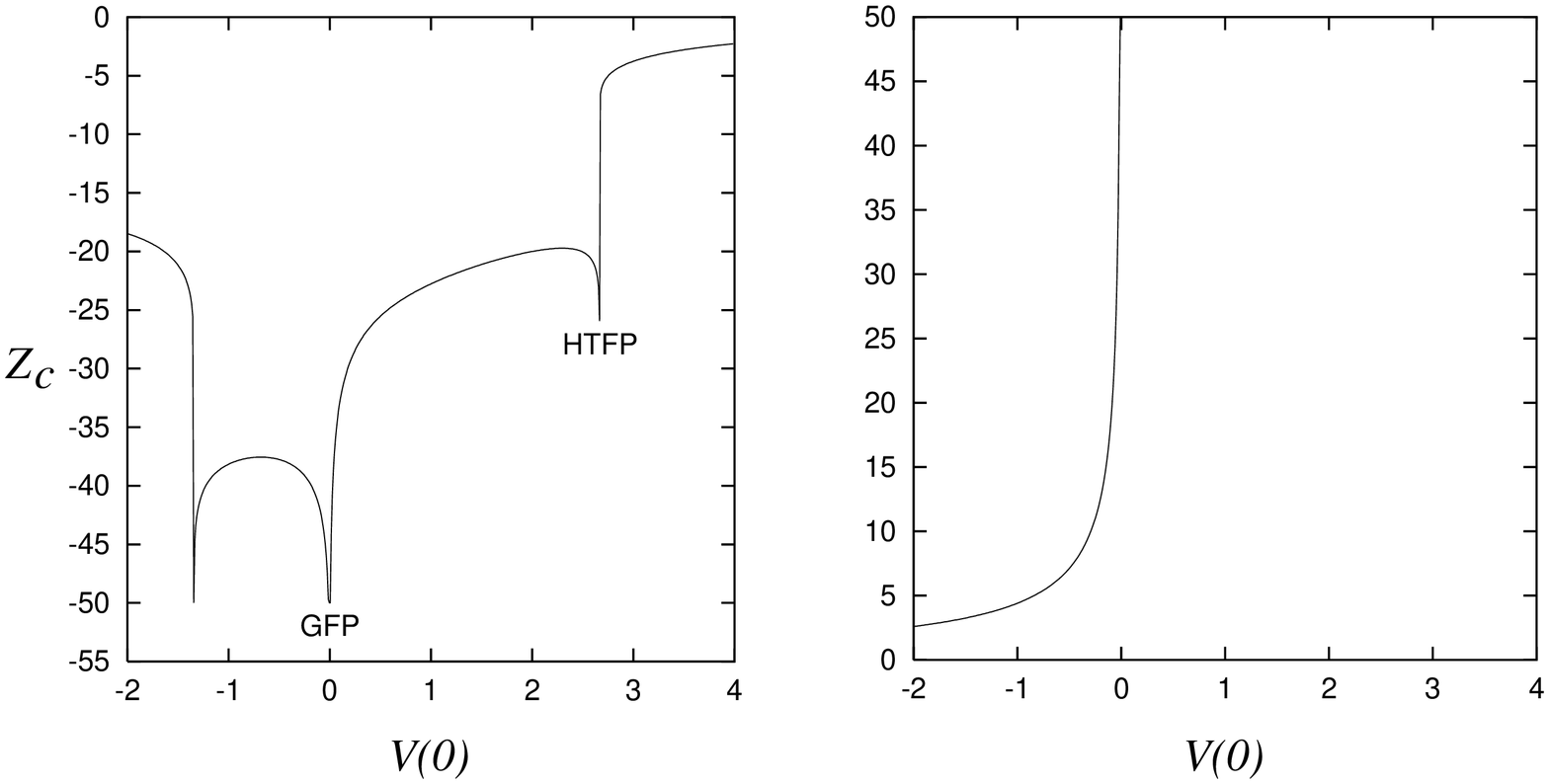}} 
\caption{$z_{c}$ versus $V(0)$ for the Fermionic Polchinski flow equation at
$N=4$ and $D=\frac{3}{2}$}
\end{figure}
\begin{figure}[H]
\vspace{-85mm}
\hspace{20mm}
\scalebox{0.36}{\includegraphics*[0pt,0pt][1000pt,1000pt]
{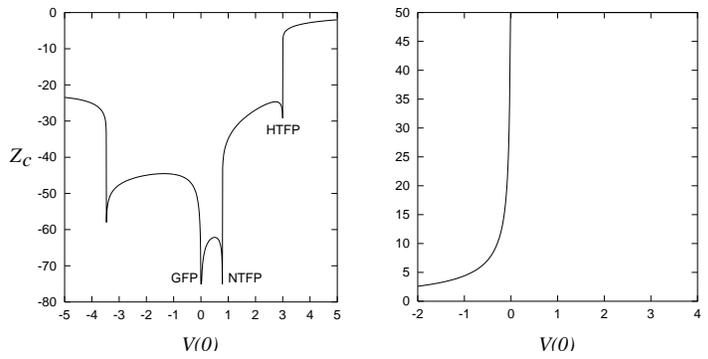}} 
\caption{$z_{c}$ versus $V(0)$ for the Fermionic Polchinski flow equation at
$N=4$ and $D=\frac{4}{3}$}  
\end{figure}
in figure 4.8. This fixed point becomes degenerate with the High
Temperature at
$D=1$, analogous to the first subleading fixed point for scalar
field theory.

\section{Discussion}

For the Polchinski flow equation, we are at liberty to introduce the cutoff in
some chirally symmetric manner, for example by insisting that $C_{UV}$ is a
function of $\frac{q^{2}}{\Lambda^{2}}$ [44]. However, if we then multiply
$C^{-1}_{UV}$ by the kinetic term in the conventional manner, we quickly
realise that 
the right side of the exact flow equation (4.68) always carries a $\sla{q}$,
and thus, is trivial within the Local Potential Approximation. This problem
has been alleviated here by allowing $C_{UV}=C_{UV}(\frac{\sla{q}}{\Lambda})$,
although it should be noted that
the non-trivial LPA to Polchinski's equation can be derived, albeit less
aesthetically, whilst keeping $C_{UV}$ a function of
$\frac{q^{2}}{\Lambda^{2}}$. However, in this case, the 
inverse cutoff function must be introduced in some non-conventional manner,
such as multiplying $\frac{q^{2}}{\Lambda}$, which is then summed with the
kinetic term\footnote{such that the right hand side of the Polchinski equation
contains derivative independent terms}. Hence we might introduce the cutoff as
\begin{equation}
\bar{\psi}\left(\sla{q}+\frac{C^{-1}_{UV}\left(\frac{q^{2}}
{\Lambda^{2}}\right)}{\Lambda}q^{2}\right)\psi,
\end{equation}
which is still afflicted with a chirally non-invariant term.
Thus, in order to obtain a non-trivial LPA
version of (4.68) we are essentially forced to break chiral symmetry with the
cutoff function. To date, there is no known method of including momentum
dependent terms within an expansion scheme for the Polchinski flow equation,
without breaking reparameterisation invariance [15].

Exact analysis in the large $N$ limit shows that only a small range of
symmetry breaking cutoffs reproduce the non-trivial fixed points with the eigenvalue spectrum
associated with the symmetry preserving cutoff. Thus it is not surprising that
non-trivial fixed points are not seen in the LPA to the Polchinski flow
equation. Generally, we interpret the results obtained from (4.80) and (4.81)
as being analogous to letting $\alpha$ become large in (4.47) and (4.48)
such that the chirally invariant component in the latter case is not
dominant. From figure 4.2 the line of fixed points associated with the large 
$N$ limit is seen to vanish for $\alpha^{2} >\frac{1}{3}$. With this
interpretation the results are seen to be entirely consistent. Thus, it has to
be concluded that the hope inspired by the successes 
of scalar field theory discussed in chapters two and three are not yet
fulfilled within Fermionic field theories.

\newpage

\chapter*{Appendices}
\addcontentsline{toc}{chapter}
{\numberline{Appendices}}

\section*{Appendix 4A: Lemma}

We present the lemma [18] used to (4.45):
\begin{equation}
\delta_{\epsilon} f(\theta(p,\Lambda),\Lambda)\rightarrow\delta(\Lambda
-p)\int^{1}_{0}dt f(t,p)
\end{equation}
as $\epsilon\rightarrow 0$, where $f(\theta_{\epsilon},\Lambda)$ is any
function whose dependence on the second argument ($\Lambda$) remains continuous
at $\Lambda = p$ in the limit $\epsilon = 0$. This follows from the identity
\begin{equation}
\delta_{\epsilon} f(\theta_{\epsilon}(p,\Lambda),\Lambda) = \left. \left\{
\frac{\partial}{\partial \Lambda} \int^{1}_{\theta_{\epsilon}(p,\Lambda)} dt
f(t,\Lambda^{\prime}) \right\} \right| _{\Lambda^{\prime}=\Lambda}
\end{equation}
where here $\delta_{\epsilon}(p,\Lambda)=-\frac{\partial}{\partial \Lambda}
\theta_{\epsilon}(p,\Lambda)$. This can be seen by noting that as
$\epsilon\rightarrow 0$ the lower limit on the integral becomes 1 for
$p>\Lambda$ and the integral vanishes, whereas it becomes 0 for
$p<\Lambda$. Thus the integral becomes a step function in $\Lambda$ but with
height $\int_{0}^{1}dtf(t,\Lambda^{\prime})$. Thus differentiating with
respect to $\Lambda$ reproduces (4.90).

\chapter{Concluding remarks}

\section{The Local Potential Approximation}

In the context of scalar field theory, there is little doubt that the Local
Potential Approximation yields reliable results, comparable with other leading
methods (table 2.1). However, we have found that the additional approximation
of truncations, whereby the potential is expanded as a series in the field,
$\phi$, and terminated at some arbitrary finite power is unreliable. In
contrast, standard techniques which involve no further approximation, such
as shooting, are found to be excellent methods of generating the physical
quantities of interest, \emph{e.g.} critical exponents. The new variational
methods presented here provide an equally reliable alternative, with the
advantage that we expect approximate solutions of LPA's for more than one
invariant to be found with relative ease. Furthermore, the $C$-function
defined using the LPA to Polchinski's flow equation provides additional
justification and support for this approximation scheme.  

In the case of Fermionic field theory the Legendre flow equation is
significantly more difficult to deal with than that for scalar field theory,
at least at finite $N$. However, within the LPA, the Polchinski flow equation
is very similar to the corresponding scalar case. Thus, it is unfortunate that
the LPA to the
Polchinski equation fails to produce non-trivial fixed points for Fermionic
field theory, due to our cutoff function breaking chiral symmetry. It should
be considered whether re-introduction of scalars might lead to non-trivial fixed
points once more. The generalisation of (2.33) and (4.74) follows immediately,
\begin{eqnarray}
\frac{\partial V}{\partial t} &=& DV -\frac{1}{2}(D-2) \phi_{i} \frac{\partial
V} {\partial \phi_{i}} - \left( \frac{\partial V}{\partial \phi_{i}}\right)^{2}
+ \frac{\partial^{2} V}{\partial \phi_{i}^{2}} \\ \nonumber 
    &-& \frac{1}{2}(D-1)\left(\psi^{\phantom{a}}_{\alpha a} \frac{\partial V}
{\partial\psi^{\phantom{a}}_{\alpha a}} +
\bar{\psi}_{a}^{\alpha}\frac{\partial V} 
{\partial \bar{\psi}_{a}^{\alpha}}\right)+\frac{\partial V}
{\partial\psi^{\phantom{a}}_{\alpha a}} 
\frac{\partial V}{\partial \bar{\psi}_{a}^{\alpha}}-\frac{\partial^{2} V}
{\partial\psi^{\phantom{a}}_{\alpha a}\partial\bar{\psi}_{a}^{\alpha}}
\end{eqnarray}
where $i$ is the flavour index for the scalar fields and the notation used for
the Fermion fields is consistent with chapter four. However the work presented
here indicates that ultimately it might be necessary to focus on other flow
equations, such as the Legendre equation.

\section{Remaining issues}

Given that the LPA has been found to be competitive, both in terms of
simplicity of implementation and the numerical estimates of the exponents, at
least 
for scalar field theories, it is hoped that ultimately Fermionic field
theories will pose nothing more than a problem of technique. An interesting
question which we have not answered is whether there exists
some method of extracting results relevant to chirally invariant theories from
those presented here. We have recently initiated some interesting work
analogous to that found in ref. [45] which is due to be published alongside
the work presented in chapter four.

However, one of the major remaining obstacles is to deal with gauge
theories. It is found that introducing a momentum cutoff destroys gauge
symmetry and ultimately leads to flow equations with gauge dependent
parameters. Clearly, it is difficult to make reliable 
physical predictions when ultimately they are gauge dependent. 
The most desirable solution would be to introduce the cutoff in some gauge
independent way.   
These problems have been addressed by many authors and recently, significant
progress has been made [46].

\section{Alternative methods}

A variety of other techniques with which we can deal with the renormalization
group have been developed by a host of other authors. Some of the
corresponding numerical estimates are compared with those due to the Local
Potential Approximation in section 2.11. Although a detailed analysis of these
methods is beyond the scope of this thesis, for completeness we include some
references to the $\epsilon$-expansion [1,16,42], lattice methods [47-49],
perturbation theory [50-53] and large $N$ expansions [1,42,43,54].

\chapter*{Bibliography}
\addcontentsline{toc}{chapter}
{\protect\numberline{Bibliography}}

\begin{tabular}{rl}

[1] & J. Zinn-Justin, \emph{Quantum Field Theory and Critical Phenomena},
Clarendon \\ 
& Press, Oxford, 1989. \\

[2] & D. Amit, \emph{Field Theory, the Renormalization Group and Critical
Phenomena}, \\
& World Scientific, 2nd edition, 1984. \\

[3] & S. Weinberg, \emph{Critical phenomena for field theorists}, in Lectures,
Erice Sub- \\
& nucl. Phys., page 1, 1976. \\

[4] & J. Cardy, \emph{Scaling and Renormalization in Statistical Physics},
Cambridge Uni- \\
& versity Press, 1996. \\

[5] & N. Goldenfeld, \emph{Lectures on phase transitions and the
renormalization group}, \\
& Addison-Wesley, 1992. \\

[6] & P. de Gennes, Phys. Lett.\textbf{38A}, (1972) 339. \\

[7] &  C. Yang and T. Lee, Phys. Rev. \textbf{87} (1952) 404. \\

[8] & L. Kadanoff, Rev. Mod. Phys. \textbf{49} (1977) 267. \\

[9] & K. Wilson, Scientific American August 1979. \\

[10] & J. F. Nicoll, T. S. Chang and H. E. Stanley, Phys. Rev. \textbf{A13}
(1976) 1251. \\

[11] & G. Zumbach, Nucl. Phys. \textbf{B413} (1994) 754;
Phys. Lett. \textbf{A190} (1994) 225. \\

[12] & T. R. Morris, Nucl. Phys. \textbf{B458} [FS] (1996) 477. \\

[13] & R. D. Ball \emph{et. al.}, Phys. Lett. \textbf{B347} (1995) 80. \\

[14] & T. R. Morris and M. Turner, Nucl. Phys. \textbf{B509} (1998) 637. \\

[15] & J. Comellas, Nucl. Phys. \textbf{B509} (1998) 662.

\end{tabular}

\begin{tabular}{rl}

[16] & K. G. Wilson and J. Kogut, Phys. Rep. \textbf{12C} (1974) 75. \\

[17] & J. Polchinski, Nucl. Phys. \textbf{A9} (1984) 269. \\

[18] & T. R. Morris, Int. J. Mod. Phys. \textbf{A9} (1994) 2411. \\

[19] & T. R. Morris, Phys. Lett. \textbf{B334} (1994) 355. \\

[20] & K-I. Aoki, K. Morikawa, W. Souma, J-I. Sumi and H. Terao,
Prog. Theor. \\
& Phys. \textbf{99} (1998) 451. \\

[21] & G. Zumbach, Nucl. Phys. \textbf{B413} (1994) 754. \\

[22] & J. Generowicz, C. Harvey-Fros and T. R. Morris,
Phys. Lett. \textbf{B407} (1997) \\
& 27. \\

[23] & M. D'Attanasio and T. R. Morris, Phys. Lett. \textbf{B409} (1997) 363. \\

[24] & A. B. Zamalodchikov, JETP Lett. \textbf{43} (1986) 730. \\

[25] & S. Forte and J. I. Latorre, Nucl. Phys. \textbf{B535} (1998) 709.\\

[26] & J. L. Cardy, Phys. Lett. \textbf{B215} (1988) 749. \\

[27] & H. Osborn, Phys. Lett. \textbf{B222} (1989) 97. \\

[28] & I. Jack and H. Osborn, Nucl. Phys. \textbf{B343} (1990) 647. \\

[29] & A. Cappelli \emph{et. al.}, Nucl. Phys. \textbf{B352} (1990) 647. \\

[30] & D. Freedman \emph{et. al.}, Mod. Phys. Lett. \textbf{A6} (1991) 531. \\

[31] & G. M. Shore, Phys. Lett. \textbf{B253} (1991) 380; \textbf{B256} (1991)
407.  \\

[32] & A. H. Castro Neto and E. Franklin, Nucl. Phys. \textbf{B400} (1993)
525. \\

[33] & P. Haagensen \emph{et. al.}, Phys. Lett. \textbf{B323} (1994) 330. \\

[34] & X. Vilasis-Cardona, Nucl. Phys. \textbf{B435} (1995) 635. \\

[35] & A. C. Petkou, Phys. Lett. \textbf{B359} (1995) 101. \\

[36] & V. Periwal, Mod. Phys. Lett. \textbf{A10} (1995) 1543. \\

[37] & A. A. Belavin, A. M. Polyakov and A. B. Zamalodchikov,
Nucl. Phys. \textbf{B241} \\
& (1984) 333. \\

[38] & See \emph{e.g.} I. S. Gradshteyn and I. M. Ryzhik, \emph{Tables of
integrals, series, and} \\
& \emph{products}, Academic press, 1980. \\

[39] & J. Gaite, Phys. Rev. Lett. \textbf{81} (1998) 3587.

\end{tabular}

\begin{tabular}{rl}

[40] & T. R. Morris, Phys. Lett. \textbf{B345} (1995) 139. \\

[41] & D. Gross and A. Neveu, Phys. Rev. \textbf{D10} (1974) 3235. \\

[42] & K-I. Aoki, K. Morikawa, W. Souma, J-I Sumi, and H. Terao,
Prog. Theor. \\
& Phys. \textbf{95} (1996) 409. \\

[43] & J. Zinn-Justin, Nucl. Phys. \textbf{B367} (1991) 105. \\

[44] & J. Comellas, Y. Kubyshin and E. Moreno, Nucl. Phys. \textbf{B490}
(1997) 653. \\

[45] & P. Ginsparg and K. Wilson, Phys. Rev. \textbf{D25} (1982) 2649. \\ 

[46] & T. R. Morris, hep-th/9810104. \\    

[47] & L. Kadanoff, W. G\"{o}tze, D. Hamblen, R. Hecht, E. Lewis,
V. Palciauskas, M. \\
& Rayl and J. Swift, Rev. Mod. Phys. \textbf{39} (1967) 395. \\

[48] & P. Bastiaansen and H. Knops, Phys. Rev. E \textbf{17} (1998) 3784. \\

[49] & I. Montvay and G. M\"{u}nster, \emph{Quantum Fields on a lattice},
Cambridge Univ- \\
& ersity Press, 1st edition, 1994. \\

[50] & E. Brezin and G. Parisi, J. Stat. Phys. \textbf{19} (1978) 269. \\

[51] & H. Kleinert, Phys. Rev. D \textbf{57} (1998) 2264. \\

[52] & G. Baker, B. Nickel and D. Meiron, Phys. Rev. B \textbf{17} (1978)
1365. \\

[53] & S. Antonenko and A. Sokolov, Phys. Rev. E \textbf{51} (1995) 1894. \\

[54] & V. Irkhin, A. Kanin and M. Katsnelson, Phys. Rev. B \textbf{54} (1996)
11953. 

\end{tabular}

\newpage
\bibliography{/home/users/chf/template/thesis}
\newpage

\pagestyle{empty} 
\begin{center} 
\vspace*{10cm} 
\emph{Ha...HaHa...Ha}
\end{center}

\end{document}